\documentclass[12pt]{report}

\usepackage{amsmath,amsfonts}
\usepackage{physics}
\usepackage{braket}
\numberwithin{equation}{chapter}

\usepackage{abstract}
\usepackage{color}
\usepackage{comment}
\usepackage[driverfallback=dvipdfm]{hyperref}
\usepackage[margin=25truemm]{geometry}
\usepackage{cite}
\usepackage{remreset}
\usepackage{authblk}
\makeatletter\@removefromreset{footnote}{chapter}\makeatother

\usepackage[pdftex]{graphicx}
\usepackage{subfigure}

\newcommand{\fix}[1]{\textcolor{red}{[#1]}}

\newcommand{\Slash}[1]{{\ooalign{\hfil/\hfil\crcr\(#1\)}}}
\newcommand{\Sch}{Schr\"{o}dinger }

\title{Study of Curved Domain-wall Fermions on a Lattice}

\author{Shoto Aoki}
\affil{Department of Physics, Osaka University, Toyonaka 560-0043, Japan}
\begin{document}
\maketitle

\begin{abstract}




In this thesis, we consider fermion systems on square lattice spaces with a curved domain-wall mass term. In a similar way to the flat case, we find massless and chiral states localized at the wall. In the case of $S^1$ and $S^2$ domain-wall embedded into a square lattice, we find that these edge states feel gravity through the induced spin connection. In the conventional continuum limit of the higher dimensional lattice, we find a good consistency with the analytic results in the continuum theory. We also confirm that the rotational symmetry is recovered automatically.

    
We also discuss the effect of a $U(1)$ gauge connection on a two-dimensional lattice fermion with the $S^1$ domain-wall mass term. We find that the gauge field changes the eigenvalue spectrum of the boundary system by the Aharanov-Bohm effect and generates an anomaly of the time-reversal ($T$) symmetry. Our numerical evaluation is consistent with the Atiyah-Patodi-Singer index, which describes the cancellation of the $T$ anomaly by the topological term on the bulk system. When we squeeze the flux inside one plaquette while keeping the total flux unchanged, the anomaly inflow undergoes a drastic change. The intense flux gives rise to an additional domain wall around the flux. We observe a novel localized mode at the flux, canceling the $T$ anomaly on the wall instead of the topological term in the bulk.


We apply the study to a problem in condensed matter physics. 
It is known that inside topological insulators, a vortex or monopole acquires a fractional electric charge and transforms into a dyon. Describing the topological insulator as a negative mass region of a Dirac fermion, we provide a microscopic description of this phenomenon in terms of the dynamical domain-wall creation. 
    
\end{abstract}

\newpage

\section*{Acknowledgment}

First I thank Prof. Hidenori Fukaya, for his invaluable guidance throughout my five years of graduate school. I appreciate his insightful perspective, constructive suggestions, unwavering support, thoughtful guidance, encouragement, and profound expertise in lattice gauge theory. 
Next, I would like to thank Prof. Tetsuya Onogi for his unwavering support throughout my graduate school journey. He not only imparted his profound knowledge of physics but also actively contributed to our daily research discussions. I was greatly inspired by his deep insights into physics. 
I am also deeply grateful to Prof. Mikito Koshino for consistently engaging in discussions and sharing his profound expertise in condensed matter physics. I take pride in having co-authored a paper with him. 
I am sincerely grateful to my collaborators, Dr. Naoto Kan, and Dr. Yoshiyuki Mastuki, for our fruitful collaboration.
I am indebted to mathematicians Prof. Mikio Furuta and Prof. Shinichiroh Matsuo for their valuable insights and opinions on mathematics.
I am grateful to the faculty members of the particle physics theory group at Osaka University, Prof. Tatsuma Nishioka, Prof. Shinya Kanemura, Prof. Satoshi Yamaguchi, Prof. Ryosuke Sato, Prof. Minoru Tanaka, Prof. Norihiro Iizuka, Prof. Kei Yagyu and Dr. Okuto Morikawa, for their support of my research activities. 
I would like to express special appreciation to the dedicated secretaries, Mika Umetani, Satomi Suzuki, Kazumi Asano, and Akiko Takao, for their significant support throughout my student life.
I received generous support from my colleagues in my Laboratory, Mr. Takanori Anegawa, Mr. Yushi Mura, Mr. Hiroki Wada and Mr. Soichiro Shimamori. Their consistent help and encouragement during challenging times have greatly enriched my college experience.
During a part of my Ph.D. program, I have been supported by JST SPRING, Grant Number JPMJSP2138 and JSPS KAKENHI Grant Number JP23KJ1459. Finally, I have come to realize more than ever how much I owe to my parents. I would like to express my heartfelt thanks to my family for their understanding and unwavering support.

\newpage

\tableofcontents

\chapter{Introduction}

Lattice gauge theory is a mathematically rigorous regularization of quantum field theory (QFT). In this framework, a Minkowski space is Wick rotated into an Euclidean space and discretized into a finite volume square lattice space with the periodic boundary condition. 
With finite degrees of freedom on a lattice, the path integral becomes a simple multiple integral. 
It enables us to deal with QFT non-perturbatively, which is an essential tool for computing a strong interaction of quarks and gluons among hadrons.

On the other hand, it is challenging to construct dynamical or background gravity on a square lattice. 
If one attempts to place a square lattice in each local inertial frame, the transition function must preserve the lattice on each frame. However, this poses difficulties in inherently curved spaces where constructing such a transition function becomes impractical. 
For instance, if a two-dimensional lattice is spread along the latitudinal and longitudinal lines on a two-dimensional sphere, the lattice shape is close to a rectangle near the equator but is distorted to a triangle around the north and south poles.


In previous works \cite{Hamber2009Quantum,Regge1961general,brower2016quantum,AMBJORN2001347Dynamicallytriangulating,Brower2017LatticeDirac,Catterall2018Topological,ambjorn2022topology}, a triangular lattice has been employed. This choice is rooted in the mathematical insight that a triangular lattice can be placed on any curved space. In the literature, they made attempts to represent background or dynamical gravity using the angles and lengths of these triangles. This methodology has also been applied in simulations of quantum gravity.


However, achieving a continuum limit in triangular lattice approaches is more challenging compared to that on a square lattice.
It is not easy to find a systematic way to make a finer lattice space from the given one, requiring the simultaneous control of many parameters: angles, areas, lengths of sites etc. Besides, restoring the original continuous symmetry in curved space is non-trivial. 
In the reference \cite{Brower2015Quantum}, it is emphasized that non-trivial and local terms are necessary to recover rotational symmetry on a two-dimensional sphere.


In this work, we investigate a domain-wall fermion system and explore the formulation of lattice theories with background gravity. Relaying on a mathematical theorem by Nash \cite{Nash1956TheImbedding}, which asserts that every curved space can be isometrically embedded into some higher-dimensional Euclidean spaces, we demonstrate an isometrical embedding of a curved space into a flat Euclidean space. In this embedding, the given metric on the curved space is equal to the metric induced from the flat metric on the Euclidean space. Additionally, the vielbein, spin structure, and spin connection are naturally induced up to a local Lorentz (or spin rotational) gauge degrees of freedom. In this framework, there is no need to consider gluing local inertial frames.

According to Einstein's equivalence principle, if particles are constrained on the embedded subspace, they experience gravity through the induced spin connection. Given that the total space is a flat Euclidean space, we can regularize it by a flat square lattice. 
Furthermore, we demonstrate that the gravitational effect can be captured from the dynamics of these localized modes. Similar to standard lattice gauge theory, the continuum limit is achieved by simply decreasing the lattice spacing. When the submanifold possesses rotational symmetry, the symmetry is automatically recovered in the continuum limit due to a $90$-degree rotational symmetry of the lattice action. 


Similar attempts have been studied and observed in condensed matter physics. In the context of a free non-relativistic quantum system, the dynamics of particles bound on a subspace were discussed in \cite{JENSEN1971586Quantummechanics,daCosta1982Quantummechanics}. This discussion was extended in the presence of an external electromagnetic field in \cite{Pershin2005Persistent,Ferrari2008Schrodinger}. The relativistic case was also considered in \cite{BRANDT20163036Diracequation,Matsutani1992Physicalrelation,Matsutani1994TheRelation,Matsutani1997Aconstant,Burgess1993Fermions}. These works found that localized modes feel a geometrical potential and a non-trivial spin connection related to the shape of the subspace. Moreover, these geometrical effects are observed on a curved graphene sheet in \cite{Szameit2010GeometricPotential,Onoe2012ObservationofRiemannian}, experimentally. 
In \cite{Lee2009Surface,Imura2012Spherical,Parente2011Spin,Takane2013UnifiedDescription,Ziesen:2023ecu}, a three-dimensional topological insulator with a curved surface was analyzed, revealing that edge-localized modes feel gravity through the induced spin connection.

A topological insulator is described by a fermion system with a domain-wall \cite{Jackiw1976Solitons,CALLAN1985427Anomalies,KAPLAN1992342AMethod}, which is a one-codimensional space separating a negative and positive mass region. Typically, the negative mass region corresponds to the topological insulator, and the positive region is interpreted as a vacuum. A localized mode on the wall is then expressed as a Jackiw-Rebbi solution \cite{Jackiw1976Solitons}.

We also consider the same $S^1$ domain-wall fermion system on a two-dimensional square lattice space but with a nontrivial $U(1)$ gauge field inside the wall. Due to the Aharanov-Bohm effect, the spectrum of the edge modes is asymmetric in the positive and negative directions, and the anomaly of the time-reversal ($T$) symmetry \cite{Alvarez-Gaume:1984zst} emerges on the wall. We demonstrate that the $T$ anomaly is canceled by the chiral anomaly \cite{Adler1969Axial-Vector,Bell1969APCACpuzzle} on the bulk and the eta invariant of the domain-wall Wilson-Dirac operator is equal to the Aityah-Patodi-Singer (APS) index \cite{Atiyah1975spectral} on the negative mass region \cite{Fukaya_2017Atiyah-Patodi-Singer, fukayaFuruta2020physicistfriendly,FukayaKawai2020TheAPS }. 
This numerical result agrees well with the so-called anomaly inflow \cite{Witten:2015Fermion} in continuum theory.



We also explore the effects induced by a strong $U(1)$ connection, where the intense gauge field manifests as a vortex in a two-dimensional space \cite{Lee:2019rfb,Khalilov:2014rka} and a magnetic monopole in a three-dimensional space \cite{Dirac:1931kp,Yamagishi:1982wpTHE,Yamagishi1983Fermion-monopole,Yamagishi1984Magnetic,WU1976365}. In such cases, the $U(1)$ gauge connection exhibits a singularity, and the Wilson term generates a positive mass region around the vortex or monopole, dynamically altering the bulk structure. However, anomaly inflow still works in a more nontrivial way. An additional zero mode emerges at the new domain-wall, canceling the anomaly on the original wall. This mode is protected by the cobordism nature of the mod two index \cite{AtiyahSinger1971TheIndex5,Witten:1982fp} or Atiyah-Singer (AS) index \cite{Atiyah1963TheIndexOfEllipticOperator,atiyah1968index1}. 

If we interpret the Dirac operator on the three-dimensional space as a Hamiltonian in a $(3+1)$-dimensional spacetime \cite{Imura2012Spherical,Takane2013UnifiedDescription,Rosenberg:2010iaWitteneffect,Qi:2008ewTopologicalFieldTheory} with the monopole, the dynamical creation of the wall provides a microscopic description of the Witten effect \cite{Witten:1979Dyons, Yamamoto2020Magneticmonopoles,Rosenberg:2010iaWitteneffect,Qi:2008ewTopologicalFieldTheory}. This effect asserts that a monopole in a topological insulator acquires a fractional electric charge and becomes a dyon.





The rest of the paper is organized as follows. In section \ref{sec:review}, we review the formulation of the fermion on a square lattice and introduce the Wilson-Dirac fermion. We also review the discussion of a flat domain-wall \cite{KAPLAN1992342AMethod}. In sec. \ref{sec:Curved_conti}, we consider a curved domain-wall fermion system in a continuum space. We find that the low-eigenvalue states are localized at the wall. They feel gravity through the induced connection and the extrinsic contribution is suppressed in the large mass limit. We also try the $S^1$ and $S^2$ domain-wall systems and show that the numerical results agree well with the continuum prediction in sec. \ref{sec:Curved_lat}. In sec. \ref{sec:Anomaly_Inflow}, we consider the $U(1)$ gauge connection in the $S^1$ domain-wall system. We show how the anomaly inflow is detected on the lattice. In sec. \ref{sec:FractionalCharge}, we put a Dirac monopole at the center of $S^2$ domain-wall system and attempt a microscopic description of the Witten effect. In sec. \ref{sec:Summary}, we summarize this article and mention an outlook of our work. In the appendix \ref{app:CLM} and \ref{app:MLM}, we give analytic results for localized modes on the $S^1$ and $S^2$ domain-wall system with a $U(1)$ gauge field.

\begin{samepage}

This thesis is based on the following papers:

\begin{itemize}


    \item S.~Aoki and H.~Fukaya,
    ``Curved domain-wall fermions'',
    PTEP \textbf{2022}, no.6, 063B04 (2022)
    doi:10.1093/ptep/ptac075
    [arXiv:2203.03782 [hep-lat]] \cite{Aoki:2022cwg}.

    \item S.~Aoki and H.~Fukaya,
``Curved domain-wall fermion and its anomaly inflow'',
PTEP \textbf{2023}, no.3, 033B05 (2023)
doi:10.1093/ptep/ptad023
[arXiv:2212.11583 [hep-lat]] \cite{Aoki:2022aezanomalyinflow}.

\item S.~Aoki, H.~Fukaya, N.~Kan, M.~Koshino and Y.~Matsuki,
``Magnetic monopole becomes dyon in topological insulators'',
Phys. Rev. B \textbf{108}, no.15, 155104 (2023)
doi:10.1103/PhysRevB.108.155104
[arXiv:2304.13954 [cond-mat.mes-hall]] \cite{Aoki:2023lqp}.

\end{itemize}
\end{samepage}

\chapter{Review of Domain-wall Fermion Systems and Anomalies}
\label{sec:review}

\section{Continuum Theory}
\label{subsec:Continuum theory}

We first consider a domain-wall fermion system in continuum. We define the partition function regularized by a Pauli-Villars field to describe the anomaly of the time reversal ($T$) symmetry.
We also review the chiral anomaly. 
The $T$ anomaly cancels between bulk and edge through the Atiyah-Patodi-Singer index theorem. This mechanism is called the anomaly inflow \cite{Witten:2015Fermion}. 




\subsection{Massive Fermion System}
\label{subsubsec:Massive}

The Euclidean action of a fermion system on a $d$-dimensional continuum theory is 
\begin{align}
    S= \int d^d x ~\bar{\psi}i\qty( \gamma^\mu D_\mu +m)\psi,
\end{align} 
where $\gamma^\mu ~(\mu=1,\cdots,d)$ are gamma matrices satisfying $\qty{\gamma^\mu, \gamma^\nu}=2 \delta^{\mu\nu}$ and $m>0$ is a fermion mass parameter. Let $A_\mu$ be a gauge field and the covariant derivative $D_\mu$ is written as
\begin{align}
    D_\mu = \pdv{}{x^\mu}- iA_\mu.
\end{align} 
In this article, we only consider a $U(1)$ gauge field. 

The partition function is written by a path integral 
\begin{align}
    Z[A]= \int D\bar{\psi} D\psi~ e^{-S}= `` \det i ( \gamma^\mu D_\mu +m )".
\end{align}
Formally, the path integral is expressed as a determinant of the massive Dirac operator. It consists of the infinite product of the eigenvalue of the operator so that the determinant is not well-defined. In this work, we regularize it by Pauli-Villars fields, which are bosonic (or fermionic) degrees of freedom. The regularized partition function is given by
\begin{align}
    Z_\text{reg}[A]=\lim_{M_{PV}\to \infty} \det \frac{ i(\gamma^\mu D_\mu +m) }{i(\gamma^\mu D_\mu +M_{PV})}=\lim_{M_{PV}\to \infty} \det \frac{ \gamma^\mu D_\mu +m}{\gamma^\mu D_\mu +M_{PV}},
\end{align}
where $M_{PV}$ is the mass term of the PV field\footnote{In fact, we need more Pauli-Villars fields to eliminate the divergence. However, we can describe the anomaly by one PV field only.}.

The correlation function of the Dirac operator is given by
\begin{align}
    \expval{ \psi (x) \bar{\psi} (y)} =\frac{1}{Z[A]} \int D\bar{\psi} D\psi~ \psi(x) \bar{\psi}(y) e^{-S}=-i ( \gamma^\mu D_\mu +m  )^{-1}(x,y).
\end{align}
The inverse of the Dirac operator denotes a propagator or a Green function of the massive Dirac operator.

Next, we consider a domain-wall fermion system on a $(2n+1)$-dimensional Euclidean space $\mathbb{R}^{2n} \times \mathbb{R}$, where our world is the $2n$-dimensional space $\mathbb{R}^{2n}$ and $\mathbb{R}$ is an extra-dimensional direction. Let $x$ and $s$ be a coordinate for $\mathbb{R}^{2n} $ and $ \mathbb{R}$, respectively. We put a domain-wall
\begin{align}
    m(s)= m_0 \text{sign}(s),~\text{sign}(s) =\left\{ \begin{array}{cc}
        +1 & (s>0) \\
        0 & (s=0) \\
        -1 & (s<0)
    \end{array}
    \right.
\end{align}
at $s=0$. Here $m_0$ is a positive parameter.

The action for a fermion system with the domain-wall mass term is given by
\begin{align}\label{eq:action DW fermion Kaplan}
    S_{\text{DW}}= \int d^{2n} x ds~ \bar{\psi} i\qty( \Slash{D} + \gamma^s D_s + m(s))\psi
\end{align} 
and the equation of motion is
\begin{align}
    \qty( \Slash{D} + \gamma^s D_s + m(s))\psi(x,s)=0.
\end{align}
Here $\Slash{D}= \sum_\mu \gamma^\mu D_\mu $ is a Dirac operator in $\mathbb{R}^{2n}$. $\gamma^s$ is anti-commute with $\Slash{D}$ so it is interpreted as a chirality operator of $\Slash{D}$. We assume that $A_\mu \neq 0$ does not depend on $s$ and $A_s=0$. The solution is written as 
\begin{align}
    \psi(x,s)= \eta_+ (x) e^{ - \abs{s} m_0 },
\end{align} 
where $ \eta_+  $ is a zero mode of $\Slash{D}$ and $\gamma^s \eta_\pm= \pm\eta_\pm$. The solution with the positive chirality is localized at the wall and the norm is finite. On the other hand, that with the negative chirality is divergent in the limit of $s\to \infty$ so it is unphysical. The wall localizes only one chiral mode. 

Assuming that the fermion filed $\psi$ is divided as $\psi= \sqrt{m_0} e^{-m_0 \abs{s}} \phi^{\text{edge}}_+ + \psi^\text{bulk}$, we have
\begin{align}
    S_{\text{DW}}= \int_{\mathbb{R}^{2n}} d^{2n}x~ \bar{\phi}^{\text{edge}}_- i\Slash{D} \phi^{\text{edge}}_+ + \int d^{2n} x ds~ \bar{\psi}^\text{bulk} i\qty( \Slash{D} + \gamma^s D_s + m(s))\psi^\text{bulk} \nonumber \\
    +(\text{Interaction terms}). 
\end{align} 
$\psi^\text{bulk}$ oscillates in the $s$ direction and has a mass $m_0$. The first term describes a Weyl spinor and the third term denotes the interaction term among the edge modes $\phi^{\text{edge}}_+ $ and bulk modes $\psi^\text{bulk}$ via nontrivial gauge field background. In the large $m_0$ limit, the bulk modes $\psi^\text{bulk}$ are decoupled from the theory. 

\subsection{Chiral Anomaly}
\label{subsubsec:Chiral anomaly}

We consider a $d=2n$-dimensional massless fermion system whose action is written as
\begin{align}\label{eq:massless fermion action 2n-dim}
    S[\psi , \bar{\psi},A]= \int d^{2n} x ~\bar{\psi}i  \Slash{D} \psi(x),
\end{align} 
where we put $\Slash{D}=\gamma^\mu D_\mu$.
The system does not change under a chiral (or axial $U(1)$) transformation
\begin{align}
    \psi \to \psi^\prime =e^{i\alpha \bar{\gamma}} \psi,~\bar{ \psi} \to \bar{\psi}^\prime=\bar{\psi} e^{i\alpha \bar{\gamma}} ,
\end{align}
where $\bar{\gamma}= (-i\gamma^1 \gamma^2) \cdots (-i\gamma^{2n-1} \gamma^{2n})$ anti-commutes with all gamma matrices and $\alpha$ is a real constant parameter. Naively, it implies that the partition function is invariant under this transformation. However, we need Pauli-Villars fields having a large mass to define a well-defined partition function and they break the chiral symmetry. 
This phenomenon is called a chiral anomaly. 

Adding a Pauli-Villars field to the massless fermion action \eqref{eq:massless fermion action 2n-dim}, we obtain the regularized action and partition function
\begin{align}
    S_\text{reg}[\psi, \bar{\psi}, \phi ,\bar{\phi},A]&=\int  d^{2n} x ~(\bar{\psi}i  \Slash{D} \psi(x) +\bar{\phi}i  (\Slash{D}+M_{PV}) \phi(x) )\\
    Z_\text{reg}[A]&= \int  D\bar{\psi} D\psi  D\bar{\phi} D\phi~ \exp[ -  S_\text{reg}[\psi, \bar{\psi}, \phi ,\bar{\phi},A] ].
\end{align}  
Note that the Pauli-Villars field $\phi$ is a bosonic fermion whose statistics is different from that of $\psi$. Thus, the Jacobian of $ D\bar{\psi} D\psi$ and $ D\bar{\phi} D\phi$ cancel with each other and the measure $ D\bar{\psi} D\psi D\bar{\phi} D\phi$ is invariant under the chiral transformation. We assume that $\alpha =\alpha(x)$ is a real-valued function and small enough, then we have
\begin{align}
    Z_\text{reg}[A]
=& Z_\text{reg}[A]\qty (1+  \int d^{2n}x ~i \alpha (x)\expval{ \partial_\mu J^\mu_{\text{reg}} - 2M_{PV} \bar{\phi} i\bar{\gamma} \phi (x)}  + \order{\alpha^2}),
\end{align}
where $J^\mu_{\text{reg}}$ is a chiral current defined by
\begin{align}
    J^\mu_{\text{reg}}= \bar{\psi}i  \gamma^\mu \bar{\gamma} \psi(x) +\bar{\phi}i \gamma^\mu \bar{\gamma} \phi(x) .
\end{align}
The divergence of the current is
\begin{align}
    \partial_\mu \expval{  J^\mu_{\text{reg}}(x) }
    =& 2 \tr \qty[\bar{\gamma} \frac{M_{PV}^2}{-\Slash{D}^2+M_{PV}^2}](x,x)\label{eq:tr gamma bar} ,
\end{align}
where the trace is taken for the gamma matrices. Taking the limit of $M_{PV}\to \infty$, we obtain
\begin{align}
    \partial_\mu \expval{  J^\mu_{\text{reg}}(x) }= 2  \frac{\epsilon^{\mu_1 \cdots \mu_{2n}}}{(4\pi)^n n!}  F_{\mu_1 \mu_2} \cdots F_{\mu_{2n-1}\mu_{2n}},
\end{align}
which is known as the chiral anomaly.

This result is related to the AS index theorem. We decompose the trace \eqref{eq:tr gamma bar} in the basis of eigenstates of $i\Slash{D}$. Let $\phi_\lambda$ be an eigenstate with eigenvalue $\lambda$ of $i\Slash{D}$, then the propagator is given by
\begin{align}
    \qty[\frac{M_{PV}^2}{-\Slash{D}^2+M_{PV}^2}]_{ab}(x,x)= \sum_{\lambda}   \frac{M_{PV}^2}{\lambda^2+M_{PV}^2}  [\phi_\lambda (x)]_a [\phi_\lambda^\dagger (x)]_b.  
\end{align}
Here the subscripts $a$ and $b$ are spinor indices. Since $\bar{\gamma} $ anti-commutes with $i\Slash{D}$, $\bar{\gamma} \phi_\lambda$ is an eigenstate with $-\lambda$. When $\lambda \neq 0$, the inner product of $ \phi_\lambda$ and $\bar{\gamma} \phi_\lambda$ is zero, then the trace is written as
\begin{align}
    \int d^{2n}x~ \tr \qty[\bar{\gamma} \frac{M_{PV}^2}{-\Slash{D}^2+M_{PV}^2}](x,x)=  \int d^{2n}x~ \sum_{\lambda =0} \frac{M_{PV}^2}{\lambda^2 +M_{PV}^2} \phi_\lambda^\dagger \bar{\gamma} \phi_\lambda (x) =n_+ -n_-, 
\end{align}
where $n_{\pm}$ is the number of the zero modes with $\bar{\gamma}=\pm 1$. The last equation is called the Atiyah-Singer index and we end up with the Atiyah-Singer index theorem:
\begin{align}
    \text{ind}(i\Slash{D})&:=n_+-n_-
    =\frac{1}{(4\pi)^n n!}\int d^{2n}x~\epsilon^{\mu_1 \cdots \mu_{2n}} F_{\mu_1 \mu_2} \cdots F_{\mu_{2n-1}\mu_{2n}} \\
    &=\frac{1}{(2\pi)^n n!}\int F^n. 
 \end{align}
Here $\frac{1}{(2\pi)^n n!} F^n$ is a Chern character, which is related to the topological structure of the fiber bundle. This theorem connects the analytic quantity and the topological invariant.

For a later convenience, we introduce the eta invariant of a Hermitian operator $A$ 
\begin{align}
    \eta(A)= \lim_{s\to 0 }\sum_{ \substack{\lambda \in \text{Spec}(A) \\ \lambda\neq 0 }} \frac{\lambda}{ \abs{\lambda}^{1+s} } + \dim \text{Ker}(A).
\end{align}
Formally, the eta invariant measures the difference between the rank of the positive eigenvalue space and that of the negative eigenvalue space. When $A$ is a finite matrix, the invariant is always an integer. However, the invariant becomes non-integral when $A$ is an operator on an infinite dimensional vector space. The AS index can be expressed by 
\begin{align}\label{eq:AS index eta}
    \text{ind}(i\Slash{D})=- \frac{1}{2} \qty(\eta(H_- )-\eta(H_+ ) ).
\end{align}
Note that $ H_{\pm}=\bar{\gamma}(\Slash{D} \pm m)$ is a Hermitian operator. Since $H_{\pm}$ anti-commutes with $\Slash{D}$, the eta invariants are related to the number of the zero modes of $\Slash{D}$ and take finite values. This massive expression works well in lattice gauge theory where we need additive mass to control the UV region. 

Finally, we address the massive fermion system
\begin{align}
    S[\psi , \bar{\psi},A]= \int d^{2n} x ~\bar{\psi}i  (\Slash{D}+m) \psi(x),
\end{align} 
and we assume that there is the chiral anomaly when $m=0$. The laws of physics are the same when $m$ is positive and when it is negative, but they belong to different phases. Namely, there is no continuous deformation from the negative mass region to the positive mass region while maintaining the gap. To emphasize the regularized partition function depends on $m$, we write it as
\begin{align}
    Z_\text{reg}[A,m]= \det \frac{ \Slash{D} +m}{\Slash{D} +M_{PV}}.
\end{align} 
Then $Z_\text{reg}[A,-m]$ is given by
\begin{align}
    Z_\text{reg}[A,-m]= \det \frac{ \Slash{D} -m}{\Slash{D} +M_{PV}}=Z_\text{reg}[A,m]e^{-i\pi \text{Ind}(i\Slash{D})}
\end{align}
for $m>0$. The ratio of $Z_\text{reg}[A,-m]$ and $Z_\text{reg}[A,+m]$ is equal to the modulo two of the index. If $e^{-i\pi \text{Ind}(i\Slash{D})}=-1$, the deformation must pass through $Z_\text{reg}=0$ corresponding to a gapless phase. 

Although not mathematically rigorous, the result of equation \eqref{eq:AS index eta} can be understood as follows. We rewrite the ratio as
\begin{align} \label{eq:ratio of partition function AS}
    \frac{Z_\text{reg}[A,-m]}{Z_\text{reg}[A,+m]}= \det \frac{ i \bar{\gamma}(\Slash{D} -m)}{i \bar{\gamma}(\Slash{D} +m)}.
\end{align}
The determinant of $i \bar{\gamma}(\Slash{D} +m) $ is formally given by
\begin{align}
\det  i \bar{\gamma}(\Slash{D} +m)= \prod i \lambda = \prod \abs{ \lambda} e^{i\frac{\pi}{2} \text{sign}(\lambda) } =\abs{ \det  i \bar{\gamma}(\Slash{D} +m)} e^{i\frac{\pi}{2} \eta(H_+) }
\end{align}
and we obtain
\begin{align}
   \frac{Z_\text{reg}[A,-m]}{Z_\text{reg}[A,+m]} =\abs{ \frac{Z_\text{reg}[A,-m]}{Z_\text{reg}[A,+m]} } e^{i\frac{\pi}{2} (\eta(H_-) - \eta(H_+))}.
\end{align}

\subsection{Time-Reversal Anomaly}
\label{subsubsec:T anomaly}

Time-reversal ($T$) anomaly or parity anomaly can arise in an odd-dimensional fermion system. We suppose that the Euclidean spacetime is the product $Y$ of $\mathbb{R} \times M^{2n-2}$ or $S^1 \times M^{2n-2}$. For simplicity, the $(2n-2)$-dimensional manifold $M^{2n-2}$ is a flat space. We take a coordinate $(t,x)$, where $x^0=t$ is interpreted as an Euclidean time direction.   

The classical action for a massless spinor is 
\begin{align}
    S[\psi,\bar{\psi},A]=\int_Y dt d^{2n-2}x~ \bar{\psi} i\Slash{D}^Y \psi 
\end{align}
and has a time-reversal symmetry corresponding to $t\to -t$. The transformation is given by
\begin{align}
    A_0(t, x) &\to A_0^T(t,x)= -A_0^T(-t,x), \\
    A_i(t, x) &\to A_i^T(t,x)= A_i^T(-t,x), \\
    \psi(t,x) &\to \psi^T (t,x)=i\gamma^0 \psi(-t,x),\\
    \bar{\psi}(t,x) &\to \bar{\psi}^T (t,x)=\bar{\psi}(-t,x) i\gamma^0.
\end{align}
The action does not change under this transformation:
\begin{align}
    S[\psi^T,\bar{\psi}^T,A^T]=S[\psi,\bar{\psi},A].
\end{align}
Naively, we expect that there is a time-reversal symmetry in the quantum system. However, a regularization can violate this symmetry.

However, the Pauli-Villars mass term is flipped under this transformation:
\begin{align}
    \int  d^{2n} x ~\bar{\phi}i  (\Slash{D}+M_{PV}) \phi(x) \to \int  d^{2n} x ~\bar{\phi}i  (\Slash{D}-M_{PV}) \phi(x),
\end{align}
and the regularized partition function changes to 
\begin{align}
    Z_\text{reg}[A^T]=
    \prod_{\lambda} \frac{ \lambda}{ \lambda -iM_{PV}}=Z[A]_\text{reg}^{\ast},
\end{align}
where $\lambda$ is an eigenvalue of $i\Slash{D}$.
The time-reversal transformation for the partition function is equivalent to the complex conjugate. The Pauli-Villars regulator breaks the time-reversal symmetry if $Z[A]\neq Z[A]^\ast$. The phase factor of $Z[A]$ is called time-reversal ($T$) anomaly. In the large mass limit, the anomaly is calculated as
\begin{align}
    Z_\text{reg}[A]=& \lim_{M_{PV} \to \infty}  \prod_{\lambda} \frac{ \lambda}{ \lambda +iM_{PV}} = \abs{Z_\text{reg}[A]} \prod_{\lambda}  \exp(-i\frac{\pi}{2} \text{sign}(\lambda)) \nonumber \\
    =& \abs{Z_\text{reg}[A]}
     \exp(-i\frac{\pi}{2} \eta(i\Slash{D}^Y)).
\end{align}

It is well-known that the $T$ anomaly is canceled by the chiral anomaly \cite{Witten:2015Fermion}. This cancellation mechanism is called anomaly inflow. We assume that $Y$ is a boundary of the one-dimensional higher space $X$ and the gauge connection $A$ on $Y$ can be extended onto $X$. On the bulk system, we consider the $\theta$ term at $\theta=\pi$ and the partition function
\begin{align}
    Z_{\text{bulk}}[A] =  \exp( i\pi \frac{1}{(2\pi)^n n!}\int_X F^n ).
\end{align}
Note that the field strength $F$ is on $X$ rather than $Y$. However, the integral depends on the configuration of $A$ on $Y$. Since $F^n$ is a closed form on $X$, the integral of $F^n$ turns into
\begin{align}
    \int_X F^n=\int_X d(AF^{n-1})=\int_Y AF^{n-1}. 
\end{align} 
by the Stokes' theorem. The integrand on the right-hand side is called a Chern-Simons term. We, therefore, simply write the field strength on $X$ as $F$. 

We defined a total partition function as
\begin{align}
    Z[X,A]= &Z_\text{reg}[A] \exp( i\pi \frac{1}{(2\pi)^n n!}\int_X F^n ) \nonumber \\
    =&\abs{Z[X,A]} \exp( i\pi \qty[ \frac{1}{(2\pi)^n n!}\int_X F^n -\frac{1}{2} \eta(i\Slash{D}^Y)]).
\end{align}
The phase factor is called the Atiyah-Patodi-Singer (APS) index of the Dirac operator $i\Slash{D}^X$ on $X$ and is defined as
\begin{align}\label{eq:APS index theorem}
    \text{Ind}_{\text{APS}}(i\Slash{D}^X)= \frac{1}{(2\pi)^n n!}\int_X F^n -\frac{1}{2} \eta(i\Slash{D}^Y).
\end{align}
Since it is an integer, the total system has the $T$-symmetry.

\subsection{APS index and Domain-wall Fermion Formulation}
\label{subsubsec:APS}

Here we review the standard and domain-wall fermion formulation of the APS index. The APS index theorem is an extension of the AS index theorem onto a manifold $X$ with boundary $Y$. The Dirac operator $i\Slash{D}^X$ on $X$ is not Hermitian unless an appropriate boundary condition is imposed. We assume that the $X$ and $Y$ are flat manifolds and the form of $X$ is $X= Y  \times \mathbb{R}_-$ for simplicity. $\mathbb{R}_-$ is a half straight line $(-\infty,0]$. For two spinors $\psi_1$ and $\psi_2$, the inner product of $\psi_1$ and $i \Slash{D}^X \psi_2$ is
\begin{align}
    (\psi_1 ,i\Slash{D}^X  \psi_2):= \int_X d^{2n}x ~ \psi_1^\dagger i \Slash{D}^X \psi_2 (x)= \eval{ \int_Y d^{2n-1}y ~ \psi_1^\dagger i \gamma^{2n} \psi_2 (y)}_{x^{2n}=0} +  ( i\Slash{D}^X \psi_1 ,  \psi_2),
\end{align}
where $x=(y,x^{2n})$ and $y$ are coordinates on $X$ and $Y$. $\gamma^{2n}$ is a gamma matrix facing the normal direction to the boundary $Y$. The first term on the right-hand side is the boundary contribution, which breaks the Hermiticity of $i\Slash{D}^X$.

We assume that the gauge connection $A$ does not depends on $x^{2n}$ and $A_1=0$ and take a representation of the gamma matrices as
\begin{align}
    \gamma^a=-\sigma_2 \otimes \tau^a~,\ \gamma^{2n}=\sigma_1 \otimes 1,~ \bar{\gamma}= \sigma_3 \otimes 1 ~(a=1,\cdots,2n-1),
\end{align}
where $\tau^a$ is a gamma matrix on $Y$. Then we have
\begin{align}
    i\Slash{D}^X=i \gamma^{2n}\qty( \gamma^{2n} \gamma^a \nabla_a+ \pdv{}{x^{2n}}   )=i \gamma^{2n}\qty( -i\sigma_3 \otimes \Slash{D}^Y + \pdv{}{x^{2n}}   ) ,
\end{align}
where $\Slash{D}^Y =\tau^a \nabla_a$ is a Dirac operator on $Y$. We put $B=-i\sigma_3 \otimes \Slash{D}^Y  $, then $B$ and $\gamma^{2n}$ anti-commute with each other. Supposing $B$ has no zero mode, we impose a boundary condition at $x^{2n}=0$
\begin{align}
    \frac{ \abs{B}-B}{2} \psi(x^{2n}=0)=0,
\end{align}
which is called the APS boundary condition. This condition kills all the negative eigenstates of $B$ at $x^{2n}=0$. $\gamma^{2n}$ flips the eigenvalue of $B$, then the boundary term vanishes. Since $B$ commutes with the chiral operator $\bar{\gamma}$, this condition is compatible with $\bar{\gamma}$. The APS index is defined as
\begin{align}
    \text{Ind}_\text{APS}(i\Slash{D}^X)= n_+ -n_-,
\end{align}
similar to the AS index. $n_{+}$ and $n_-$ are the number of the zero modes with the positive and negative chirality, respectively. By the APS index theorem, this integer is written as the eq. \eqref{eq:APS index theorem}. Note that there is no normalizable edge mode and the eta invariant comes from bulk modes. However, the APS boundary condition is non-local due to $\abs{B}$. A non-local condition could violate a causality and its physical meaning is unclear. Moreover, it is difficult to extend this formulation onto a lattice theory due to chiral symmetry breaking.

Fukaya, Onogi and Yamaguchi reformulated the APS index by a domain-wall fermion \cite{Fukaya_2017Atiyah-Patodi-Singer}. They embed $X$ into a closed manifold $\hat{X} =Y \times \mathbb{R}$ and consider a domain-wall fermion system
\begin{align}
    S_{\text{DW}}[\psi , \bar{\psi},A]= \int_{ \hat{X}} d^{2n} x ~\bar{\psi}i  (\Slash{D}^{\hat{X}}+\epsilon(x^{2n}) m) \psi(x),
\end{align} 
where $\epsilon(x^{2n})= \text{sign}(x^{2n})$. The negative mass region is the target manifold $X$, which is interpreted as a topological insulator. As well as the previous argument, we expect that the phase factor of the ratio of the regularized partition function 
\begin{align} 
    \frac{Z_\text{reg}[A,\epsilon m]}{Z_\text{reg}[A,+m]}= \det \frac{ i \bar{\gamma}(\Slash{D}^{\hat{X}} + \epsilon m)}{i \bar{\gamma}(\Slash{D}^{\hat{X}} +m)}= \abs{ \frac{Z_\text{reg}[A,\epsilon m]}{Z_\text{reg}[A,+m]} } e^{i\frac{\pi}{2} (\eta(H_\epsilon)- \eta(H_+) )}
\end{align}
is related to the APS index. Let us evaluate
\begin{align}
    -\frac{1}{2} (\eta(H_\epsilon)- \eta(H_+) )
\end{align}
by the zeta function regularization. Here the Hermitian operator $H_\epsilon $ is written as
\begin{align}
    H_\epsilon:= \bar{\gamma}(\Slash{D}^{\hat{X}} + \epsilon m)= \bar{\gamma} \gamma^{2n}\qty( \pdv{}{x^{2n}} + \gamma^{2n} m\epsilon ) + i \sigma_1 \otimes \Slash{D}^Y .
\end{align}
We solve the eigenvalue problem of $H_\epsilon \psi=\Lambda \psi $ to compute the eta invariant. We take an ansatz $\psi^{\text{edge}}(x)=e^{-m \abs{x^{2n}}} u_+ \otimes \phi(y)$, where $u_+$ is a positive eigenvalue mode of $\sigma_1$. This localized mode eliminates the first term of $H_\epsilon $ and we have
\begin{align}
    H_\epsilon \psi= e^{-m \abs{x^{2n}}} u_+ \otimes (i \Slash{D}^Y \phi(y)).
\end{align} 
It means that the domain-wall fermion system describes the boundary system. Let $\lambda$ and $\phi_\lambda$ be an eigenvalue of $i\Slash{D}^Y$ and the corresponding eigenstate, then we find a boundary localized state
\begin{align}
    \psi_{\Lambda=\lambda}^{\text{edge}}(x)= e^{ -m \abs{x^{2n}}} u_+ \otimes \phi_{\lambda} (y).
\end{align}
By using the basis of eigenstates, the eta invariant of $H_\epsilon$ is obtained
\begin{align}
    \eta(H_\epsilon )= \frac{1}{(2\pi)^n n!}\int_{\hat{X}} \epsilon(x^{2n}) F^n +\eta(i\Slash{D}^Y) .
\end{align}
The first term comes from the bulk modes and the eta invariant of $i\Slash{D}^Y $ is generated by the edge modes. On the other hand, the eta invariant of $H_+$ is given by
\begin{align}
    \eta(H_+ )= \frac{1}{(2\pi)^n n!}\int_{\hat{X}}  F^n .
\end{align}
Thus the phase factor of the partition function is equal to the APS index:
\begin{align}\label{eq:APS index eta}
    - \frac{1}{2} \qty(\eta(H_\epsilon )-\eta(H_+ ) )=  \frac{1}{(2\pi)^n n!}\int_{x^{2n}<0}  F^n -\frac{1}{2}\eta(i\Slash{D}^Y) =\text{ind}_{\text{APS}}(i\Slash{D}^X).
\end{align}
This expression is similar to the equation \eqref{eq:AS index eta} and can be extended onto the lattice gauge theory, directly in \cite{FukayaKawai2020TheAPS}.  


\section{Lattice Theory}
\label{subsec:Lattice theory}

In this section, we construct a lattice fermion action. It is known that a naive difference operator has 
$2^d -1$ doublers. We show how a Wilson term eliminates them. We also put a flat domain-wall on a lattice and show that there appear edge-localized states \cite{KAPLAN1992342AMethod}.

\subsection{Wilson Dirac fermion}
\label{subsec:WilsonDirac}
\begin{figure}
\centering
\includegraphics[scale=0.5]{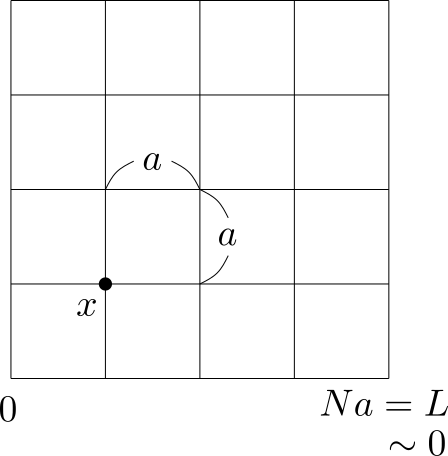}
\caption{A two-dimensional square lattice with a lattice spacing $a$ and the length of one edge $L$. We impose a periodic boundary condition identifying $x^\mu +L$ and $x^\mu$.}
\label{fig:sqaurelattice}
\end{figure}

We introduce a $d$-dimensional square lattice with a lattice spacing $a$ and the length of one edge is $L$. We assume that the ratio $\frac{L}{a}$ is an integer $N$ and impose a periodic boundary condition for all directions (see Figure \ref{fig:sqaurelattice}). The lattice point is denoted by $x=\qty(x^1,\cdots , x^d)$ and $x^\mu~(\mu=1,\cdots ,d)$ takes a discrete value in $\qty{0,a,\cdots, a(N-1)}$. Then, the square lattice is represented by $ \qty(a\mathbb{Z}/ Na\mathbb{Z})^d$.

We put $A_\mu=0$ for simplicity. 
The forward difference operator acts on a spinor field $\psi$ as 
\begin{align}
    (\nabla_\mu \psi)(x)= \psi(x+a \hat{\mu}) -\psi(x) 
\end{align}
for $x\in \qty(a\mathbb{Z}/ Na\mathbb{Z})^d$. $\hat{\mu}$ is a unit vector for $\mu$-direction. The Hermitian conjugate  $\nabla_\mu^\dagger $ is given by
\begin{align}
    (\nabla_\mu^\dagger \psi)(x)= \psi(x-a \hat{\mu}) -\psi(x). 
\end{align}

Since $\pdv{}{x^\mu}$ is an anti-Hermitian operator, we approximate $\pdv{}{x^\mu}$ by $\frac{\nabla_\mu - \nabla_\mu^\dagger}{2a}$ and the naive fermion action is 
\begin{align}
    S_\text{lat}^\prime= a^d \sum_{x\in \qty(a\mathbb{Z}/ Na\mathbb{Z})^d} \bar{\psi} i\qty[ \gamma^\mu \frac{\nabla_\mu- \nabla_\mu^\dagger}{2a} +m] \psi(x).
\end{align}

In the large volume limit $L=Na\to \infty$, an eigenfunction of $\frac{\nabla_\mu- \nabla_\mu^\dagger}{2a}$ is given by $e^{ip_\mu x^\mu} \tilde{\psi}(p)$ and the eigenvalue is $i \sin(p_\mu a)$, where $p_\mu$ is the momentum in the $\mu$-direction. We extend $\psi$ as $\psi(x)= \int_{-\frac{\pi}{a} }^{\frac{\pi}{a}} \qty(\frac{dp}{2\pi} )^d e^{-ipx} \tilde{\psi}(p) $ and obtain
\begin{align}
    S_\text{lat}^\prime={\int_{-\frac{\pi}{a} }^{\frac{\pi}{a}}} \qty(\frac{dp}{2\pi} )^d \bar{\tilde{\psi}} (-p)i\qty( \frac{1}{a} \gamma^\mu \sin (a p_\mu) +im) \tilde{\psi}(p).
\end{align}

We put $s_\mu= \frac{1}{a} \sin(ap_\mu)$, and the propagator for the lattice fermion is given by
\begin{align}
    G^\prime(p)=\frac{1}{ \gamma^\mu s_\mu+im}= \frac{ \gamma^\mu s_\mu-im}{s^2+m^2}. 
\end{align}
The pole near $s^\mu=0$ describes the physical on-shell state with the mass $m$. But we find multiple zeros $s(\bar{p})^2=0$ at
\begin{align}
    \bar{p}_\mu=0 ,~\frac{\pi}{a}.
\end{align}
The other poles but $\bar{p}=0$ have momenta of about $1/a$. 
Such unphysical particles are called ``doublers" and we must eliminate them in the continuum limit.



According to the Nielsen-Ninomiya theorem \cite{Nielsen:1980rz,Nielsen:1981xu}, this fermion doublering occurs when the fermion action satisfies all of the following conditions:
\begin{itemize}
    \item translational symmetry on a lattice,
   \item chiral symmetry of the Dirac operator in the massless limit, or anti-commutation relation with $\bar{\gamma}$,
   \item Hermiticity,
   \item bilinear form of fermions,
   \item locality of interactions.
\end{itemize}
These conditions should be naturally satisfied in a quantum field theory on continuous spacetime, but doublers cannot be avoided in lattice theory that holds these conditions. This is called the doubling problem.


In ref. \cite{Wilson:1975id}, Wilson introduced an additive mass term violating a chiral symmetry:
\begin{align}
   K_W^d=\frac{ra}{2} \sum_{\mu=1}^d \nabla^\dagger_\mu \nabla_\mu= -\frac{ra}{2} \sum_{\mu=1}^d\qty( \nabla^\dagger_\mu+ \nabla_\mu).
\end{align}
This term is called the Wilson term and $r$ is the Wilson parameter. The additional term does not affect the physical pole but it gives an infinite mass to the doublers.

To confirm this, let us write the Wilson fermion action in the momentum space,
\begin{align}
    S_\text{lat}=&\sum_{x\in (a\mathbb{Z}/ L \mathbb{Z})^d}  a^d \bar{\psi} i\qty( \sum_{\mu=1}^d\qty(\gamma^\mu  \frac{\nabla_\mu-\nabla_\mu^\dagger}{2a} +{\frac{r}{2a} \nabla_\mu \nabla_\mu^\dagger})+m)\psi(x) \nonumber \\
    =&\int_{-\frac{\pi}{a} }^{\frac{\pi}{a}} \qty(\frac{dp}{2\pi} )^d \bar{\tilde{\psi}} (p) \qty( \sum_{\mu=1}^d \qty(\frac{1}{a} \gamma^\mu \sin (a p_\mu) +i r\frac{1-\cos (ap_\mu)}{a}) +im ) ,\tilde{\psi}(p)
\end{align}
and propagator
\begin{align}
    G(p)=\frac{1}{ \gamma^\mu s_\mu+im(p)}= \frac{ \gamma^\mu s_\mu-im(p)}{s^2+m(p)^2}. 
\end{align}
Here, $m(p)= r\sum_{\mu=1}^d\frac{1-\cos (ap_\mu)}{a} +m$ gives a mass for zeros $\bar{p}$:
\begin{align}
    m(\bar{p})= \left\{ \begin{array}{cc}
        m & (\bar{p}=0) \\
        \sim \frac{1}{a} & (\bar{p}\neq 0)
    \end{array} \right. .
\end{align}

In the lattice gauge theory, the $U(1)$ gauge field $A_\mu$ is realized as a link variable:
\begin{align}
    U_\mu (x)= \exp[i \int_{x+ a\hat{\mu}}^x A_\nu dx^\nu]. 
\end{align}
Under a gauge transformation 
\begin{align}
    \psi(x) &\to e^{i\phi(x)} \psi(x), \\
    A_\nu (x) &\to A_\nu(x) + \partial_\nu \phi (x),
\end{align}
the link variable changes to
\begin{align}
    U_\mu (x) \to e^{i\phi(x)} U_\mu (x) e^{-i\phi(x+a\hat{\mu})}.
\end{align}

The covariant differential operator is given by
\begin{align}
    (\nabla_\mu ^A\psi)(x) &= U_\mu (x) \psi(x+a \hat{\mu}) -\psi(x) , \\
    (\nabla_\mu^{A\dagger} \psi)(x)&= U_\mu^\dagger (x-a\hat{\mu}) \psi(x-a \hat{\mu}) -\psi(x). 
\end{align}

Therefore, the gauge-invariant lattice action is obtained as
\begin{align}
    S_\text{lat}=&\sum_{x\in (a\mathbb{Z}/ L \mathbb{Z})^d}  a^d \bar{\psi}i\qty( \sum_{\mu=1}^d\qty(\gamma^\mu  \frac{\nabla_\mu^A-\nabla_\mu^{A\dagger}}{2a} +{\frac{r}{2a} \nabla_\mu^A \nabla_\mu^{A\dagger}})+m)\psi(x).
\end{align} 
When there is no fear of confusion, we simply write $\nabla_\mu$ as the covariant derivative $\nabla_\mu^A$ and assume $r=1$ throughout the rest of this article.

\subsection{Flat Domain-wall Fermion on Lattice}
\label{subsec:FlatDW}


In this section, we review Kalpan's discussions \cite{KAPLAN1992342AMethod} for a flat domain-wall and construct the lattice action related to \eqref{eq:action DW fermion Kaplan} on a square lattice space with the lattice spacing $a$. We assume that the length of the $s$-direction is infinity long, but the others are $L$. This lattice space is expressed by $(a\mathbb{Z}/ L \mathbb{Z})^{2n} \times a\mathbb{Z} $.

We consider a flat domain-wall mass term 
\begin{align}
    m(s)= \frac{\sinh(\mu_0 a)}{a} \text{sign}(s)
\end{align}   
at $s=0$, where $\mu_0$ is a positive constant. The lattice action is given by
\begin{align}
    S_\text{lat}=&\sum_{x,s}  a^{2n+1} \bar{\psi} i\left[ \sum_{\mu=1}^{2n}\qty(\gamma^\mu  \frac{\nabla_\mu-\nabla_\mu^\dagger}{2a} +{\frac{1}{2a} \nabla_\mu \nabla_\mu^\dagger}) \right. \nonumber \\
   &\left. + \gamma^{s} \frac{\nabla_{s}-\nabla_{s}^\dagger}{2a} + {\frac{1}{2a} \nabla_{s} \nabla_{s}^\dagger}  +m(s)\right]\psi(x,s),
\end{align}
and the equation of motion is obtained by
\begin{align}\label{eq:eom_flatdw}
    \Slash{D}^{2n}\psi(x,s)=-\left[K_W^{2n}+ \gamma^{s} \frac{\nabla_{s}-\nabla_{s}^\dagger}{2a} + {\frac{1}{2a} \nabla_{s} \nabla_{s}^\dagger}  +m(s)\right]\psi(x,s),
\end{align}
where $\Slash{D}^{2n}=\sum_{\mu=1}^{2n}\gamma^\mu  \frac{\nabla_\mu-\nabla_\mu^\dagger}{2a} $ represents the naive Dirac operator on $(a\mathbb{Z}/ L \mathbb{Z})^{2n}$. 

Let us solve the Dirac equation with $A_\mu=0$ and $A_s=0$ and find the zero mode of $\Slash{D}^{2n}$. 
Putting $\psi(x,s)= e^{ipx} \tilde{\psi}(p,s)$, $\tilde{\psi}(p,s)$ satisfies
\begin{align}
   \qty[ \frac{\gamma^s-1}{2a} \nabla_s -\frac{\gamma^s+1}{2a} \nabla_s^\dagger +m(s)+  \frac{1}{a}F(p) ] \tilde{\psi}(p,s)=0, 
\end{align}
where $F(p) =\sum_\mu\qty(1- \cos(p_\mu a)) \geq 0$. We take an ansatz $\tilde{\psi}(p,s)= \tilde{\phi}_\pm (p) u_\pm (s)$, where $\gamma_s \tilde{\phi}_\pm (p)=\pm \tilde{\phi}_\pm (p)$ and $u_\pm (s)$ are functions of $s$. Then we obtain simple equations 
\begin{align}
    u_+(s- a)= & (1+a m(s) +F(p)) u_+(s), \\
    u_-(s+ a)= & (1+a m(s) +F(p)) u_-(s) 
\end{align}
and solutions
\begin{align}
    u_+(s)=&(1+a m(s) +F(p))^{-s/a} u_+(0), \\
    u_-(s)=&(1+a m(s) +F(p))^{s/a} u_-(0).
\end{align}
In order for these solutions to be square integrable,
\begin{align}\label{eq:con+}
    \abs{ 1 - \sinh(\mu_0 a) +F(p)  } <1 \quad \text{and} \quad \abs{ 1+\sinh(\mu_0 a) +F(p)  } >1,
\end{align}
must be satisfied for $u_+(s)$ and 
\begin{align}\label{eq:con-}
    \abs{ 1 - \sinh(\mu_0 a) +F(p)  } >1\quad \text{and} \quad \abs{ 1+\sinh(\mu_0 a) +F(p)  } <1,
\end{align}
is imposed for $u_-(s)$. Note that both $\sinh(\mu_0 a)= \order{a}$ and $F(p)$ are positive. 

Since $F(p)$ is larger than two for doubler poles, all doublers are decoupled from the low energy theory in the continuum limit.
For the physical pole, $F(p) =\order{a^2}$. Since $ \sinh(\mu_0 a) \simeq \mu_0 a$, only one chiral mode with $\gamma^s=+1$ is localized at the wall. 

This method looks successful in extracting the physical state with $\gamma^s=+1$ in a $2n$-dimensional square lattice space. Practically, we deal with a lattice space with an infinite volume as the limit of a lattice with a finite volume. Since the domain-wall is flat, we need at least two domain-walls. Then the chiral modes appear on each wall, but their chirality is opposite to each other. By the tunneling effect, they are mixed slightly and it is not easy to extract only one of the states with the definite chirality.


\chapter{Curved Domain-wall on Continuum Space}
\label{sec:Curved_conti}


In this section, we consider a fermion system with curved domain-walls in an Euclidean continuum space. Just as the same as the flat case, we will see that massless modes are localized at the curved domain-wall. Moreover, these edge modes feel gravity through the induced connection and the extrinsic contribution is suppressed in the large mass limit.  
We give analytic solutions in the case of the $S^1$ and $S^2$ domain-walls.
These results are compared to the lattice results in the next section. We also discuss the embedding of a Euclidean Schwarzschild space as the junction of two domain-walls.

\section{General Curved Domain-wall}
\label{subsec:Curved_conti_gene}


According to the Nash embedding theorem \cite{Nash1956TheImbedding}, any curved $n$-dimensional manifold $Y$ is isometrically identified as a subspace of a higher-dimensional Euclidean space $X=\mathbb{R}^m$ ($m\gg n$)\footnote{This theorem holds for $m\geq \frac{1}{2}(n+2)(n+5)$ \cite{Gromov1970Embeddingsand}.}. Namely, there is an embedding function $\phi: Y \to X$ such that $Y$ is identified as a subspace defined by
\begin{align} \label{eq:Nash}
    x^I=\phi^I (y^1,\cdots ,y^n)~(I=1, \cdots, m)
\end{align}
and the given metric $g$ on $Y$ is equal to
\begin{align} \label{eq:InducedMetric}
    g_{ij}= \sum \delta_{IJ} \pdv{x^I}{y^i } \pdv{x^J}{y^j}.
\end{align}
Here, $(x^I)$ is the standard coordinate on $X=\mathbb{R}^m$ and $\delta_{IJ}$ is the Kronecker's delta, which is a flat metric on $X$. $(y)$ is a local coordinate on $Y$. If one can constrain a fermion field on $X$ at the subspace, it feels gravity through the induced gravity by the equivalence principle.

In this article, we consider the domain-wall mass term as a potential that confines a particle to the subspace. 
The domain-wall is a boundary where a sign of mass is flipped and a $n$-dimensional (or one-codimensional) manifold in $\mathbb{R}^{n+1}$. 
We take the normal vector in the direction of increasing mass and it defines an orientation of the domain-wall. 

 Not all curved space is constructed as a single domain-wall in a flat space. Some curved space is implemented as a higher codimensional subspace. In this case, we need to introduce multiple domain-walls and the junction is identified as the curved space. 




%



Let $f: X \to \mathbb{R}$ be a smooth function such that 
\begin{align}
    Y=\Set{x \in \mathbb{R}^{n+1} | f(x) =0} \neq \emptyset.
\end{align}
We assume that the total derivative $df$ does not disappear on $Y$. Then $Y$ is an $n$-dimensional hypersurface, which is a boundary separating the region of $f>0$ and $f<0$. Namely, $Y$ is a domain-wall. Capital letters used as a superscript (or subscript) denote the indices in the total space, while lowercase letters indicate those on the domain-wall.

At $p \in Y \subset X$, a tangent vector space $T_p X$ is split as 
\begin{align}
    T_p X \simeq T_p Y \oplus N_p,
\end{align}
where $N_p$ is a normal vector space to $T_p Y$. A vector $\pdv{}{y^a}$ in $T_p Y$ is written as
\begin{align}
    \pdv{}{y^a}=\sum_{I=1}^{n+1} \pdv{\phi^I}{y^a} \pdv{}{x^I}
\end{align}
in this identification. Then, a metric $g$ on $Y$ is induced by the equation \eqref{eq:InducedMetric}. Note that $\pdv{f}{y^a}=0$. On the other hand, $N_p$ is spanned by one vector
\begin{align}
    \text{grad}(f)= \sum_{I=1}^{n+1}\pdv{f}{x^I} \pdv{}{x^I},
\end{align}
which is orthonormal to $\pdv{}{y^a}$.

Vielbein is an orthogonal basis determined by the metric. We can choose a good vielbein of $T_p X$:
\begin{align}
    \{\underbrace{e_1,\cdots,  e_n}_{\text{vielbein of $Y$ }} ,\underbrace{e_{n+1} }_{\text{normal vector}}\}, \label{eq:vielbein}
\end{align}
where $e_I=e_I^{\ J} \pdv{}{x^J}$. The component $e_I^{~J}$ is determined by The Gram-Schmidt process:
\begin{align}
    \begin{aligned}
        e_1&= \frac{e_1^\prime}{\norm{e_1^\prime}} ,~\qty(e_1^\prime= \pdv{}{y^1}) \\
        e_2&=\frac{e_2^\prime}{\norm{e_2^\prime}},~\qty(e_2^\prime= \pdv{}{y^2} -\delta\qty(e_1,\pdv{}{y^2}) e_1) \\
        e_3&= \cdots,
    \end{aligned}
\end{align}
and 
\begin{align}
    e_{n+1}=\frac{1}{ \norm{\text{grad}(f)} } \text{grad}(f),
\end{align}
where $\norm{v}=\sqrt{\delta(v,v)}=\sqrt{\sum_I v^I v^I}$ for a vector field $v=v^I \pdv{}{x^I}$ on $X$. Since $X$ is an Euclidean space, $e_I^{\ J}$ is a matrix in $SO(n+1)$. We write $(e^{-1})_I^{\ J}$ as the inverse matrix of $e_I^{\ J}$. We extend these vectors to a neighborhood of $Y$ while keeping
\begin{align}
    e_a(f)=0 \quad \text{and} \quad \delta(e_I,e_J)=\delta_{IJ}.
\end{align}
In this construction, there is an ambiguity in the $SO(n)$ gauge transformation corresponding to a rotation about the axis $e_{n+1}$. For simplicity, we assume 
\begin{align}
    e_a(  \norm{\text{grad}(f)})=0
\end{align}
around the domain-wall $Y$.

$Y$ is a spin manifold so is $X$. Let $\tilde{\gamma}^1, \cdots , \tilde{\gamma}^n$ be $n$ gamma matrices, then $n+1$ gamma matrices related to $X$ is given by
\begin{align}
    \gamma^a=-\sigma_2 \otimes \tilde{\gamma}^a,\ \gamma^{n+1}=\sigma_1 \otimes 1. 
\end{align}
Note that such gamma matrices are a two-fold irreducible representation of $n+1$-dimensional Clifford algebra and there is a $\mathbb{Z}_2$ grading operator
\begin{align}
    \bar{\gamma}=\sigma_3 \otimes 1,
\end{align}
which anti-commutes with all the gamma matrices. 
The gamma matrix in the normal to the domain-wall $Y$ is written as 
\begin{align}
    \gamma_{\text{normal}}=e_{n+1}^ I \gamma_I =\frac{1}{ \norm{\text{grad}(f)} } \pdv{f}{x^I} \gamma^I .
\end{align}
Here, we use the Kronecker delta to raise or lower indices. In this thesis, the chirality for a spinor $\psi$ is defined as
\begin{align*}
    \matrixel{\psi}{ \gamma_\text{normal}}{\psi}=\int_{\mathbb{R}^{n+1}} d^{n+1}x ~\psi^\dagger  \gamma_\text{normal} \psi .
\end{align*}

The Hermitian Dirac operator on $X$ with the curved domain-wall $Y$ is defined as
\begin{align}
    H=\bar{\gamma}\qty( \Slash{D}+ m\text{sign}(f))=\bar{\gamma}\qty( \sum_{I=1}^{n+1}\gamma^I \pdv{}{x^I}+m\text{sign}(f))
\end{align}
in the standard frame $ \qty(\pdv{}{x^I})$. Replacing the derivative operator $(\pdv{}{x^I})$ with the vielbein $(e_I)$, we obtain
\begin{align}
    H=\bar{\gamma}\qty( \sum_{I}\gamma^I (e^{-1})_I^{~J} e_J +m\text{sign}(f)).
\end{align} 
Here there exists an element  $s= \exp( \frac{1}{4} \sum_{IJ} \alpha_{IJ} \gamma^I \gamma^J ) \in \text{Spin}(n+1)$ satisfied
\begin{align}
    s^{-1} \gamma^J s=  \gamma^I e_{I}^{~J},
\end{align} 
where $\alpha_{IJ}=-\alpha_{JI}$ are real parameters. By the local Lorentz (spin) transformation of $s$, the Dirac operator becomes
\begin{align}
    H \to  H^s= s^{-1}Hs= \bar{\gamma}\qty(  \sum_{I}\gamma^I \qty(e_I + s^{-1}e_I(s))+ m \text{sign}(f)), 
\end{align}
and $\gamma_{\text{normal}}$ also changes to
\begin{align}
    \gamma_{\text{normal}}\to s^{-1}\gamma_{\text{normal}}s= \gamma_{n+1}.
\end{align}

For particles not constrained by the domain-wall, this spin connection $s^{-1} e_I (s) $  is simply a pure gauge field. We put $ \omega_K=s^{-1} e_K (s)=\frac{1}{4}\sum_{KJ}\omega_{IJ,K} \gamma^I \gamma^J $. The component $\omega_{IJ,K}$ is expressed as
\begin{align}
{\omega}_{I J,K}=-\frac{1}{2}\qty( C^I_{J,K}+C^J_{K,I} -C^K_{I,J}),
    \label{eq:connection expressed by a frame}
\end{align}
from the Koszul formula, where $C^{K}_{I,J}$ is a coefficient of the commutator $[e_I,e_J]=e_K C^K_{I,J}$. These are expressed as
\begin{align}
     C^{K}_{I,J}=\delta^{NM} e^{\ N}_K \qty( e_I^L \pdv{e_J^{\ M}}{x^L} - e_J^L \pdv{e_I^{\ M}}{x^L} ).
\end{align}
The Dirac operator becomes
\begin{align}
    \begin{aligned}
        H^s
        =\bar{\gamma} \gamma^a\qty(e_a+\frac{1}{4}\sum_{bc} \omega_{bc,a} \gamma^b\gamma^c+ \frac{1}{2}\sum_{b}\omega_{b\ n+1,a}\gamma^b \gamma^{n+1}) \\
       +\bar{\gamma}\qty(\gamma^{n+1}\qty(e_{n+1}+\frac{1}{4}\sum_{bc}\omega_{bc,n+1} \gamma^b\gamma^c+\frac{1}{2} \sum_{b}\omega_{b\ n+1,n+1} \gamma^b \gamma^{n+1} )+m\text{sign}(f)).
    \end{aligned}
\end{align}
$ \frac{1}{4}\sum_{bc} \omega_{bc,a} \gamma^b\gamma^c$ corresponds to a non-trivial spin connection on $Y$ to be induced. The other components $D^s$ also contains $\frac{1}{2}\sum_{b}\omega_{b\ n+1,a}\gamma^b \gamma^{n+1}$, $ \frac{1}{4}\sum_{bc}\omega_{bc,n+1} \gamma^b\gamma^c$ and $\frac{1}{2} \sum_{b}\omega_{b\ n+1,n+1} \gamma^b \gamma^{n+1}$ describe the extrinsic information. 

First, $ \frac{1}{4}\sum_{bc}\omega_{bc,n+1} \gamma^b\gamma^c$ vanishes by a spin rotation about the axis $e_{n+1}$. We take a new coordinate $t$ along an integral curve of $e_{n+1}$ so that $t$ represents the distance from $Y$.
We simply write
\begin{align}
    e_{n+1}=\pdv{}{t}
\end{align}
and $\text{sign}(f)=\text{sign}(t)$. 
\begin{align}
    L(y,t) &= P\exp\left[\frac{1}{4}\sum_{bc}\gamma^b\gamma^c
        \int_0^t dt'\omega_{bc,n+1}(y,t')  \right],
\end{align}
where $P$ is the path-ordered product, $\omega_{bc,n+1}$ is absent in $L(y,t)H^s L(y,t)^{-1}$. This spin rotation also changes the other terms. However, $L$ is equal to an identity operator on $Y$ and does not change the intrinsic connection on the wall.

Next, we compute $\frac{1}{2} \sum_{b}\omega_{b\ n+1,n+1} \gamma^b \gamma^{n+1}$. The commutator of a horizontal vector $e_b$ and the normal vector $e_{n+1}$ is expressed as
\begin{align}
    [e_b,e_{n+1}]&=C^c_{b,n+1}e_c+C_{b,n+1}^{n+1}e_{n+1}.
\end{align}
Since $e_b(f)=0$, $e_{n+1}(f)=\norm{\text{grad}(f)}$ and $e_b \qty(  \norm{\text{grad}(f)})$, we have 
\begin{align}
0=C^{n+1}_{b,n+1} \norm{\text{grad}(f)}.
\end{align}  
Note that $\norm{\text{grad}(f)}$ is nonzero around $Y$. Then the component is obtained by 
\begin{align}
    \omega^{b}_{\ n+1,n+1}&=-\frac{1}{2} \qty(C^{b}_{n+1,n+1}+C^{n+1}_{n+1,b}-C^{n+1}_{b,n+1})=C^{n+1}_{b,n+1} =0.
\end{align}

Finally, we consider $\frac{1}{2}\sum_{b}\omega_{b\ n+1,a}\gamma^b \gamma^{n+1}$. $\omega_{b\ n+1,a}$ is a symmetric under exchanging of $a$ and $b$:
\begin{align}
    \omega_{a~n+1,b}-\omega_{b~n+1,a}=C^{n+1}_{a,b}=0.
\end{align}
The symmetric tensor $h_{ab}=\omega_{b\ n+1,a}$ is known as the second fundamental form or shape operator\footnote{$h_{ab}$ is an extrinsic quantity determined by the way of embedding. However, the determinant of $h_{ab}$ only includes the intrinsic information when $Y$ is a two-dimensional surface. This fact is called Gauss's Theorema Egregium (Latin for ``Remarkable Theorem") and motivates modern geometry including Riemannian geometry.}. $\frac{1}{n}\tr h = \frac{1}{n} \sum_a h_{aa}$ is called the mean curvature or extrinsic curvature\footnote{In general relativity, we add this term to the Einstein-Hilbert action as Gibbons-Hawking-York boundary term \cite{York:1972sj,Gibbons:1976ue}.}. Now we obtain
\begin{align}
    H^s
    =& \bar{\gamma}\gamma^a\qty( e_a+\frac{1}{4}\sum_{bc}\omega_{bc,a} \gamma^b\gamma^c    ) \nonumber \\
    &+\bar{\gamma} \qty(\gamma^{n+1} \qty(e_{n+1} -\frac{1}{2} \tr h  ) +m\text{sign}(f))
\end{align}
This expression is consistent with the previous works \cite{BRANDT20163036Diracequation,Matsutani1992Physicalrelation, Matsutani1994TheRelation,Matsutani1997Aconstant,Burgess1993Fermions}. The first term is the Dirac operator on $Y$ and the second term denotes an operator for $e_{n+1}$ or $t$ direction. A spinor field on $Y$ is only defined locally but $s$ determines a transition function and spin structure. In this case, this spinor field is induced by that on $X$ so the spin structure corresponds to a trivial element of the spin bordism group. 

$H^{s}$ can be written as
\begin{align}
    H^{s}
    =&\mqty(m \text{sign}(t) & i\Slash{D}^Y+\pdv{}{t}-\frac{1}{2} \tr h \\i\Slash{D}^Y-\pdv{}{t}+ \frac{1}{2} \tr h & - m \text{sign}(t) ),
\end{align}
where $\Slash{D}^Y=\sum_{a} \tilde{\gamma}^a \qty(e_a + \frac{1}{4}\sum_{bc}\omega_{bc,a} \tilde{\gamma}^b \tilde{\gamma}^c)$ is the Dirac operator on $Y$. 
In this expression, a spinor field $\psi$ on $X$ is written by two spinor fields $\chi_1 $ and $\chi_2$: 
\begin{align}
    \psi^{s}=\mqty( \chi_1 \\ \chi_2).
\end{align}

Let us consider the eigenvalue problem of $H \simeq H^{s}$ in the large $m$ limit. Since the mass term is a function of $t$, a low-energy state must be a zero mode of 
\begin{align}
    H_{\text{normal}}^{s}=\bar{\gamma} \qty(\gamma^{n+1}\qty( e_{n+1} - \frac{1}{2} \tr h ) +m \text{sign}(t))=\mqty(m \text{sign}(t) & +\pdv{}{t} - \frac{1}{2} \tr h \\-\pdv{}{t}+  \frac{1}{2} \tr h & - m \text{sign}(t) ).
\end{align}
The solution, which is also the eigenstate of $\gamma_{n+1}=\sigma_1 \otimes 1$, is given by
\begin{align}
    \psi^{s}= e^{-m \abs{t}}\left[\exp( \int_0^t dt^\prime   \frac{1}{2} \tr h (y,t^\prime)) \mqty(\chi(y) \\ \chi(y))+\order{t}\right].
\end{align}

If $\chi(y)$ is an eigenstate of $i \Slash{D}^Y|_{t=0}$ with the eigenvalue $\lambda$, the eigenstate and eigenvalue of $H\simeq H^{s}$ converge to this localized mode and eigenvalue $\lambda$. To see this, we take the difference between $H^{s} \psi^{s}$ and $\lambda \psi^{s}$:
\begin{align}
    (H^{s}-\lambda) \psi^{s}=  e^{-m\abs{t}}e^{\int_0^t dt^\prime \frac{1}{2} \tr h (y,t^\prime)} \underbrace{\qty( \frac{1}{2} \int_0^t dt^\prime \gamma^a e_a(    \tr h (y,t^\prime)) )}_{\order{t}} \mqty( \chi(y) \\ \chi(y) ).
\end{align}
In the  limit of $m\gg \abs{ \tr h }$, we can estimate the residual as
\begin{align}
    \norm{(H^{s}-\lambda)\psi^{s}}  \leq \frac{C}{m},
\end{align}
where $\norm{\ast}$ is a norm of the spinor on $X$. Since $C$ is a positive real number and independent of $m$, this error disappears in the large $m$ limit. Thus, the eigenstate and eigenvalue of $H$ are approximated by
\begin{align}
    \psi= s^{-1}  e^{-m \abs{t}}\left[\exp( \int_0^t dt^\prime  \frac{1}{2} \tr h (y,t^\prime)) \mqty(\chi(y) \\ \chi(y))+\order{t}\right]
\end{align} 
and $\lambda$ in the large $m$ limit. This mode is also an eigenstate of $\gamma_\text{normal}=+1$.

Remarkably, the effective Dirac operator for the edge modes does not depend on the extrinsic curvature. This is in contrast with the \Sch \cite{JENSEN1971586Quantummechanics,daCosta1982Quantummechanics}, Maxwell \cite{Exner1989, Szameit2010GeometricPotential} and Klein-Gordon \cite{Jalalzadeh:2004uv} equations, where the extrinsic curvature appears as the geometrical potential even for fields constrained to a cured space. The reason is that the geometrical potential comes from the second derivative term in the normal direction \cite{BRANDT20163036Diracequation}.




In the presence of a gauge field, the Hermitian Dirac operator is defined as
\begin{align}
    H= \bar{\gamma} D = \bar{\gamma} \qty[\sum_{I} \gamma^I \qty(\pdv{}{x^I}-iA_I) + m\text{sign}(f)]
\end{align}  
and the edge modes obey the Dirac operator on the domain-wall $Y$
\begin{align}
    \Slash{D}^Y=\sum_{a} \tilde{\gamma}^a \qty(e_a -iA_a + \frac{1}{4}\sum_{bc}\omega_{bc,a} \tilde{\gamma}^b \tilde{\gamma}^c),
\end{align}
where $A_a= e_{a}^I A_I$.

Solving the Hermitian Dirac operator $H$ with a curved domain-wall mass term in a flat Euclidean space, we can detect the gravitational effect on the edge-localized fermion at the wall. The spin structure of them corresponds to the trivial element of the spin bordism group. 
The spectrum of the edge modes reflects the spin connection. 
In the following two examples, we demonstrate how to extract the gravity on edge-localized modes. 

\section{$S^1$ Domain-wall}
\label{subsec:Curved_conti_S1}

We analytically solve the eigenvalue problem of the $S^1$ domain-wall fermion system embedded into a two-dimensional Euclidean space with a $U(1)$ connection $A$ :
\begin{align}
    H&=\sigma_3 \qty(\sigma_1 \qty(\pdv{}{x}-iA_1) +\sigma_2 \qty(\pdv{}{y}-iA_2) + m\epsilon) ,
\end{align}
where $\epsilon=\text{sign}(r-r_0)$ assigns a domain-wall and $m$ is a positive constant. We set a $U(1)$ gauge field
\begin{align} \label{eq:U1 flux on S1}
    A= \left\{ \begin{array}{ll}
        \alpha \qty(-\frac{y}{r_1^2} dx+ \frac{x}{r_1^2} dy)=\alpha
\frac{r^2}{r_1^2}d\theta & (r<r_1) \\
    \alpha \qty(-\frac{y}{r^2} dx+ \frac{x}{r^2} dy) =\alpha d\theta& (r>r_1)
    \end{array} \right. 
\end{align}
around the center of the domain-wall. It makes a field strength $F_{12}= \frac{ 2\alpha }{r_1^2}$ for $r <r_1$ and zero for $r>r_1$. The real constant $\alpha$ denotes the total flux divided by $2\pi$. We assume $r_1\ll r_0$. The edge-modes have little support of wave function in the region $r<r_1$ but they feel the Aharanov-Bohm effect. The Hermitian Dirac operator becomes 
\begin{align}
    H=\sigma_3 \mqty(m\epsilon & e^{-i\theta}(\pdv{}{r}-\frac{i}{r}\pdv{}{\theta} - \frac{\alpha}{r}) \\
    e^{i\theta}(\pdv{}{r}+\frac{i}{r}\pdv{}{\theta}+ \frac{\alpha}{r} )& m\epsilon
    ) \label{eq:Hermitian Dirac operator for S^1 in R^2}
\end{align}
in the polar coordinate. 

$H$ commutes with the total angular momentum operator $J=-i\pdv{}{\theta}+\frac{1}{2} \sigma_3$ whose eigenvalue $j$ is a half-integer. Assuming the functional form $\psi= \mqty( f(r)e^{i(j-\frac{1}{2})\theta } \\ g(r) e^{i(j+\frac{1}{2})\theta })$, the eigenvalue problem $H\psi=E \psi$ becomes
\begin{align}
    E\mqty( f\\ g)=  \mqty(m\epsilon  & \pdv{}{r} + \frac{j+\frac{1}{2}- \alpha}{r} \\ -(\pdv{}{r} -\frac{j-\frac{1}{2}- \alpha}{r}) & -m\epsilon ) \mqty( f\\ g).
\end{align}
The edge-localized mode with $H=E~(\abs{E}<m)$ and $J=j$ is given by 
\begin{align}\label{eq:edgeS1}
    \psi^{E,j}&=\left\{
\begin{array}{ll}
A \mqty(\sqrt{m^2-E^2} I_{ \abs{j-\frac{1}{2}-\alpha}} (\sqrt{m^2-E^2} r)e^{i(j-\frac{1}{2})\theta }\\ 
         (m+E) I_{\abs{j+\frac{1}{2}-\alpha}} (\sqrt{m^2-E^2} r)e^{i(j+\frac{1}{2})\theta }) & (r<r_0) \\
B\mqty((m+E)K_{j-\frac{1}{2}-\alpha} (\sqrt{m^2-E^2} r)e^{i(j-\frac{1}{2})\theta }\\ 
         \sqrt{m^2-E^2} K_{j+\frac{1}{2}-\alpha} (\sqrt{m^2-E^2} r)e^{i(j+\frac{1}{2})\theta }) & (r>r_0),
\end{array}
\right.
\end{align}
where $A$ and $B$ are constant, $I_\nu$ and $K_\nu$ denote a first and second kind of modified Bessel function, respectively. 


The continuation condition at $r=r_0$ leads to an equation
\begin{align}\label{eq:condition of E S1}
    \frac{I_{\abs{j-\frac{1}{2}-\alpha}}}{I_{\abs{j+\frac{1}{2}-\alpha}}}\frac{K_{j+\frac{1}{2}-\alpha}}{K_{j-\frac{1}{2}-\alpha}}(\sqrt{m^2-E^2}r_0)=\frac{m+E}{m-E}, 
\end{align}
which determines the eigenvalue $E$. In the large $m$ limit, $E$ converges to 
\begin{align}
    E\simeq \frac{j-\alpha}{r_0}\ \qty(j=\pm\frac{1}{2},\pm\frac{3}{2},\cdots) \label{eq:S1 eigenvalue} .
\end{align}
$\psi^{E,j}$ converges to a positive eigenstate of
\begin{align}\label{eq:gamma normal S1}
    \gamma_\text{normal}=\frac{1}{r}(x\sigma_1+y\sigma_2)
\end{align}
and is localized at $r_0$: $\psi^{E,j} \sim e^{-m \abs{r-r_0}}$. The remaining constants $A$ and $B$ are also determined up to a normalization factor. 

Next, we use the method of the previous section to construct the effective Dirac operator. In the standard frame $\qty(\pdv{}{x}, \pdv{}{y} )$, the gamma (Pauli) matrices in the normal direction and the tangent direction are given by
\begin{align}
    \sigma_r = \sigma_1 \cos\theta +\sigma_2 \sin \theta,\ 
    \sigma_\theta = \sigma_2\cos\theta -\sigma_1 \sin\theta.
\end{align}
We change the frame $\sigma_r \to e^{-i\frac{\theta}{2}\sigma_3}\sigma_r e^{i\frac{\theta}{2} \sigma_3}=\sigma_1$ and $\sigma_\theta \to e^{-i\frac{\theta}{2}\sigma_3}\sigma_\theta e^{i\frac{\theta}{2}\sigma_3}=\sigma_2$ by a $\text{Spin}(2)$ rotation. 

Next, however, that the above transformation loses $2\pi $ periodicity of $\theta$. This reflects that we need multiple patches and their transition functions to make the $\theta$ dependence well-defined. Instead, we introduce the additional $U(1)$ transformation or $\text{Spin}^c(1)$ transformation $\psi \to \psi^\prime= e^{i\frac{\theta}{2}\sigma_3} e^{ -i \frac{\theta}{2}} \psi $ to obtain
\begin{align}
    H^{\prime}=&e^{i\frac{\theta}{2} \sigma_3} e^{-i\frac{\theta}{2}} H e^{-i\frac{\theta}{2} \sigma_3} e^{+i\frac{\theta}{2}} \nonumber\\
    =& \sigma_3 \qty(\sigma_1 \qty(\pdv{}{r}+\frac{1}{2r})  + m\epsilon) -i\sigma_1 \frac{1}{r} \qty(\pdv{}{\theta} +i\frac{1}{2} -i\alpha ),
\end{align}
which maintains the periodicity.
$-i \alpha$ expresses the Aharanov-Bohm effect on the edge modes from the background $U(1)$ gauge field. On the other hand, $i \frac{1}{2}$ is a $\text{Spin}^c(1) \simeq U(1)$ connection and comes from the spin structure induced at the $S^1$ domain-wall\footnote{In mathematically, we should distinguish $U(1)$ and $\text{Spin}^c(1)$. Because the spin bordism group with $U(1)$-bundle is $\Omega^{\text{spin}}_{1}(BU(1)) \simeq \mathbb{Z}_2$ but the spin$^c$ bordism group is $\Omega^{\text{spin}^c}_{1} (\qty{pt}) \simeq 0$.}.  

In the large $m$ limit, edge-localized modes are the zero modes of the operator $\sigma_3\qty(\sigma_1 \qty(\pdv{}{r}+\frac{1}{2r})  + m\epsilon)$ expressed by
\begin{align}
    \psi^\prime= \frac{e^{-m \abs{r-r_0}}}{\sqrt{r}} \mqty( \chi(\theta) \\ \chi(\theta)),
\end{align}
where $\chi(\theta)$ is a periodic function on the circle. The effective operator for $\chi(\theta)$ is
\begin{align}\label{eq:effective Dirac operator S1}
H^{S^1}_{\text{eff}}=  \frac{1}{r_0} \qty( -i\pdv{}{\theta} +\frac{1}{2} -\alpha ).
\end{align}
By comparing to \eqref{eq:S1 eigenvalue}, $-i\pdv{}{\theta} +\frac{1}{2} $ corresponds to the total angular momentum operator $J$. In the above analysis, we have assumed that $r_1$ is sufficiently small. For finite $r_1$, we need more subtle treatment around the flux as presented in our appendix \ref{app:CLM}.

\section{$S^2$ Domain-wall}
\label{subsec:Curved_conti_S2}

We consider a $S^2$ domain-wall fermion system in a three-dimensional Euclidean space with a $U(1)$ gauge field $A$. The radius of the domain-wall is $r_0$ and the Hermitian Dirac operator is given by
\begin{align} 
    H= \bar{\gamma }\qty(\sum_{j=1}^3\gamma^j D_j+ m\epsilon )=\mqty(m \epsilon & \sigma^j (\partial_j-iA_j) \\ -\sigma^j (\partial_j-iA_j) & -m\epsilon) \label{eq:Hermitian Dirac operator for S^2 in R^3},\\ 
    ( \bar{\gamma}=\sigma_3 \otimes 1,\ {\gamma}^j=\sigma_1\otimes \sigma_j) ,
\end{align}
where $\epsilon=\text{sign}(r-r_0)$ and $m$ is a positive mass parameter. Here, we need a two-flavor spinor to define $\bar{\gamma}$.

We assume that the $U(1)$ gauge field $A$ respects the rotational symmetry. Such a gauge field is generated by the magnetic monopole:
\begin{align} \label{eq:gauge conn n-monopole}
    A=n\frac{1-\cos \theta}{2} d\phi
\end{align}
whose topological charge is $n $. This expression is not well-defined around $\theta=\pi$, so we need a transition function $e^{-i n\phi}$ and $n$ is quantized to an integer. The field strength $F=dA= \sum F_{ab} \frac{1}{2}dx^a \wedge dx^b$ is written as
\begin{align}
    F_{ab}=\partial_a A_b- \partial_b A_a= \frac{n}{2} \epsilon_{abc}\frac{x_c}{r^3} 
\end{align}
and it is parallel to the unit vector in the radius direction. The integral of $F$ over the $S^2$ domain-wall is given as 
\begin{align}
    \frac{1}{2\pi} \int_{S^2} F=n.
\end{align}

We solve the eigenvalue problem by using the rotational symmetries. We first discuss angular momentum operators in a non-relativistic system in the presence of the monopole. To make angular momentum operators a gauge covariant, we replace a derivative operator with the covariant derivative operator. Now we get modified angular momentum operators 
\begin{align}
    L_a=-i\epsilon_{abc} x_b (\partial_c -iA_c)-\frac{n}{2} \frac{x_a}{r},
\end{align}
which satisfy commutation relations
\begin{align}
    [L_a,L_b]=i\epsilon_{abc} L_c .
\end{align}
The last term $-\frac{n}{2} \frac{x_a}{r}$ comes from the Lorentz force around the monopole. 
In the polar coordinate, ladder operators and $L_3$ are written as
\begin{align}
    L_{\pm}=&L_1\pm iL_2 =e^{ \pm i\phi } \qty(\pm \pdv{}{\theta} +i \frac{\cos \theta}{ \sin \theta} \pdv{}{\phi} +\frac{n}{2} \frac{\cos \theta -1}{ \sin \theta} ) \\
    L_3=& -i\pdv{}{\phi} -\frac{n}{2}.
\end{align}

The Casimir operator $L^2= \sum_a L_a L_a$ commutes with all angular momentum operators, so there exist simultaneous eigenstates $u_{l,l_3}$ of $L^2$ and $L_3$:
\begin{align}
    L^2 u_{l,l_3}= l(l+1) u_{l,l_3} ,~L_3 u_{l,l_3}=l_3 u_{l,l_3},
\end{align}
where $l>0$ and $l_3=-l,-l+1,\cdots,l$. $l$ and $l_3$ are called the azimuthal quantum number and the magnetic quantum number of $L_a$, respectively. Since the angular momentum operators act on a periodic function, $l_3$ takes a value in $\mathbb{Z}+\frac{n}{2}$. $l$ also has some constraints. The Casimir operator is written by
\begin{align}
    L^2=  (-i\epsilon_{abc} x_b (\partial_c -iA_c))^2 + \qty(\frac{n}{2} )^2
\end{align}
which requires $l(l+1)- \qty(\frac{n}{2})^2 \geq 0$. Thus $l$ takes 
\begin{align}
    l=& \abs{\frac{n}{2}},~ \abs{\frac{n}{2}}+1,~ \abs{\frac{n}{2}}+2,~\cdots.
\end{align}
Note that $l$ and $l_3$ can be a half-integer in the presence of the monopole.

We return to the relativistic system \eqref{eq:Hermitian Dirac operator for S^2 in R^3}. The derivative operator in the off-diagonal component is written as
\begin{align}\label{eq:sigma dot nabla S2}
    \sigma^j (\partial_j -iA_j) =\frac{\sigma^j x^j}{r} \qty(\pdv{}{r} +\frac{1}{r} -\frac{D^{S^2}}{r} ),
\end{align}
where $D^{S^2}$ denotes a derivative operator
\begin{align}\label{eq:DS2}
    D^{S^2}=\sigma^a \qty(L_a +\frac{n}{2} \frac{x_a}{r} )+1.
\end{align}
$D^{S^2}$ and $\frac{x^j \sigma^j}{r}$ anti-commute with each other. 
Let us introduce the total angular momentum
\begin{align}
    J_a=L_a+\frac{1}{2} \sigma_a,
\end{align}
where the second term is a spin operator. The commutation relations are 
\begin{align}
    [J_a, \sigma^j (\partial_j -iA_j ) ]=0, \\
    [J_a ,D^{S^2}]=0, \\
    [J_a, \frac{x^j \sigma_j}{r}]=0, \\
    [J_a,J_b]=i\epsilon^{abc} J_c. 
\end{align}

Let $j$ be an azimuthal quantum number of $J_a$, $j$ takes
\begin{align}
    j=& \abs{\frac{n}{2}}-\frac{1}{2},~ \abs{\frac{n}{2}}+\frac{1}{2},~ \abs{\frac{n}{2}}+\frac{3}{2},~\cdots.
\end{align}
Since $j$ must always be non-negative, a state corresponding to $j= \abs{\frac{n}{2}}-\frac{1}{2}$ does not exist when $n=0$. The degeneracy of the state with $j= \abs{\frac{n}{2}}-\frac{1}{2}$ is $2j+1=\abs{n}$, but that of the others is $2\times (2j+1)$.


We construct a simultaneous eigenstate of $J^2, J_3$ and $D^{S^2}$. When $j\neq \abs{\frac{n}{2}}-\frac{1}{2}$, we find a simultaneous eigenstate $\chi_{j,j_3, \pm}$ satisfying
\begin{align}
J^2 \chi_{j,j_3,\pm}&=j(j+1) \chi_{j,j_3,\pm} ,\\
J_3 \chi_{j,j_3,\pm}&=j_3 \chi_{j,j_3,\pm},\\
D^{S^2} \chi_{j,j_3,\pm}&= \pm \sqrt{\qty(j+\frac{1}{2})^2 -\frac{n^2}{4} } \chi_{j,j_3,\pm},\\
\frac{\sigma^j x^j}{r} \chi_{j,j_3,\pm}&= \chi_{j,j_3,\mp},
\end{align}
where $j_3$ takes a value in $-j ,-j+1,\cdots ,j$. $\pm$ denotes a sign of the eigenvalue of $D^{S^2} $ and is flipped by $\frac{\sigma^j x^j}{r}$. It causes $2\times(2j+1)$-fold degeneracy. When $j=\abs{\frac{n}{2}}-\frac{1}{2}$, the eigenstate $\chi_{j,j_3,0}$ satisfies
\begin{align}
   D^{S^2} \chi_{j,j_3,0}&=0,\\
        \frac{\sigma^j x^j}{r} \chi_{j,j_3,0}&= \text{sign}(n) \chi_{j,j_3,0}.
\end{align}
It gives $2j+1= \abs{n}$-fold degeneracy.

We put $\psi= \frac{1}{\sqrt{r}}\mqty( f(r) \chi \\ g(r) \frac{\sigma^j x^j}{r} \chi)$ ($\chi=\chi_{j,j_3,\pm},~\text{or}~\chi_{j,j_3,0}$) and the equation reduces to
\begin{align}
    E\mqty( f\\ g)=  \mqty(m\epsilon  & \pdv{}{r} + \frac{ \pm \nu +\frac{1}{2}}{r} \\ -(\pdv{}{r} -  \frac{ \pm \nu -\frac{1}{2} }{r}) & -m\epsilon ) \mqty( f\\ g),
\end{align}
where $\nu= \sqrt{\qty(j+\frac{1}{2})^2 -\frac{n^2}{4} }$ is the eigenvalue of $D^{S^2}$. 
For a state with $j\neq \abs{\frac{n}{2}}-\frac{1}{2}$, $\chi=\chi_{j,j_3,\pm}$ and the edge-localized mode is obtained as
\begin{align}\label{eq:EdgeMode S2 }
    \psi_{j,j_3,\pm }&=\left\{
\begin{array}{ll}
     \frac{A}{\sqrt{r}}\mqty( \sqrt{m^2-E^2} I_\abs{\pm \nu -\frac{1}{2} }(\sqrt{m^2-E^2}r) \chi_{j,j_3,\pm} \\(m+E) I_{\abs{\pm \nu +\frac{1}{2}}}(\sqrt{m^2-E^2}r) \frac{\sigma\cdot x}{r}\chi_{j,j_3,\pm} )  & (r<r_0) \\
     \frac{B}{\sqrt{r}}\mqty( (m+E) K_{{\pm \nu -\frac{1}{2} }} (\sqrt{m^2-E^2}r)\chi_{j,j_3,\pm} \\\sqrt{m^2-E^2} K_{{\pm \nu +\frac{1}{2}}}(\sqrt{m^2-E^2}r) \frac{\sigma\cdot x}{r}\chi_{j,j_3,\pm} ) & (r>r_0)
\end{array}
    \right. .
\end{align}
From the continuation condition at $r=r_0$, one has
\begin{align}\label{eq:condition of E S2}
    \frac{I_{\abs{ \pm \nu-\frac{1}{2}}}}{I_{\abs{ \pm \nu+\frac{1}{2}}}}\frac{K_{\pm \nu+\frac{1}{2}}}{K_{\pm \nu-\frac{1}{2}}}(\sqrt{m^2-E^2}r_0)=\frac{m+E}{m-E}, 
\end{align}
which determines the eigenvalue $E$. In the large $mr_0$ limit, the eigenvalue converges to
\begin{align}
    E\simeq \pm \nu= \pm \sqrt{\qty(j+\frac{1}{2})^2 -\frac{n^2}{4} }.
\end{align}
For a state with $j= \abs{\frac{n}{2}}-\frac{1}{2}$, the eigenvalue $E$ is exactly zero in the large $mr_0$ limit. We find an edge-localized state
\begin{align}\label{eq:EdgeMode S2 zero mode}
    \psi_{j,j_3, 0 }&=
     \frac{A}{r} e^{-m\abs{r-r_0}} \mqty(  \chi_{j,j_3,0} \\ \text{sign}(n) \chi_{j,j_3,0} ),
\end{align}
where $A$ is a normalization factor. The edge modes are also an eigenstate of 
\begin{align}
    \gamma_{\text{normal}}= \frac{\gamma^i x^i}{r} =\sigma_1 \otimes \frac{\sigma^i x^i}{r}
\end{align}
with the eigenvalue is unity.

In order to derive the effective two-dimensional dimensional Dirac operator on the edge modes. We perform a gauge transformation by
\begin{align}
    R(\theta,\phi)=\exp(\theta[\gamma^3,\gamma^1]/4)\exp(\phi[\gamma^1,\gamma^2]/4)=1\otimes\exp(i\theta\sigma_2/2)\exp(i\phi\sigma_3/2),
\end{align}
which changes gamma matrices in the $\theta$, $\phi$ and $r$ directions to $\gamma^1$, $\gamma^2$ and $\gamma^3$. $R(\theta,\phi)$ is anti-periodic for $\phi \to \phi+2\pi$. In order to make it a periodic function, we perform a $U(1)$ gauge transformation by $e^{-i\phi}$. Then we get a transformed Hermitian Dirac operator
\begin{align}
    H\to H^\prime=&e^{-i\frac{\phi}{2}} R(\theta,\phi)   H R(\theta,\phi)^{-1} e^{i\frac{\phi}{2}} \nonumber \\
    =&\mqty( \epsilon m & \sigma_3 \qty( \pdv{}{r} +\frac{1}{r}+ \frac{1}{r}\sigma_3 \Slash{D}^{S^2})\\
    -\sigma^3 \qty( \pdv{}{r} +\frac{1}{r}+  \frac{1}{r} \sigma^3\Slash{D}^{S^2}) & -\epsilon m),
\end{align}
where the effective Dirac operator on $S^2$ is given by 
\begin{align}\label{eq:Dirac op on S2}
      \Slash{D}^{S^2}=\qty(\sigma_1 \pdv{}{\theta} +\frac{\sigma_2}{\sin \theta } \qty(  \pdv{}{\phi}+ \frac{i}{2} -\frac{\cos\theta}{2 } \sigma_1 \sigma_2 -i n\frac{1-\cos \theta}{2} ) ) .
\end{align}
$-\frac{\cos\theta}{2 } \sigma_1 \sigma_2  $ is the induced spin connection on $S^2$. The last term $i n\frac{1-\cos \theta}{2} $ is a $U(1)$ gauge connection generated by the monopole. Under the transformation, $D^{S^2}$ and $\gamma_{\text{normal}} $ change to
\begin{align}
    D^{S^2} &\to - \sigma_3 \Slash{D}^{S^2}, \\
    \gamma_{\text{normal}} &\to \sigma_1 \otimes \sigma_3.
\end{align}

\section{Higher Codimensional Space}
\label{subsec:Higher_Codim}

We have shown that the non-trivial spin connection is induced to a single domain-wall. If we install more domain-walls into the flat space, we can construct a more complex and interesting manifold realized by the junction of the walls. In this subsection, we consider a four-dimensional Euclidean Schwarzschild space embedded into a flat space $\mathbb{R}^6$ rather than $\mathbb{R}^{5}$ \cite{Kasner1921Impossibility,Kasner1921}.  

Let $f_a: \mathbb{R}^{n+2} \to \mathbb{R}~(a=1,2)$ be two real-valued functions defining two domain-walls $DW_a= \Set{ x \in \mathbb{R}^{n+2}| f_a(x)=0 }$. Then the junction $Y:=DW_1 \cap DW_2 $ is a $n$-dimensional manifold if $df_1$ and $df_2$ are linearly independent on $Y$. The Hermitian Dirac operator is given by
\begin{align}
    H= i \sum_{I=1}^{n+2} {\gamma}^I\pdv{}{x^I} +\bar{\gamma}_1 m_1\text{sign}(f_1)  +\bar{\gamma}_2 m_2\text{sign}(f_2),  
\end{align} 
where $\gamma^I= \sigma_1 \otimes \sigma_1 \otimes \tilde{\gamma}^I~(I=1,\cdots ,n),~\gamma^{n+1}= \sigma_1 \otimes \sigma_2 \otimes 1,~ \gamma^{n+2}=\sigma_2 \otimes 1 \otimes 1,~\bar{\gamma}_1= \sigma_1\otimes \sigma_3 \otimes 1 $ and $\bar{\gamma}_2= \sigma_3\otimes 1 \otimes 1$ are gamma matrices anti-commuting with each other and $m_1$ and $m_2$ are positive mass parameters. Here $\tilde{\gamma}^i$ is a gamma matrix on $Y$. The gamma matrix in the normal to $DW_a$ is defined by
\begin{align}
    \gamma_{\text{normal}}^a= \frac{1}{ \norm{\text{grad}(f_a)} } \pdv{f_a}{x^I} \gamma^I ~(a=1,2).
\end{align}

In the large $m_1$ and $m_2$ limit, the low-energy modes are localized at $Y$ and are also simultaneous eigenstates of $i\bar{\gamma_a} \gamma_{\text{normal}}^a $
as well as the single domain-wall system. When both of the domain-walls are flat: $f_a=x_{n+a}$, the localized mode is obtained by
\begin{align}
    \psi(x)=e^{-m_1 \abs{x^{n+1}} -m_2 \abs{x^{n+2}}} \mqty(1 \\1 ) \otimes \mqty(1 \\1 ) \otimes {\chi}(x^1 ,\cdots ,x^n),
\end{align}  
where ${\chi}$ is a spinor on $Y$.
This method can be extended to the fermion system with any number of domain-walls and any dimensions.


Finally, we try to embed a four-dimensional Euclidean Schwarzschild space with the inverse temperature $\beta$ into a six-dimensional Euclidean space $\mathbb{R}^6$ \cite{Kasner1921}. The flat metric in $\mathbb{R}^6$ is denoted by
\begin{align}
    ds^2=dx^2+dy^2+dz^2+ dX^2+dY^2+dZ^2,
\end{align}
where $X$, $Y$ and $Z$ are the extra coordinates including the Euclidean time $\tau$ and radius $r= \sqrt{x^2+y^2+z^2}$. Taking a new coordinate $(R,\tau)$ inside $\sqrt{X^2 + Y^2} < \frac{\beta}{2\pi} $:
\begin{align}
    X= \frac{\beta}{2\pi} \frac{ R }{ \sqrt{ R^2 + \qty(\frac{\beta}{2\pi} )^2 } } \sin 2\pi \frac{ \tau }{\beta},~ Y= \frac{\beta}{2\pi} \frac{ R }{ \sqrt{ R^2 + \qty(\frac{\beta}{2\pi} )^2 } } \cos 2\pi \frac{ \tau }{\beta},
\end{align}
then the metric is expressed by
\begin{align*}
    ds^2=dx^2+dy^2+dz^2+ \frac{ R^2 }{  R^2 + \qty(\frac{\beta}{2\pi} )^2 } d\tau^2 + \qty(\frac{ \qty(\frac{\beta}{2\pi} )^2 }{  R^2 + \qty(\frac{\beta}{2\pi} )^2 } )^3 dR^2  +dZ^2.
\end{align*}
Here $ \tau $ has a periodicity with $\beta$: $\tau \sim \tau +\beta $. 
We define a real-valued function
\begin{align}
    F(s):= \int^s_0 ds^\prime~ \sqrt{ 1- \qty(\frac{ \qty(\frac{\beta}{2\pi} )^2 }{  s^{\prime 2} + \qty(\frac{\beta}{2\pi} )^2 } )^3 }
\end{align}
and consider a hypersurface as the first domain-wall
\begin{align}
    f_1:=Z- F(R)=0 \quad \qty( R=\frac{\beta}{2\pi} \frac{\sqrt{ X^2+Y^2} }{ \sqrt{ \qty( \frac{\beta}{ 2\pi})^2 -X^2-Y^2 } }).
\end{align}
Then the metric on the hypersurface $ DW_1=\qty{f_1=0}$ is given by
\begin{align}
    \eval{ ds^2}_{DW_1} = dx^2+dy^2+dz^2+ \frac{ R^2 }{  R^2 + \qty(\frac{\beta}{2\pi} )^2 } d\tau^2 + dR^2.
\end{align}
We introduce the second domain-wall by the following equation
\begin{align}
    f_2:= r- \frac{\beta}{4\pi} - \frac{\pi }{\beta} R^2=r- \frac{\beta}{4\pi} \frac{ \qty(\frac{\beta}{2\pi})^2 }{\qty(\frac{\beta}{2\pi})^2 -X^2-Y^2}=0.
\end{align}
The metric induced to the junction of $DW_1=\qty{f_2=0}$ and $DW_2=\qty{f_2=0}$ is acquired by
\begin{align}
    \eval{ ds^2}_{ DW_1 \cap DW_2} = \qty(1- \frac{\beta}{ 4\pi r}) d\tau^2 + \frac{1}{ 1- \frac{\beta}{ 4\pi r}} dr^2 +r^2 d\Omega^2,
\end{align}
where $d\Omega^2$ is a metric on the unit two-dimensional sphere. Taking $\beta= 8\pi M$, the metric coincides with the Euclidean Schwarzschild metric around a black hole with the mass $M$. 

\chapter{Curved Domain-wall on Square Lattice}
\label{sec:Curved_lat}

We consider a fermion system with a curved domain-wall mass on a square lattice. We study the Wilson Dirac fermion with the $S^1$ domain-wall and $S^2$ domain-wall embedded into one-dimensional higher flat lattice space with trivial link variables. We calculate the eigenvalue problem, numerically. We show that the lattice results are consistent with the continuum prediction and the rotational symmetry is automatically recovered in the continuum limit. 


\section{$S^1$ Domain-wall}
\label{subsec:Curved_lat_S1}

\begin{figure}
\centering
\includegraphics{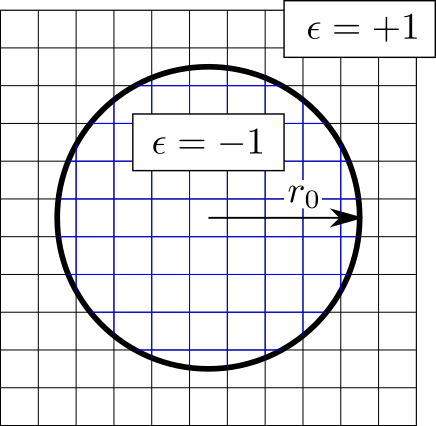}
\caption{A $S^1$ domain-wall with the radius $r_0$ in the two-dimensional square lattice. The blue grid represents the negative mass region and the black one represents the positive mass region.}
\label{fig:DWS1}
\end{figure}

We construct a square lattice space in the same way in sec. \ref{subsec:WilsonDirac}. Let $a$ and $L$ be lattice spacing and the length of the one edge, then the lattice space is identified as $(a \mathbb{Z}/ L \mathbb{Z})^2 $. We assume $N=L/a$ is an even number and the flux $\alpha$ is zero. The Hermitian Wilson-Dirac operator \eqref{eq:Hermitian Dirac operator for S^1 in R^2} is given by
\begin{align}\label{eq:Hermitian Wilson Dirac op of S^1 in R^2}
    H =\frac{1}{a}\sigma_3 \qty(\sum_{i=1,2}\qty[\sigma_i\frac{\nabla_i-\nabla^\dagger_i}{2} +\frac{1}{2}\nabla_i \nabla^\dagger_i ]+\epsilon am ), 
\end{align}
where the $S^1$ domain-wall with the radius $r_0$ by the sign function
\begin{align}\label{eq:S1domain-wall}
    \epsilon= \left\{  \begin{array}{cc}
        -1 & ( (x-x_0)^2+ (y-y_0)^2<r_0^2  ) \\
        +1 & ( (x-x_0)^2+ (y-y_0)^2 \geq r_0^2  ) 
    \end{array}
    \right. ,
\end{align}
where $(x_0,y_0)=( a(N-1)/2, a(N-1)/2)$ is the center of the wall. Since $x_0$ and $y_0$ are half-integers, the center point $(x_0, y_0)$ is located inside one plaquette. The $S^1$ domain-wall is depicted in Figure \ref{fig:DWS1}.

The gamma matrix in the normal direction \eqref{eq:gamma normal S1} on the square lattice is given by
\begin{align}\label{eq:S1 gamma matrix on lattice}
    \gamma_{\text{normal}}= \frac{(x-x_0)\sigma_1 + (y-y_0) \sigma_2}{\sqrt{ (x-x_0)^2+ (y-y_0)^2}},
\end{align}
which is well-defined on all lattice points.
Since the domain-wall is curved, $\gamma_\text{normal}$ depends on a lattice point in $(a \mathbb{Z}/ L \mathbb{Z})^2 $. Note, however, that $\gamma_{\text{normal}}$ is not smooth at the periodic boundaries. 
We expect that for the edge-localized modes, the effect from the boundary is exponentially small when the boundary is far from the domain-wall. 

\begin{figure}
\centering
\includegraphics[bb=0 0 501 309,width=\textwidth]{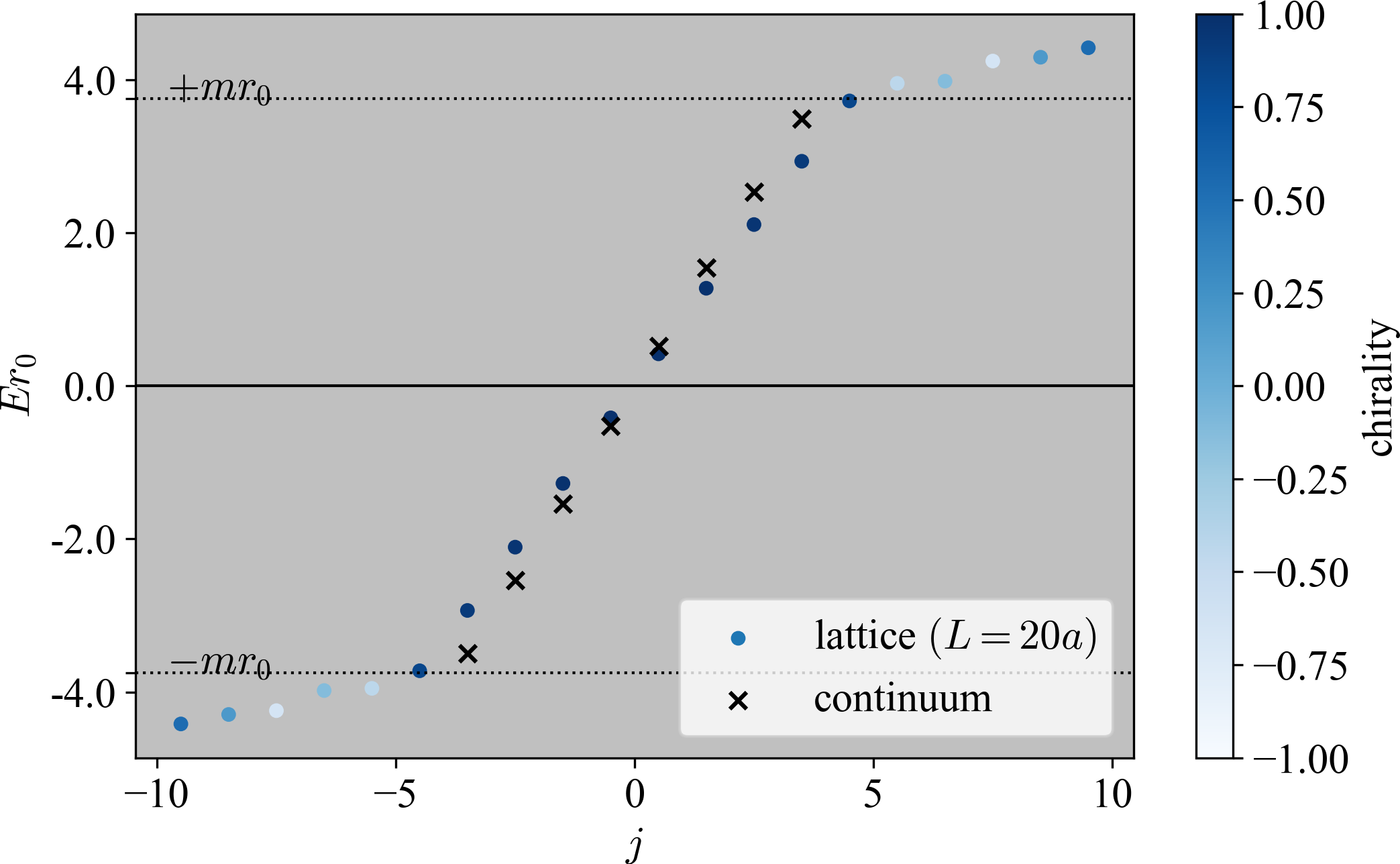}
\caption{The eigenvalue spectrum of the Hermitian Wilson-Dirac operator \eqref{eq:Hermitian Wilson Dirac op of S^1 in R^2} at $L=20a,~r_0=5a$ and $m=15/L$. The color gradation shows the expectation value of \eqref{eq:S1 gamma matrix on lattice}. The filled circles denote the lattice data and cross symbols express the corresponding continuum results.}
\label{fig:S1_eigenvalue}
\end{figure}

We compute the eigenvalue spectrum of the Hermitian Wilson-Dirac operator \eqref{eq:Hermitian Wilson Dirac op of S^1 in R^2} with $L=20a,~r_0=5a=L/4$ and $m=15/L$.  We focus on the edge-localized mode around $E=0$. We plot the lattice data by circle symbols in Figure \ref{fig:S1_eigenvalue} and the cross symbols denote continuum predictions \eqref{eq:condition of E S1}. The color gradation represents the expectation value of \eqref{eq:S1 gamma matrix on lattice}. Here, we label the eigenvalue with a half-integer $j$ like
\begin{align}
    \cdots\leq E_{-\frac{3}{2}}\leq E_{-\frac{1}{2}} \leq 0 \leq E_{\frac{1}{2}} \leq E_{\frac{3}{2}} \leq \cdots.
\end{align}
For the edge modes, $j$ corresponds to the total angular momentum.

\begin{figure}[]
    \begin{center}
     \subfigure{	
     \includegraphics[bb=0 0 448 301, height=0.3 \textheight]{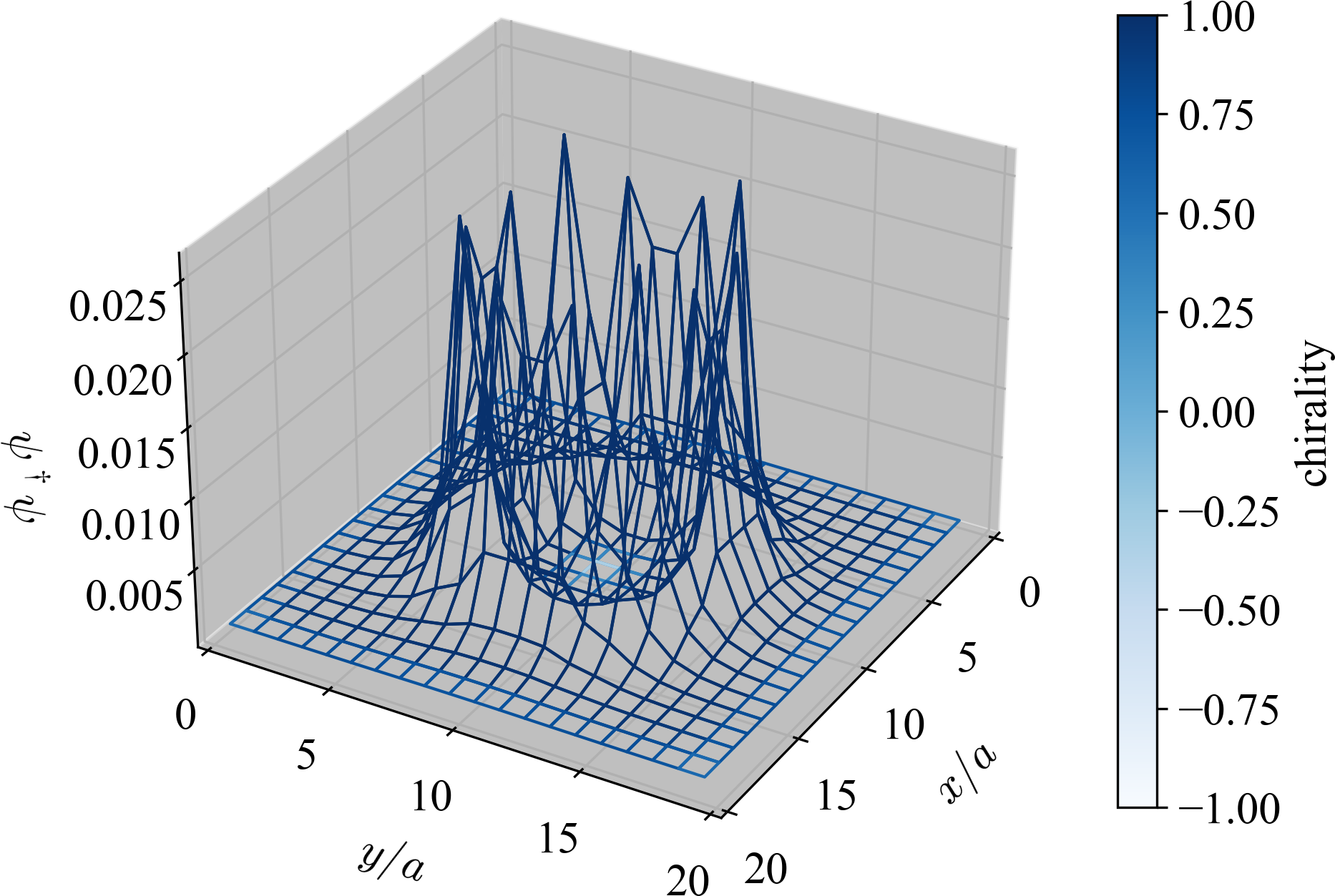}
     }\\ 
     \subfigure{
        \includegraphics[bb=0 0 448 301, height=0.3 \textheight]{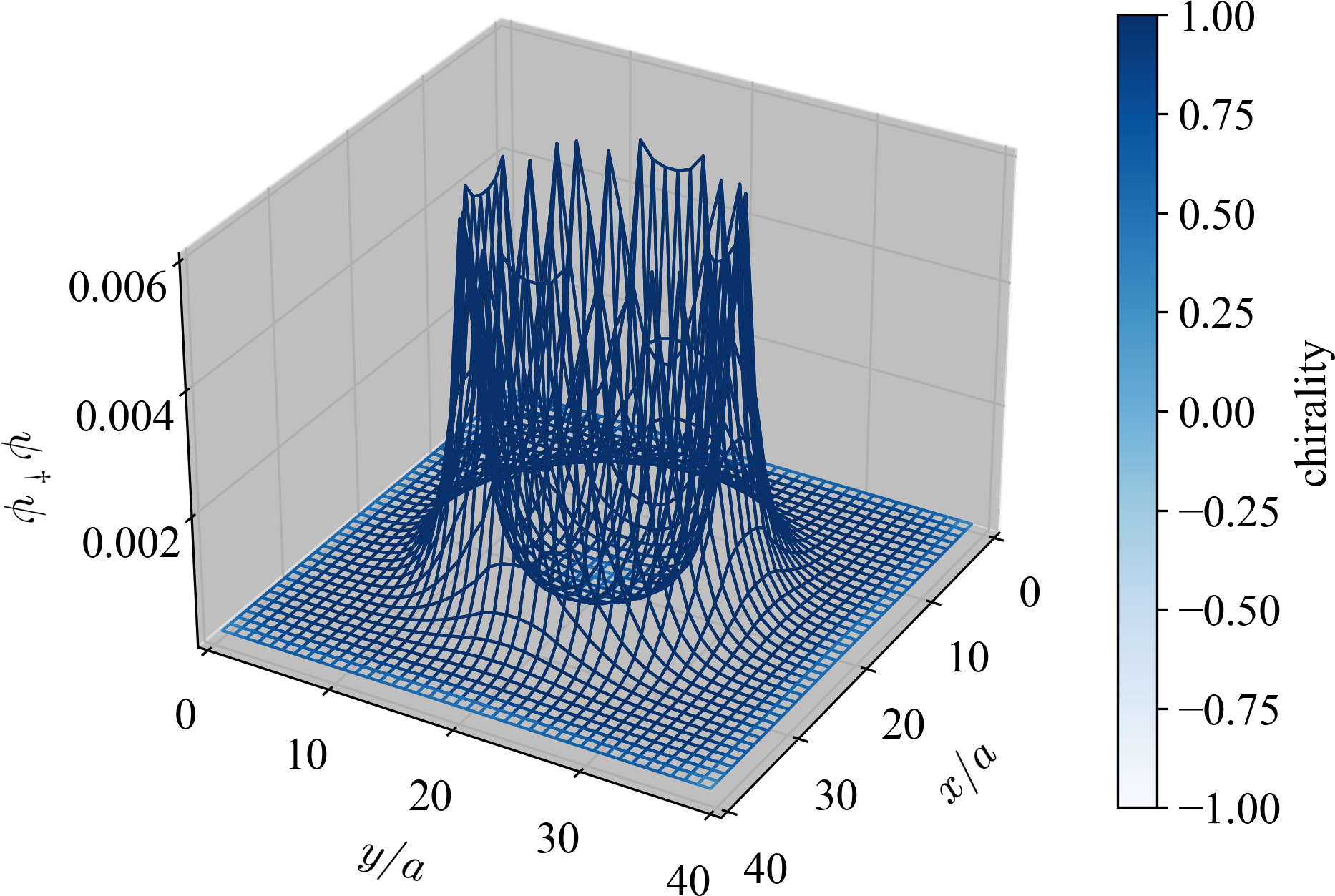}
     }\\
     \subfigure{
        \includegraphics[bb=0 0 448 301, height=0.3 \textheight]{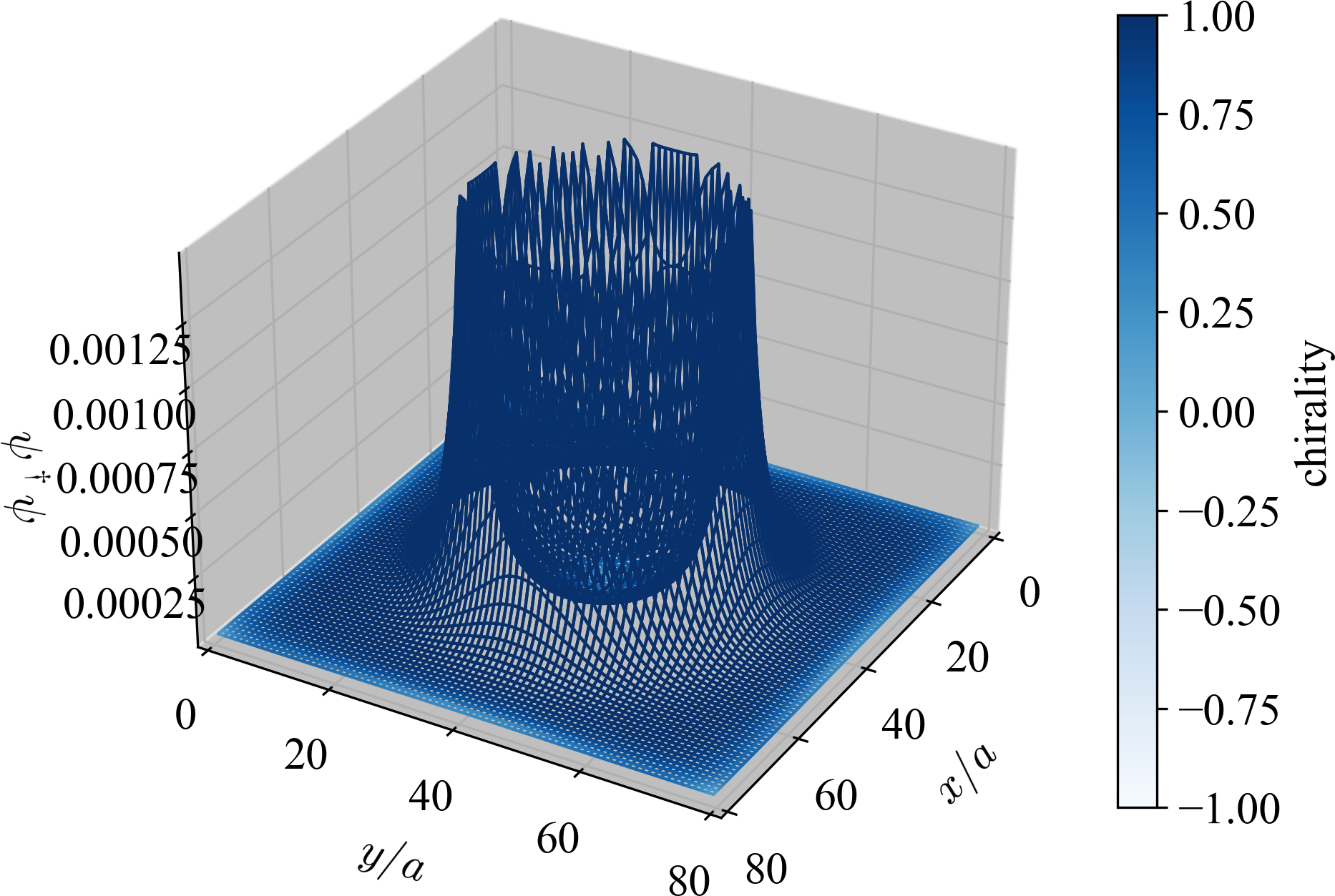}
     }
     \caption{Top panel: The amplitude of the first excited mode with $r_0=L/4$ and $a=L/N=L/20$. Middle: The same plot with $a=L/40$. Bottom: The same plot with $a=L/80$. The color gradation denotes the chirality at each lattice point \eqref{eq:colorgradation}.} 
     \label{fig:S1_eigenstate}
    \end{center}
\end{figure}

The eigenstates whose eigenvalue is between $-m$ and $m$ are almost chiral. Their eigenvalues agree well with the corresponding continuum prediction. We can see a gap from $E=0$ as the geometrical effect caused by the spin structure. In Figure \ref{fig:S1_eigenstate}, we show the amplitude of the first excited state with $r_0 E_{1/2}= 0.4180$ and $ \expval{\gamma_{\text{normal}}}=0.9893$, which shows that these eigenmodes are localized at the wall. The color gradation shows the chirality at each lattice point 
\begin{align} \label{eq:colorgradation}
    \psi^\dagger \gamma_{\text{normal}} \psi (x,y)/ \psi^\dagger  \psi(x,y).
\end{align}

\begin{figure}
\begin{minipage}[b]{0.45\linewidth}
\centering
\includegraphics[bb=0 0 409 293,width=\textwidth]{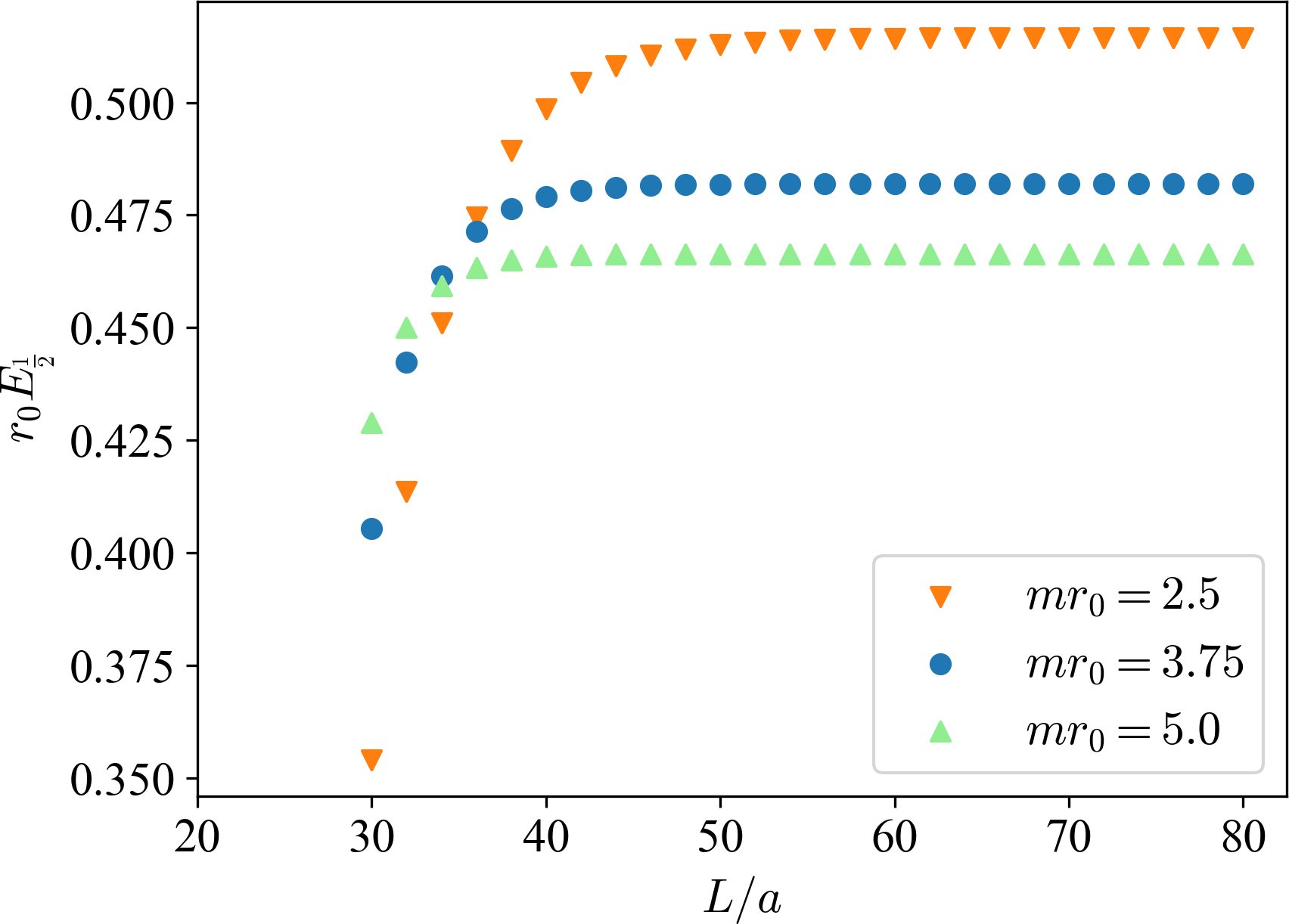}
\end{minipage}
\hfill
\begin{minipage}[b]{0.45\linewidth}
    \centering
    \includegraphics[bb=0 0 397 297,width=\textwidth]{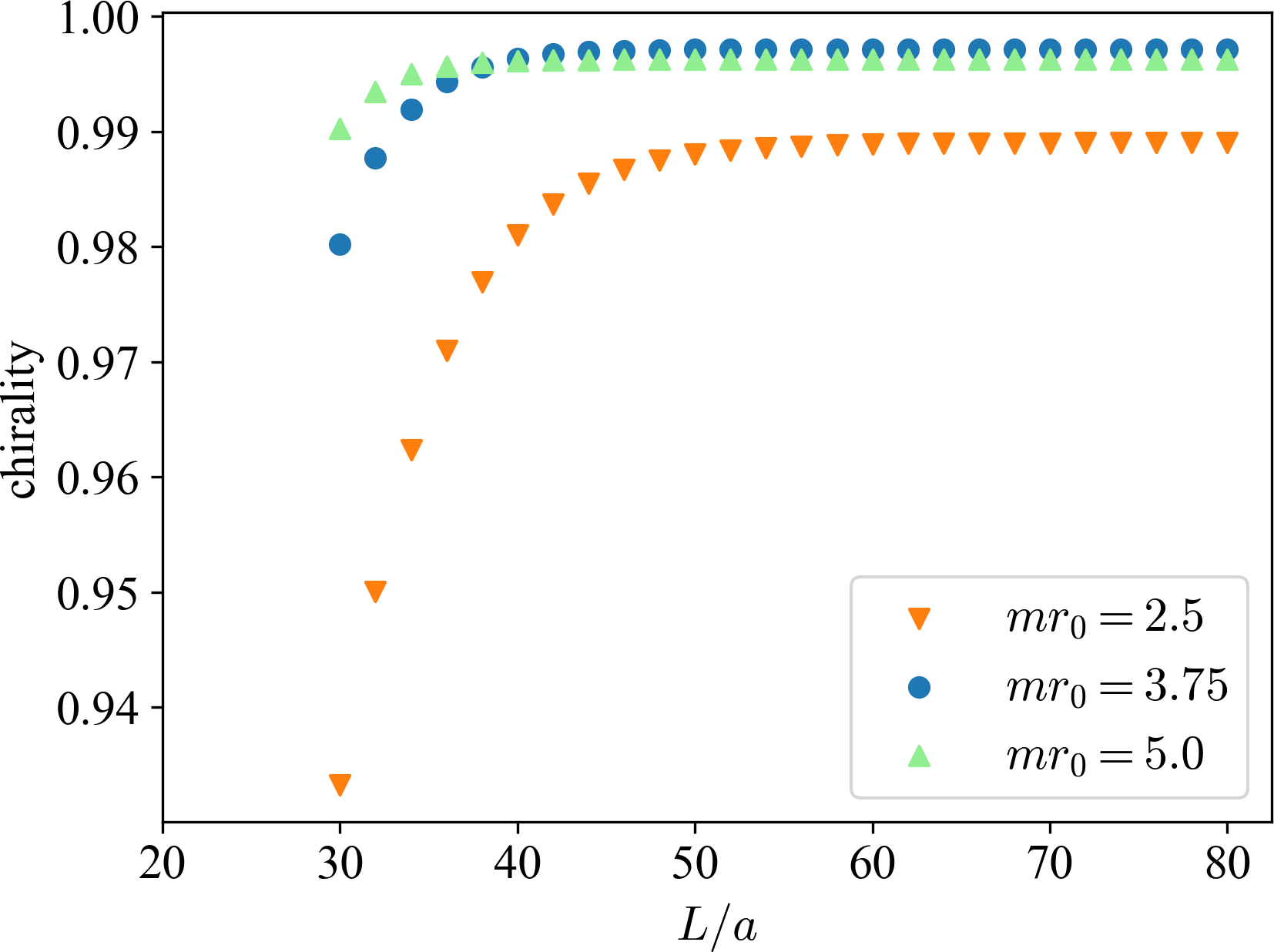}
    \end{minipage}
\caption{Left panel: The finite volume effect for the eigenvalue of the first excited mode as a function of the lattice length $L$. Right: That for the chirality. We fix $r_0=10 a$ and choose three mass parameters $mr_0=2.5,~3.75 $ and $5.0$.}
\label{fig:finitevolume}
\end{figure}

Let us estimate the finite volume effect. In the numerical computation, we take a periodic boundary condition in all directions with the finite size of $L$. 
We plot $r_0 E_\frac{1}{2}$ and the chirality as a function of $L$ in the left and right panels in Figure \ref{fig:finitevolume}, respectively. Here, we fix the domain-wall radius $r_0=10a$ and the mass parameter $mr_0 =2.5,~3.75$ or $5.0$. As we can see, the finite volume effect with the heavier mass is more quickly suppressed as $L$ increases. For $ma= 10/40$, the effect can be ignored when $L>50a =5 r_0$ and for $ma= 15/40,~20/40$, it is small enough when $L>40a =4 r_0$.

\begin{figure}

        \centering
        \includegraphics[bb=0 0 409 301,width=\textwidth]{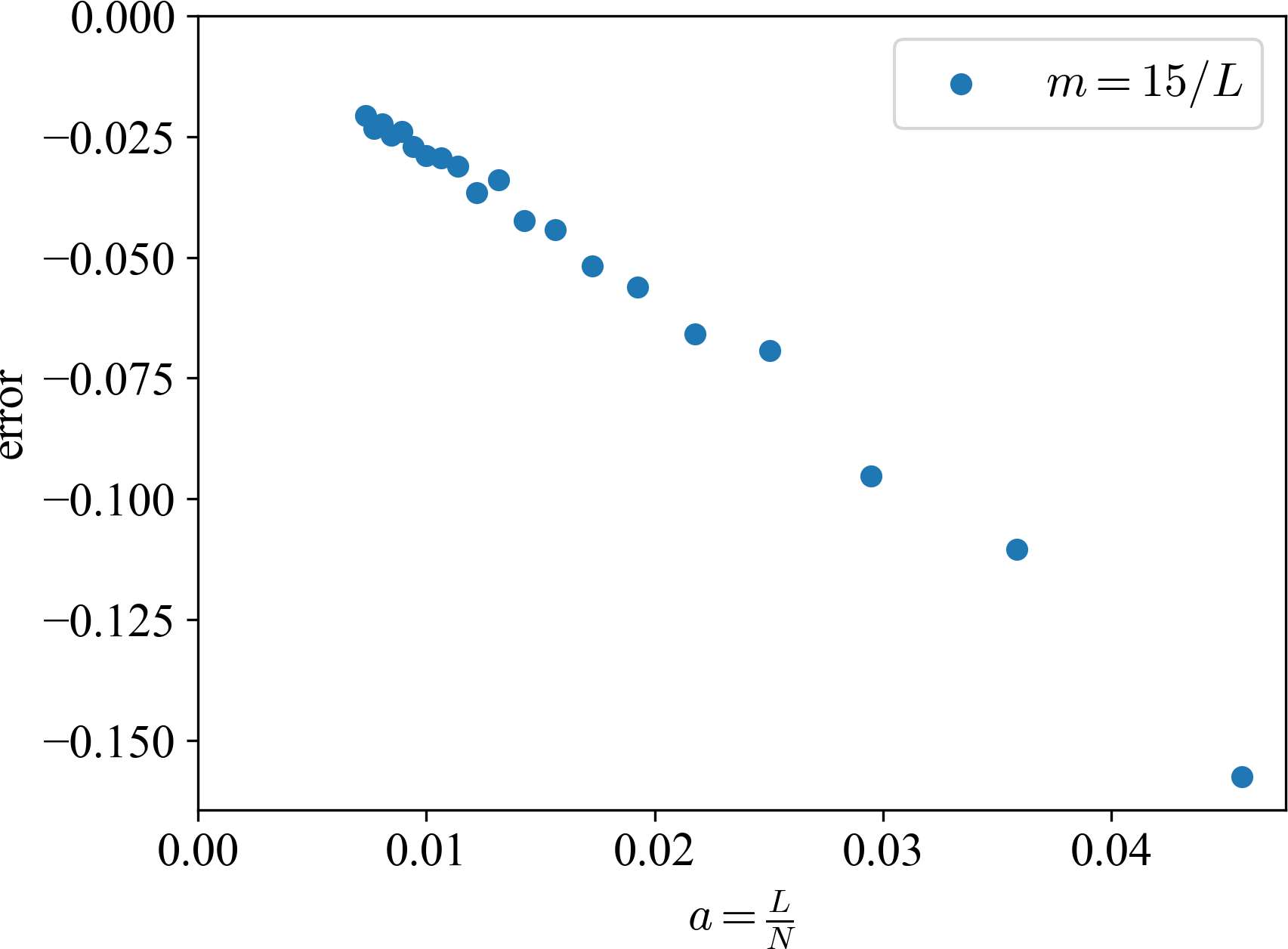} 

    \caption{The relative deviation of the eigenvalue of the first excited state \eqref{eq:error} as a function of the lattice spacing $a$ with $m=15/L$ and $r_0=L/4$. Three neighborhood points are averaged.}
    \label{fig:continuumlimit}
\end{figure}

\begin{figure}
\centering
\includegraphics[bb=0 0 411 301,width=\textwidth]{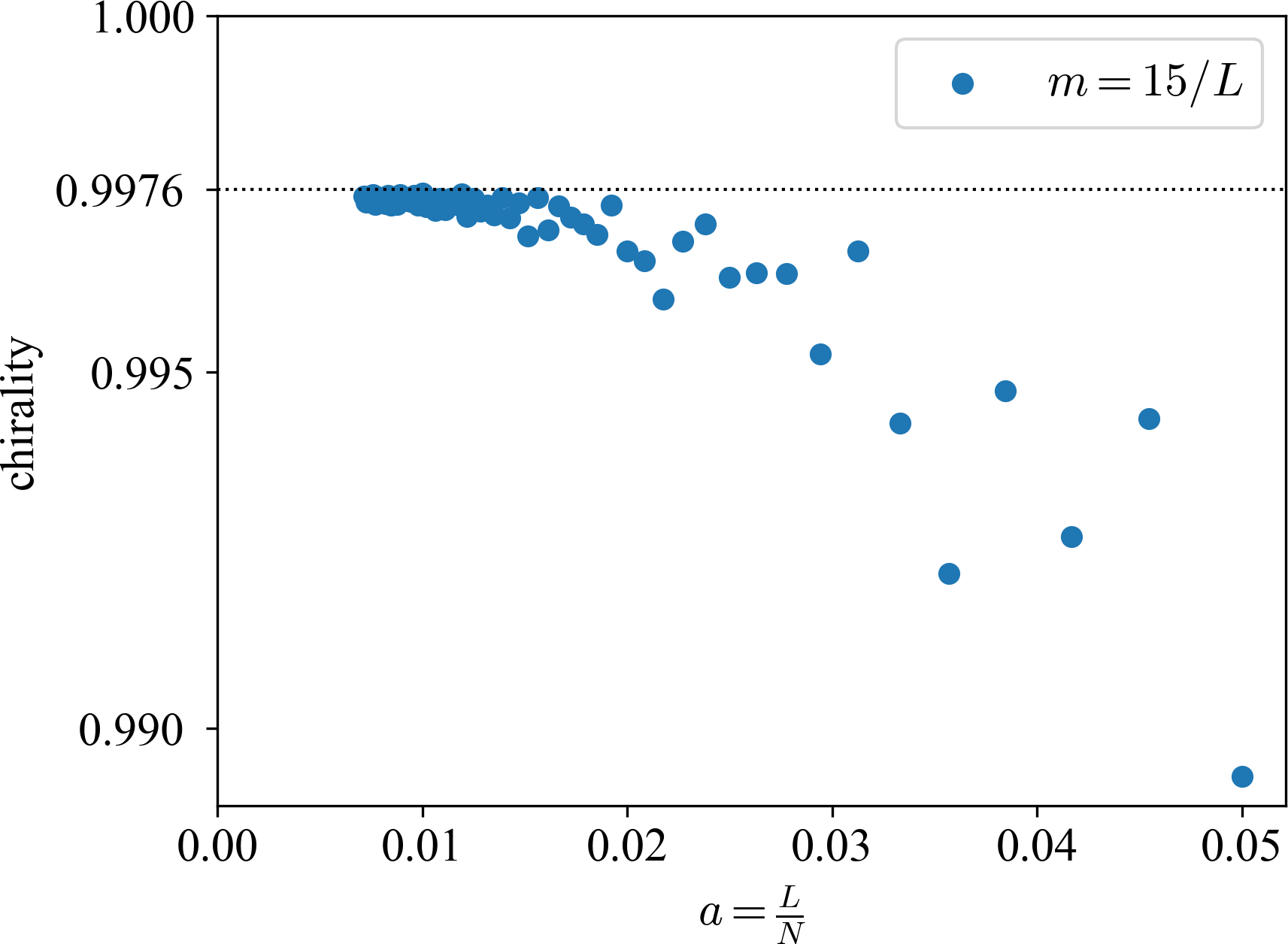} 
\caption{The chirality of the eigenvalue of the first excited state as a function of the lattice spacing $a$ with $m=15/L$ and $r_0=L/4$. The continuum prediction of the chirality is $0.9976$ computed by \eqref{eq:edgeS1}}
\label{fig:continuumlimit_chirality}
\end{figure}

Next, we discuss the continuum limit $a \to 0$ with $m=15/L$ and $r_0=L/4$ fixed. In the setup with $mr_0=3.75$, the finite volume effect is negligible. We plot the relative deviation of $E_\frac{1}{2}$ from the continuum counterpart $E_\text{conti}$ in Figure \ref{fig:continuumlimit} defined by
\begin{align}\label{eq:error}
    \text{error}=\qty(E_{1/2}-E_\text{conti})/E_\text{conti},
\end{align}
as a function of the lattice spacing $a=L/N$. 
In the plot, we take the average of three neighborhood points to cancel oscillating behavior. The error converges almost linearly to zero. 
We also plot the chirality of the first excited mode in the Fig.~\ref{fig:continuumlimit_chirality}. The chirality converges to the continuum prediction $ 0.9976$\footnote{The continuum prediction is computed in the finite $mr_0$ and is slightly shifted from unity.}. Therefore, the convergence indicates that the effect of the boundary on the lattice is negligible, as is expected. 

\begin{figure}
\centering
\includegraphics[bb=0 0 381 297,width=\textwidth]{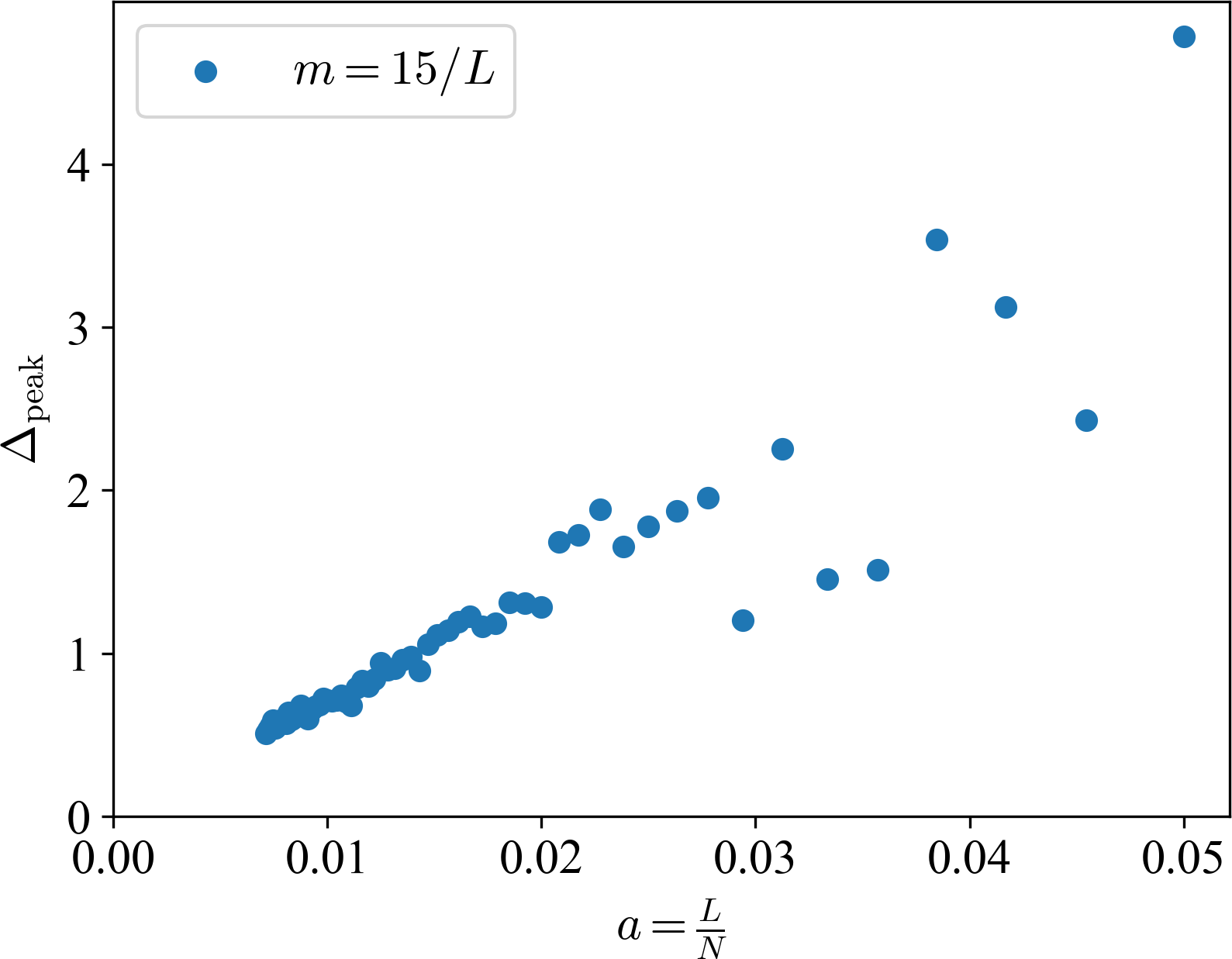}
\caption{The difference between the maximal and minimal peak around the domain-wall \eqref{eq:difference of peaks} is plotted as a function of the lattice spacing $a$ when $m=15/L$ and $r_0=L/4$.} 
\label{fig:RotationalSymmetry}
\end{figure}

Finally, we show that the rotational symmetry is restored naturally in the continuum limit. Figure \ref{fig:S1_eigenstate} represents the amplitude of the first excited mode when $a=L/20,~L/40$ and $L/80$. On a coarser lattice, there are spiky peaks that violate the rotational symmetry. However, these peaks become milder as the lattice is finer. Let $P$ be a set of the peaks around the domain-wall, then $P$ is equivalent to
\begin{align}
    P&= \Set{ \max_{y }( \psi^\dagger\psi(x,y) ) |  -r_0<x<r_0} =\Set{ \max_{x }( \psi^\dagger\psi(x,y) )|  -r_0<y<r_0}.
\end{align}
We take the difference between the maximal and minimal peak 
\begin{align}\label{eq:difference of peaks}
    \Delta_{\text{peak}}=(\max(P)-\min(P))/a^2,
\end{align} 
which gives a measure for the violation of the rotational symmetry. We plot it in Figure \ref{fig:RotationalSymmetry} when $m=15/L$ and $r_0=L/4$. The numerical data indicate that the rotational symmetry is restored in the continuum limit. This is presumably related to the recovery of rotational symmetry of the higher dimensional square lattice.

\section{$S^2$ Domain-wall}
\label{subsec:Curved_lat_S2}

Next, we consider the $S^2$ domain-wall embedded into a flat three-dimensional square lattice space $(a\mathbb{Z}/ L\mathbb{Z} )^3$. The Dirac operator we consider is
\begin{align}\label{eq:Hermitian Wilson Dirac op of S^2 in R^3}
    H =\frac{1}{a}\bar{\gamma} \qty(\sum_{i=1}^3\qty[\gamma_i\frac{\nabla_i-\nabla^\dagger_i}{2} +\frac{1}{2}\nabla_i \nabla^\dagger_i ]+\epsilon am ), 
\end{align}
where we put the center of the spherical domain-wall at $(x_0,y_0,z_0)= (a(N-1)/2, a(N-1)/2,a(N-1)/2)$ with radius $r_0$. Concretely, the domain-wall is assigned by a step function
\begin{align}
    \epsilon= \left\{  \begin{array}{cc}
        -1 & ( (x-x_0)^2+ (y-y_0)^2 + (z-z_0)^2<r_0^2  ) \\
        +1 & ( (x-x_0)^2+ (y-y_0)^2 + (z-z_0)^2 \geq r_0^2  ) 
    \end{array}
    \right. .
\end{align}
The gamma matrix normal to $S^2$ is defined as
\begin{align}\label{eq:S2 gamma matrix on lattice}
    \gamma_{\text{normal}}= \frac{(x-x_0)\gamma_1 + (y-y_0) \gamma_2 + (z-z_0) \gamma_3}{\sqrt{ (x-x_0)^2+ (y-y_0)^2 +(z-z_0)^2}}.
\end{align}

\begin{figure}
    \centering
    \includegraphics[bb=0 0 501 309,width=\textwidth]{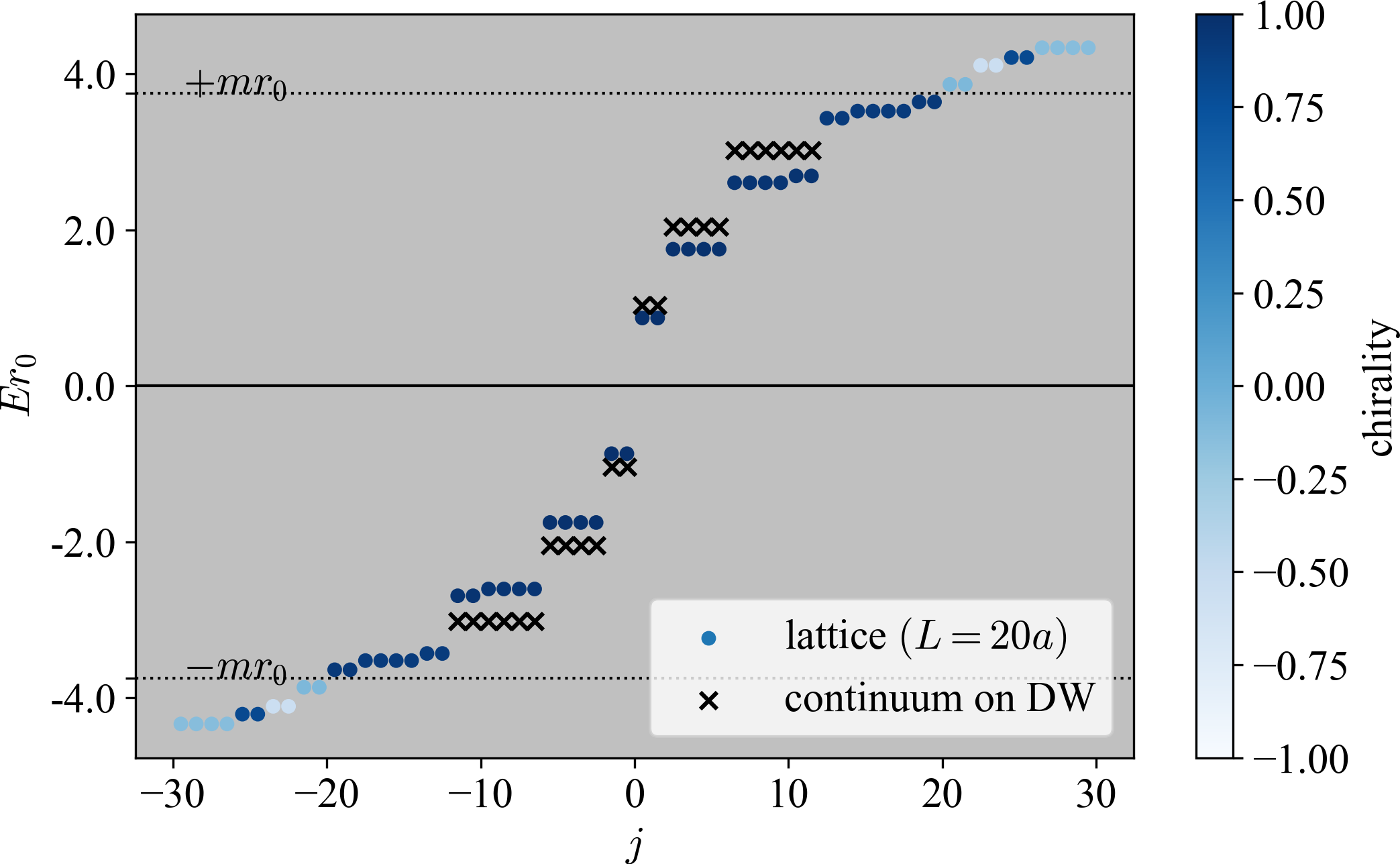} 
    \caption{The eigenvalue spectrum of the three-dimensional Hermitian Wilson-Dirac operator \eqref{eq:Hermitian Wilson Dirac op of S^2 in R^3} at $L=20a,~r_0=5a$ and $m=15/L$. The color gradation shows the expectation value of the chirality operator \eqref{eq:S2 gamma matrix on lattice}. The filled circles denote the lattice data and cross symbols express the corresponding continuum results.}
    \label{fig:S2_eigenvalue}
\end{figure}

Figure \ref{fig:S2_eigenvalue} shows the eigenvalue spectrum of the Dirac operator \eqref{eq:Hermitian Wilson Dirac op of S^2 in R^3} around $E=0$ with $L=20a,~r_0=5a=L/4$ and $m=15/L$. The eigenvalues are labeled by a half-integer $j$ and the gradation represents their chirality, which is the expectation value of \eqref{eq:S2 gamma matrix on lattice}.

\begin{figure}[]
    \begin{center}
     \subfigure{	
     \includegraphics[bb=0 0 448 301, width=.45\columnwidth]{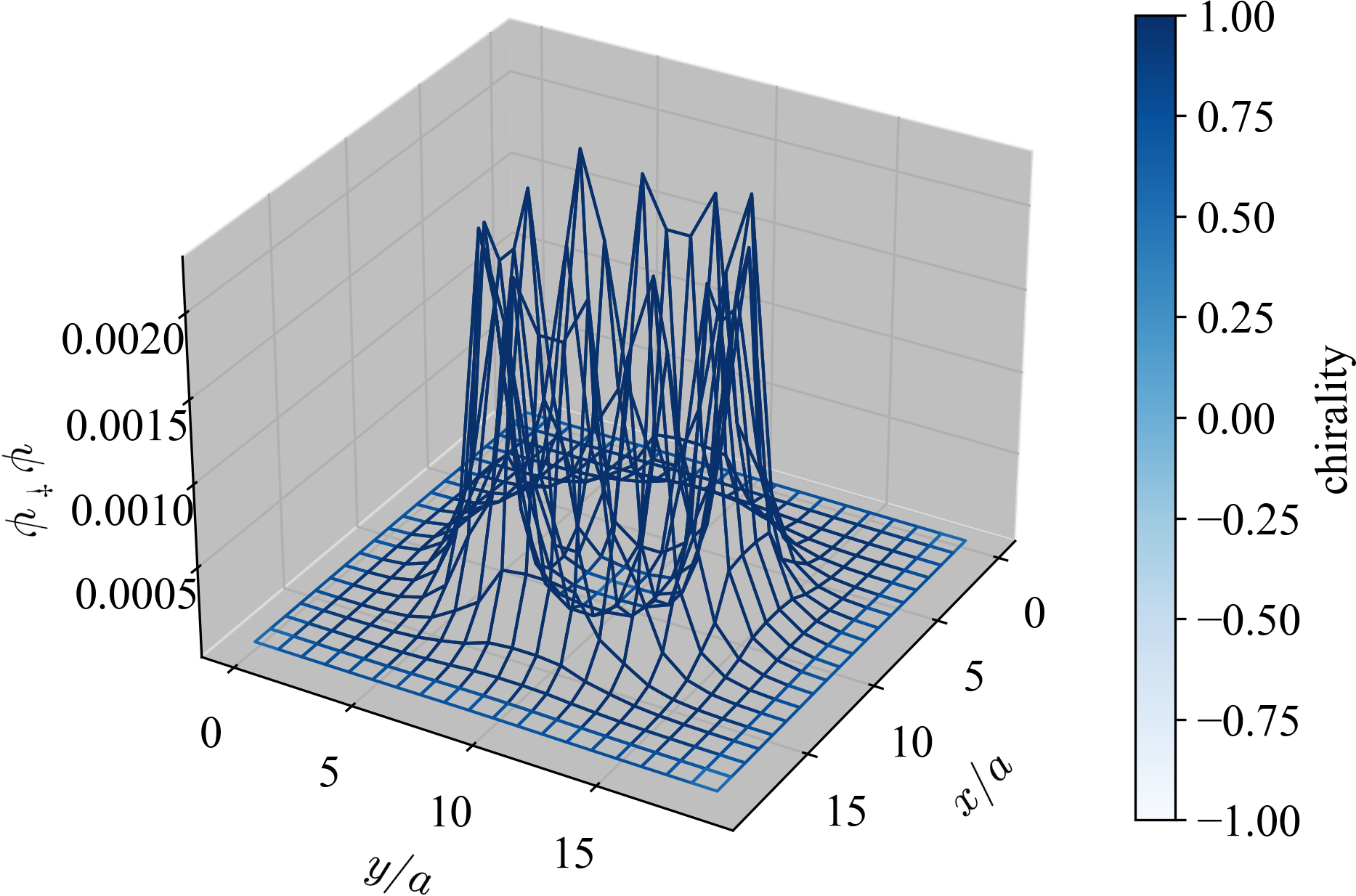}
     }
     \\ 
     \subfigure{	
     \includegraphics[bb=0 0 448 301, width=.45\columnwidth]{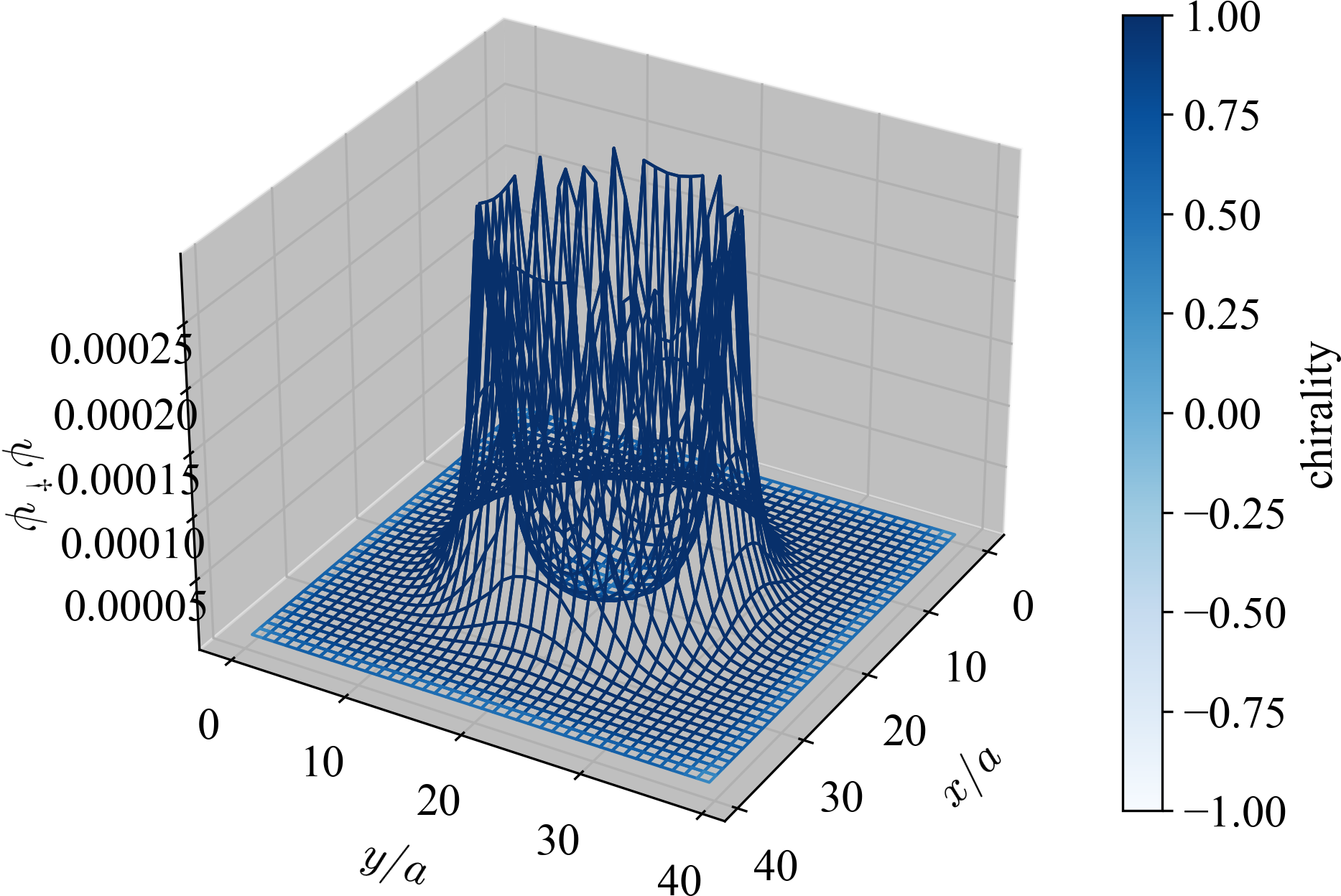}
     }
     \\
     \subfigure{	
     \includegraphics[bb=0 0 448 301, width=.45\columnwidth]{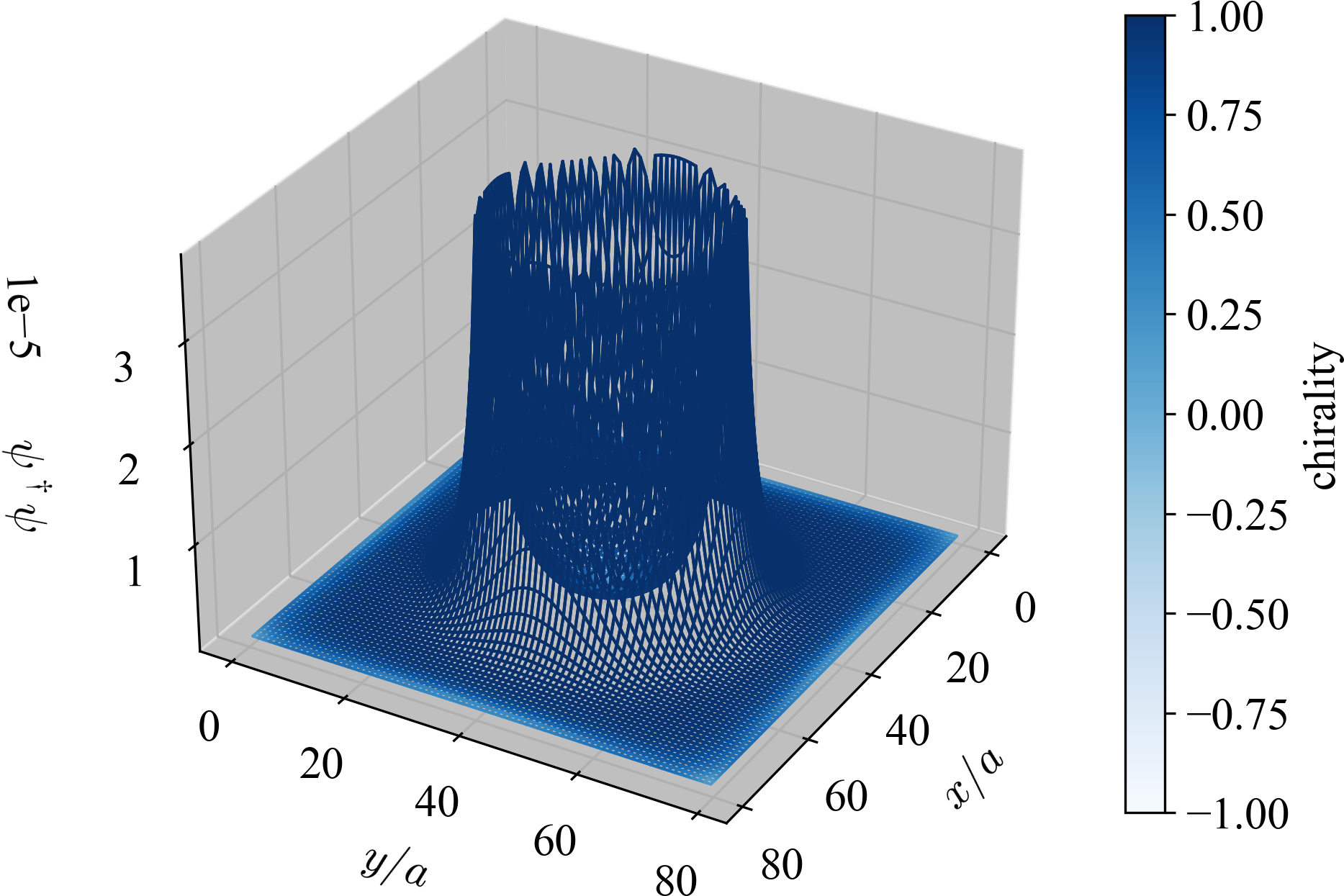}
     }
     \caption{Top panel: The amplitude of the first excited mode at $r_0=L/4$ and $a=L/N=L/20$ on the $z=aN/2$ plane. The vertical axis is the magnitude. Middle: The same plot as the top panel at $a=L/40$. Bottom: The same plot as the top panel at $a=L/80$. }
     \label{fig:S2_eigenstate}
    \end{center}
\end{figure}

We can see that the lattice result (filled circle symbols) is in a good agreement with the continuum prediction \eqref{eq:condition of E S2} (cross symbols). There is a gap from $E=0$ generated by the positive scalar curvature of the domain-wall. The dark blue color of low energy modes indicates that their chirality is close to unity. They are localized at the wall as shown in the top panel of Figure \ref{fig:S2_eigenstate} depicting the amplitude of the first excited mode on the $z= aN/2$ plane, which is the closest one to the center of the $S^2$ domain-wall. 

\begin{figure}
    \begin{minipage}[b]{0.45\linewidth}
    \centering
    \includegraphics[bb=0 0 402 293,width=\textwidth]{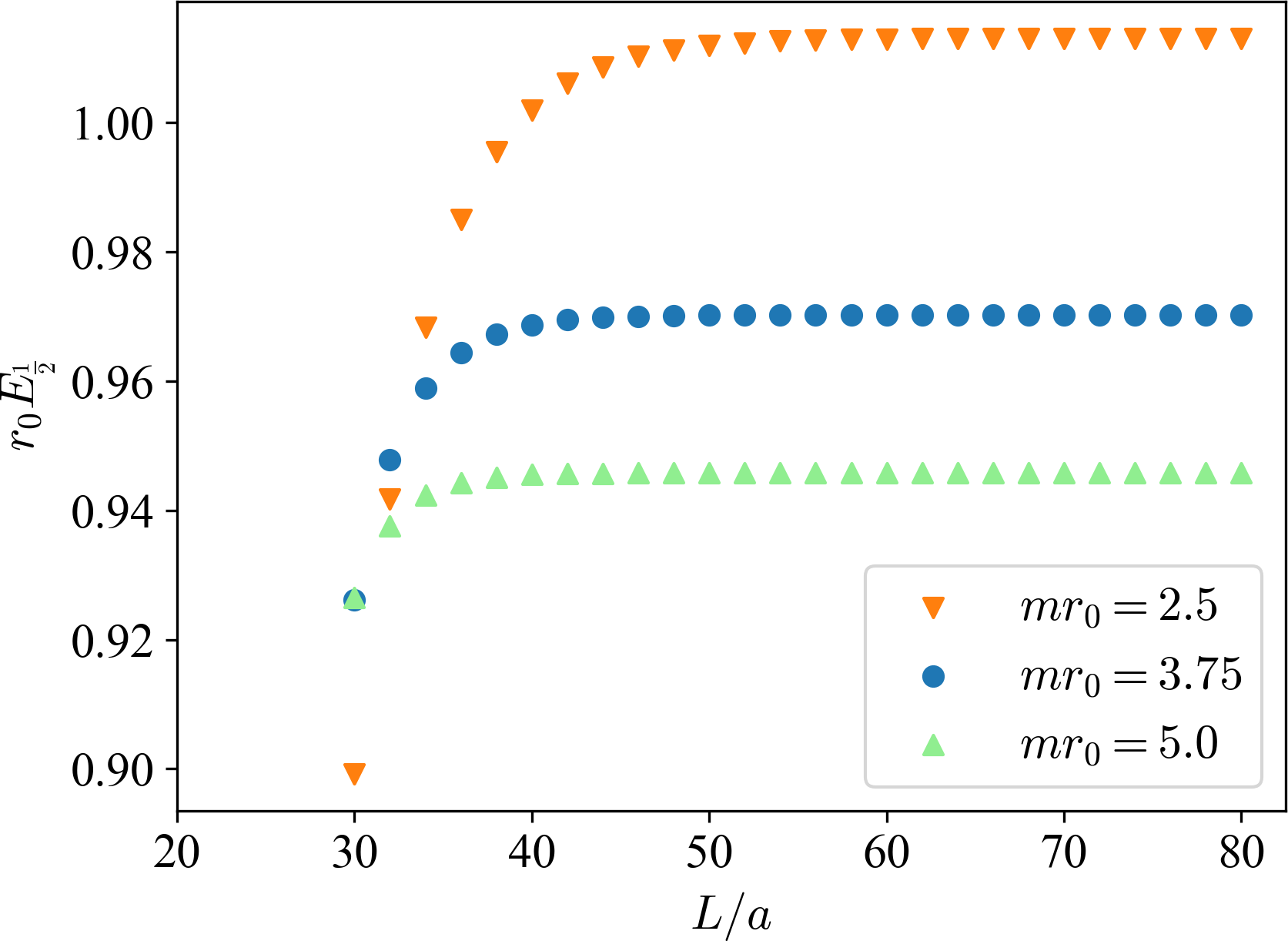} 
    \end{minipage}
    \hfill
    \begin{minipage}[b]{0.45\linewidth}
        \centering
        \includegraphics[bb=0 0 402 293,width=\textwidth]{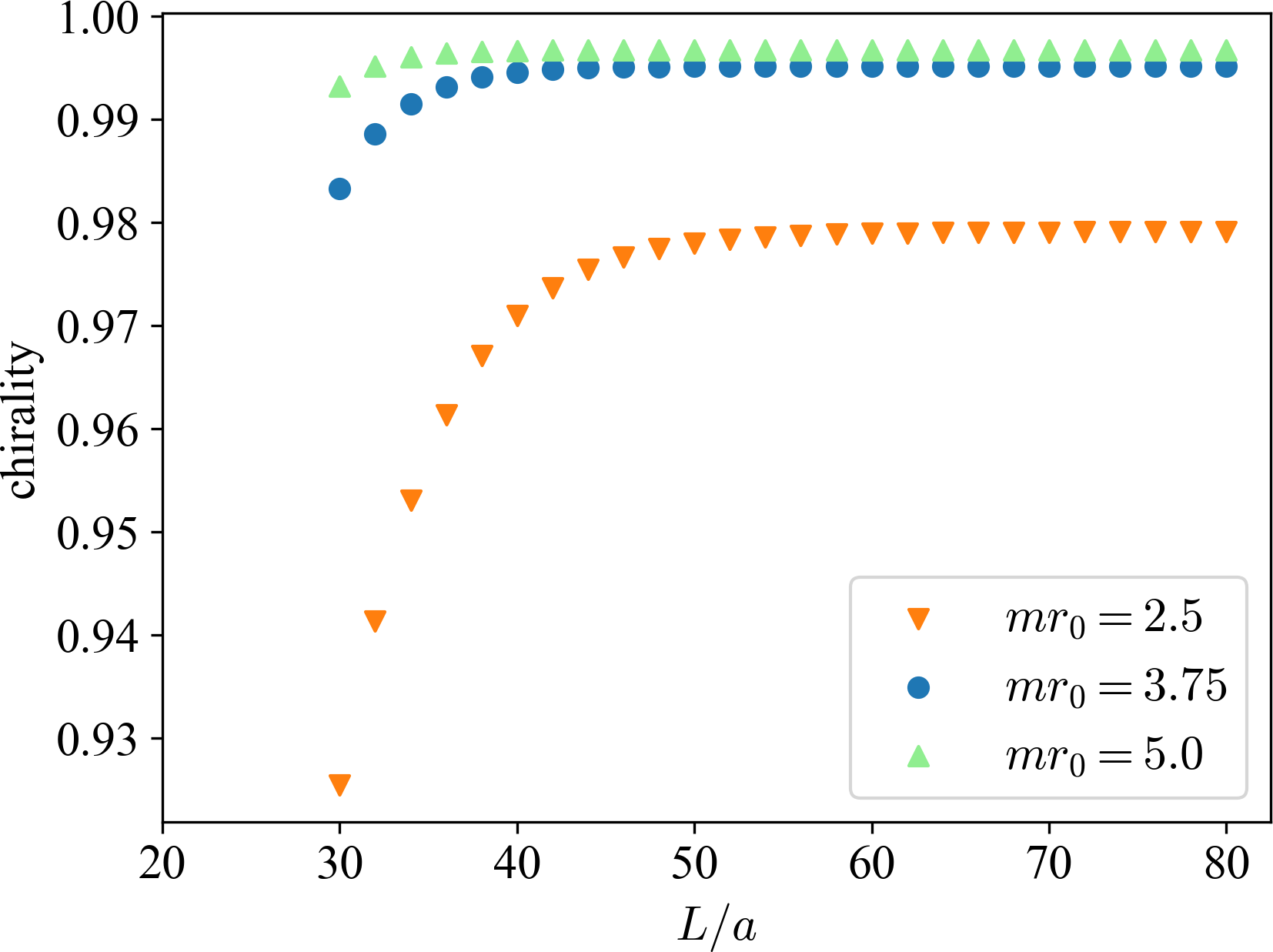}
        \end{minipage}
    \caption{Left panel: Volume dependence of the first excited eigenvalue is shown as a function of the lattice length $L$. Right: That of the chirality. We fix $r_0=10 a$ and choose three mass parameters $mr_0=2.5,~3.75 $ and $5.0$.}
    \label{fig:S2_finitevolume}
\end{figure}

We discuss the finite volume effect on the eigenvalue and the chirality. As in the previous chapter, we plot $r_0 E_\frac{1}{2}$ and their chirality as a function of $L/a$ in the left panel and right panel in Figure \ref{fig:S2_finitevolume}. We fix the domain-wall radius $r_0=10a$ and mass parameter $mr_0 =2.5,~3.75$ or $5.0$. The effect is saturated in the large volume limit $L\to \infty$. For $mr_0=2.5$, the effect is small enough
if $L$ is larger than $5r_0$. For $mr_0 =3.75$ and $5.0$, it is negligible when $L$ is larger than $4r_0$.

\begin{figure}
    \begin{minipage}[b]{0.45\linewidth}
    \centering
    \includegraphics[bb=0 0 409 301,width=\textwidth]{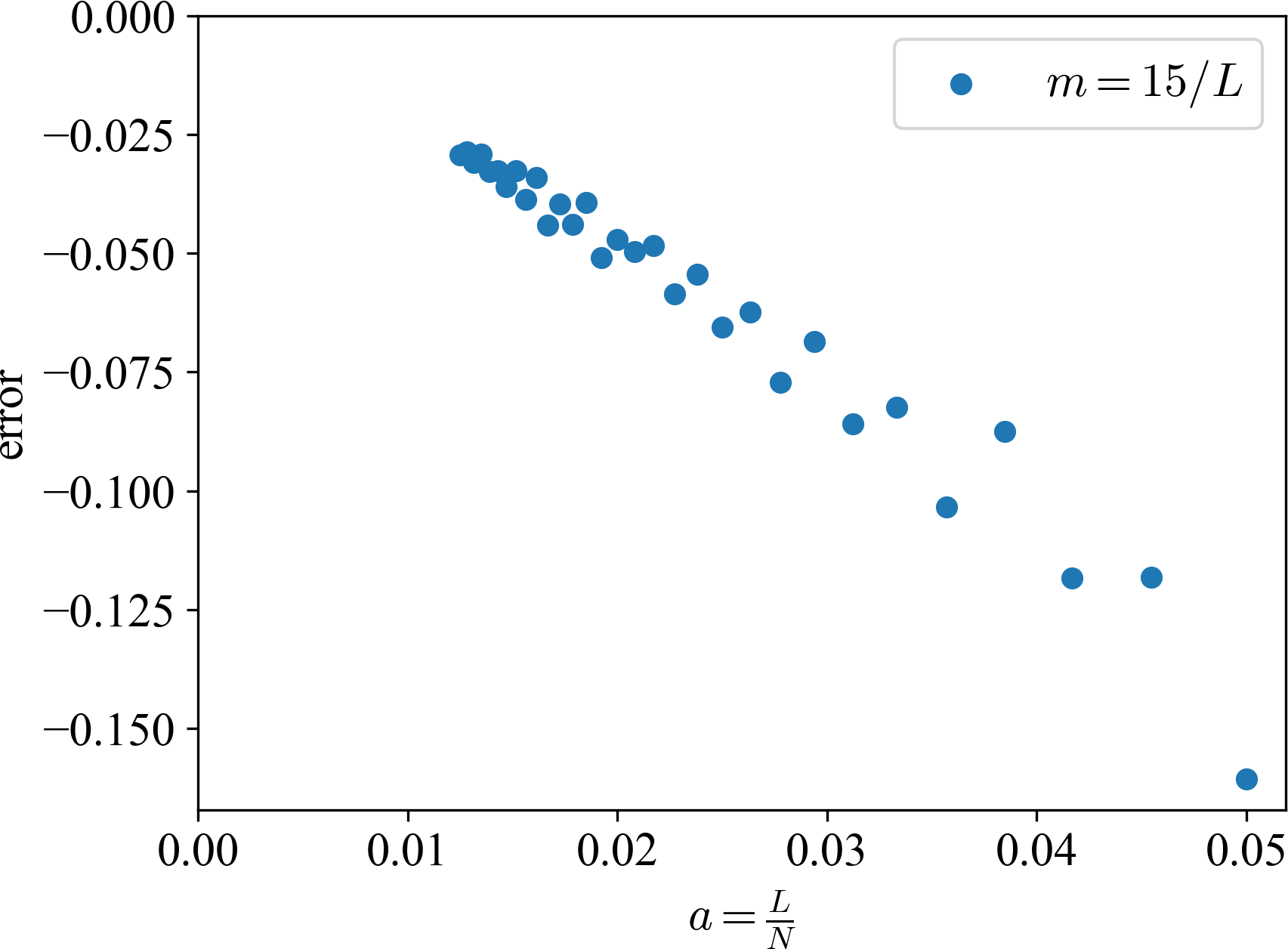}
    \end{minipage}
    \hfill
    \begin{minipage}[b]{0.45\linewidth}
        \centering
        \includegraphics[bb=0 0 409 301,width=\textwidth]{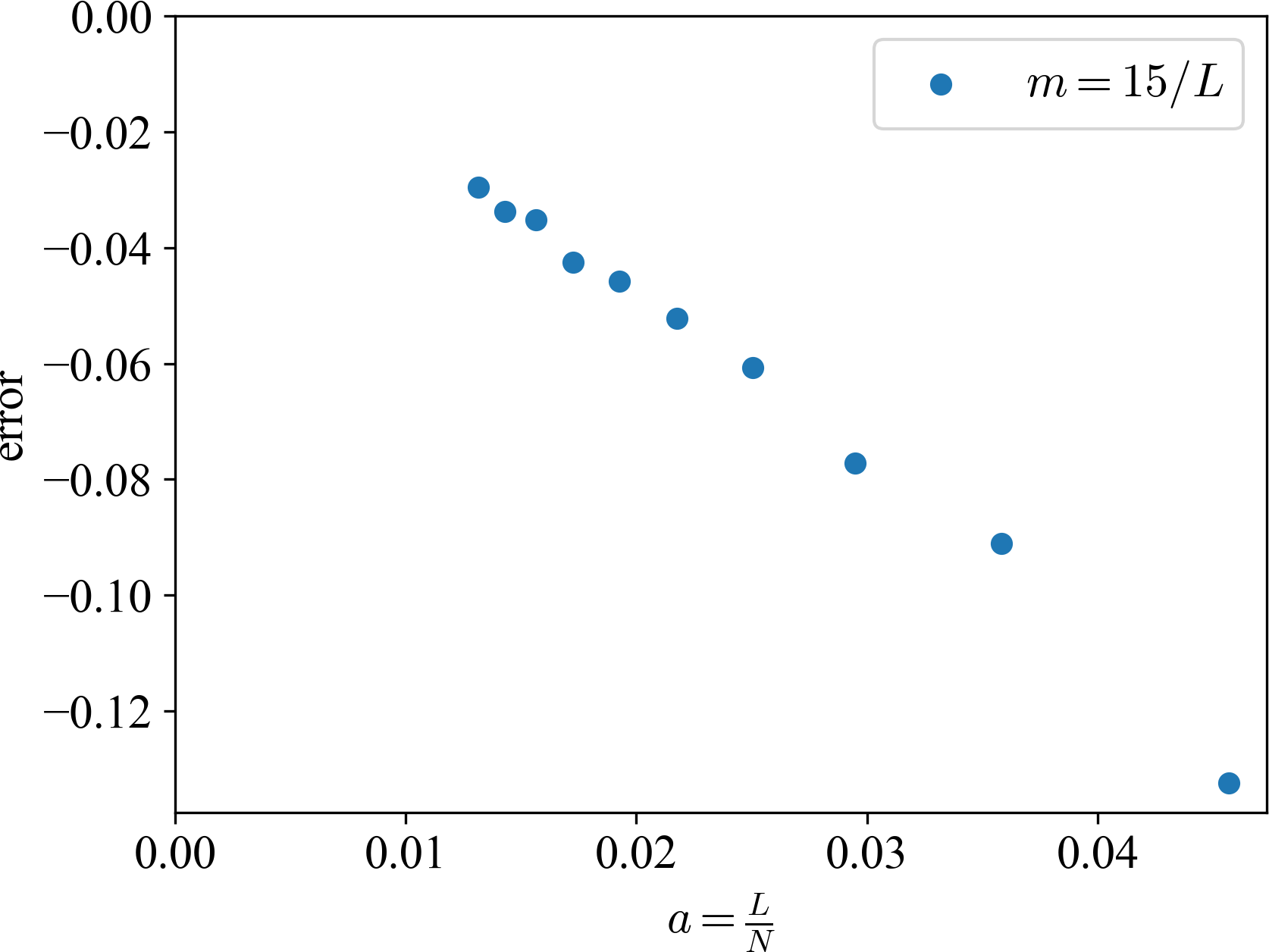}
        \end{minipage}
    \caption{Left panel: The relative deviation of the eigenvalue of the first excited state \eqref{eq:error} from the continuum prediction as a function of the lattice spacing $a$. Right: Three neighborhood points binning of the left panel. We put $m=15/L$ and $r_0=L/4$.}
    \label{fig:S2_continuumlimit}
\end{figure}

Let us consider the systematics due to the lattice spacing, too. In Figure \ref{fig:S2_continuumlimit}, we plot the deviation of the first excitation eigenvalue $E r_0$ from the continuum prediction as a function of $a$ when $m=15/L$ and $r_0=L/4$. This figure shows that the deviation linearly depends on the lattice spacing $a$. We also plot their chirality. The lattice data converge to the continuum prediction much faster than a linear function.

\begin{figure}
\centering
\includegraphics[bb=0 0 409 301,width=\textwidth]{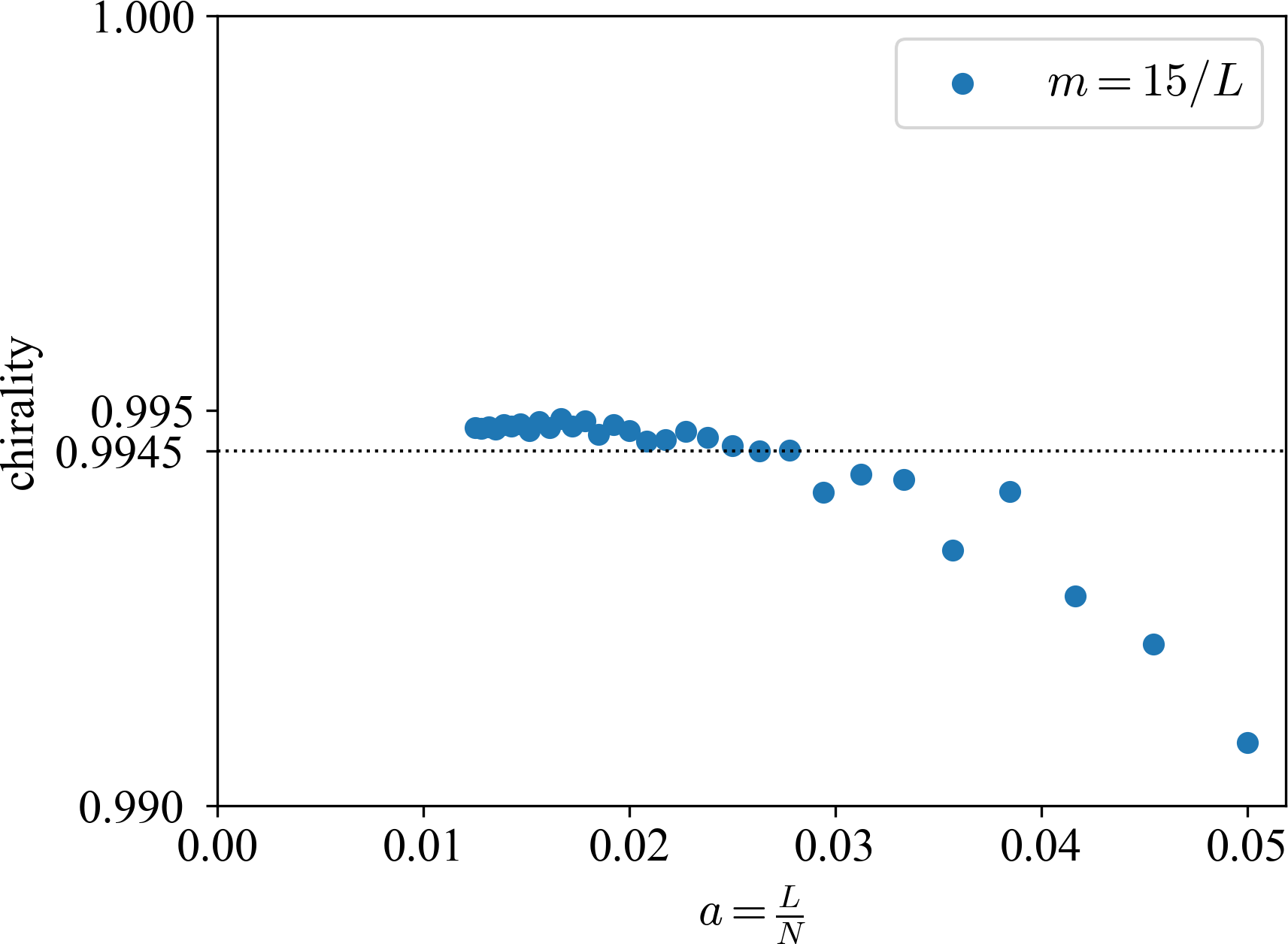}
\caption{The chirality of the eigenvalue of the first excited state as a function of the lattice spacing $a$ with $m=15/L$ and $r_0=L/4$. The continuum prediction of the chirality is $0.9945$ computed in \eqref{eq:EdgeMode S2 }.}
\label{fig:S2_continuumlimit_chirality}
\end{figure}

\begin{figure}
    \centering
    \includegraphics[bb=0 0 381 297,width=\textwidth]{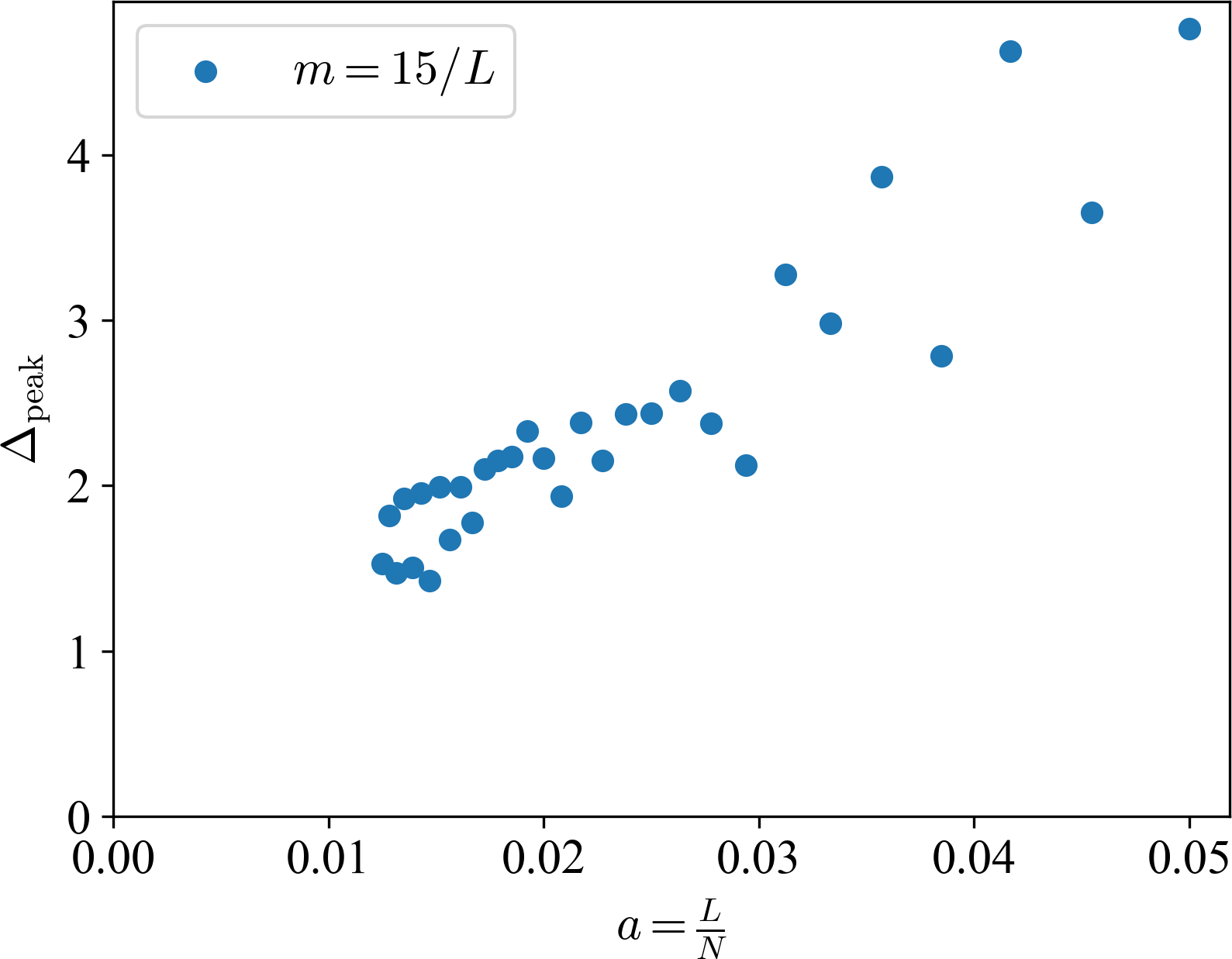} 
    \caption{The difference between the maximal and minimal peak around the domain-wall is plotted as a function of the lattice spacing $a$ when $m=15/L$ and $r_0=L/4$.}
    \label{fig:S2_RotationalSymmetry}
\end{figure}

The rotational symmetry is also recovered in the naive continuum limit of the three-dimensional lattice. As shown in Figure \ref{fig:S2_eigenstate}, the finer the lattice, the smoother the edge spikes are.
To measure the violation of the rotational symmetry, we define $\Delta_\text{peak}$, which is the difference between the highest and lowest peak of the amplitude around the wall divided by the three-dimensional cube. Figure \ref{fig:S2_RotationalSymmetry} shows that the rotational symmetry is recovered automatically in the continuum limit.

\chapter{Anomaly Inflow}
\label{sec:Anomaly_Inflow}

In this section, we consider the same $S^1$ domain-wall system in the previous chapter with a $U(1)$ gauge field. We first assign a weak field and show that the gauge field makes changes to the Dirac spectrum on the edge modes on the wall as the Aharanov-Bohm effect. Consequently, the time-reversal ($T$) anomaly emerges on the wall. However, the $T$ anomaly is canceled by the chiral anomaly inside the circle and the total system is free from the $T$ anomaly. We illustrate how the cancellation, which is called anomaly inflow, is described on the two-dimensional lattice space. Next, we squeeze the flux in one plaquette. We show that the flux generates a new domain-wall around the singular point and a new localized mode arises. The new mode cancels the $T$ anomaly instead of the chiral anomaly.


\section{Anomaly Inflow on Lattice}
\label{subsec:Anomaly_Inflow}

We assign a weak $U(1)$ gauge connection on a two-dimensional square lattice with the $S^1$ domain-wall. The gauge field affects the edge-localized modes and makes their eigenvalue spectrum asymmetric in the positive and negative directions. It gives the time-reversal ($T$) anomaly \cite{Alvarez-Gaume:1984zst} on the boundary system. However, the whole lattice system has a time-reversal symmetry and the $T$ anomaly must be canceled by the contribution from the bulk system \cite{Witten:2015Fermion}. We show how the anomaly is canceled on the lattice, which is equivalent to the Atiyah-Patodi-Singer index theorem on the two-dimensional disk.

\begin{figure}
\centering
\includegraphics[width=\textwidth]{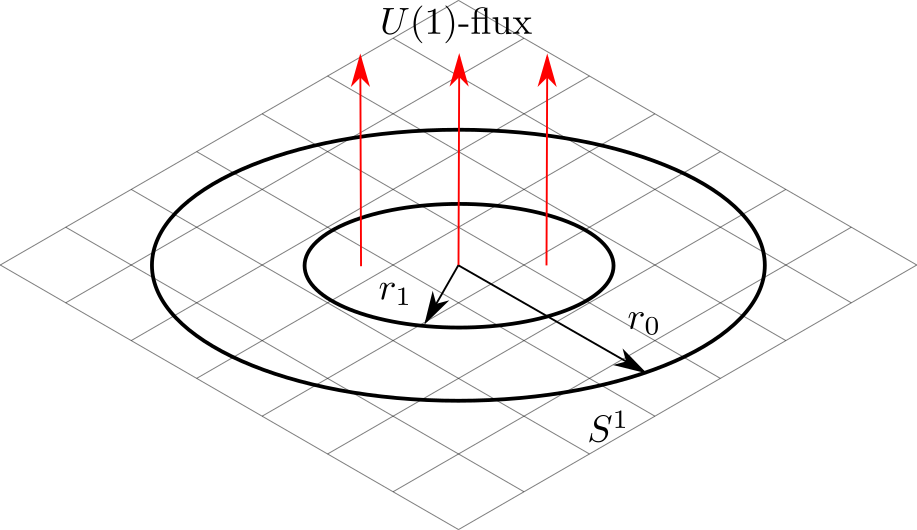}
\caption{$S^1$ domain-wall with the radius $r_0$ on a two dimensional square lattice. The $U(1)$ gauge field $A$ has field strength $F_{12}=2\alpha/r_1^2$ for $r<r_1$ and zero for $r>r_1$. $\alpha$ denotes the total flux (divided by $2\pi$) of the entire system \cite{Aoki:2022aezanomalyinflow}.}
\label{fig:S1DW_Flux}
\end{figure}

We set the $U(1)$ gauge connection 
\begin{align} 
    A_1= &-\alpha\frac{y-y_0}{r_1^2},~  A_2= \alpha\frac{x-x_0}{r_1^2} ~(\text{for } r\leq r_1), \\
    A_1=& -\alpha\frac{y-y_0}{r^2},~  A_2= \alpha\frac{x-x_0}{r^2} ~(\text{for } r> r_1) ,
\end{align}
where we take the radial coordinate, $x-x_0=r \cos \theta$ and $y-y_0=r \sin \theta$ taking $(x_0,y_0)$ as the origin. 
The field strength is $F_{12}=2\alpha/r_1^2$ for $r<r_1$ and zero for $r>r_1$. $\alpha$ denotes the total flux (divided by $2\pi$) of the entire system. We assume $0\leq \alpha \leq 1$. The covariant difference operator in the $\mu$-direction is given by
\begin{align}
    (\nabla_\mu \psi)(x) &= \exp[i \int_{x+ a\hat{\mu}}^x A_\nu dx^\nu] \psi(x+a \hat{\mu}) -\psi(x) \\
    (\nabla_\mu^\dagger \psi )(x) &= \exp[i \int_{x- a\hat{\mu}}^x A_\nu dx^\nu] \psi(x-a \hat{\mu}) -\psi(x).
\end{align}
The Hermitian Dirac operator we consider is 
\begin{align}\label{eq:Hermitian Wilson Dirac op of S^1 in R^2 with flux}
    H =\frac{1}{a}\sigma_3 \qty(\sum_{i=1,2}\qty[\sigma_i\frac{\nabla_i-\nabla^\dagger_i}{2} +\frac{1}{2}\nabla_i \nabla^\dagger_i ]+\epsilon am ), 
\end{align}
where $\epsilon $ is a step function defined by \eqref{eq:S1domain-wall}. We set the radius $r_0$ is larger than $r_1$. This setup is depicted in Figure \ref{fig:S1DW_Flux}.

\begin{figure}
\centering
\includegraphics[bb=0 0 517 373,width=\textwidth]{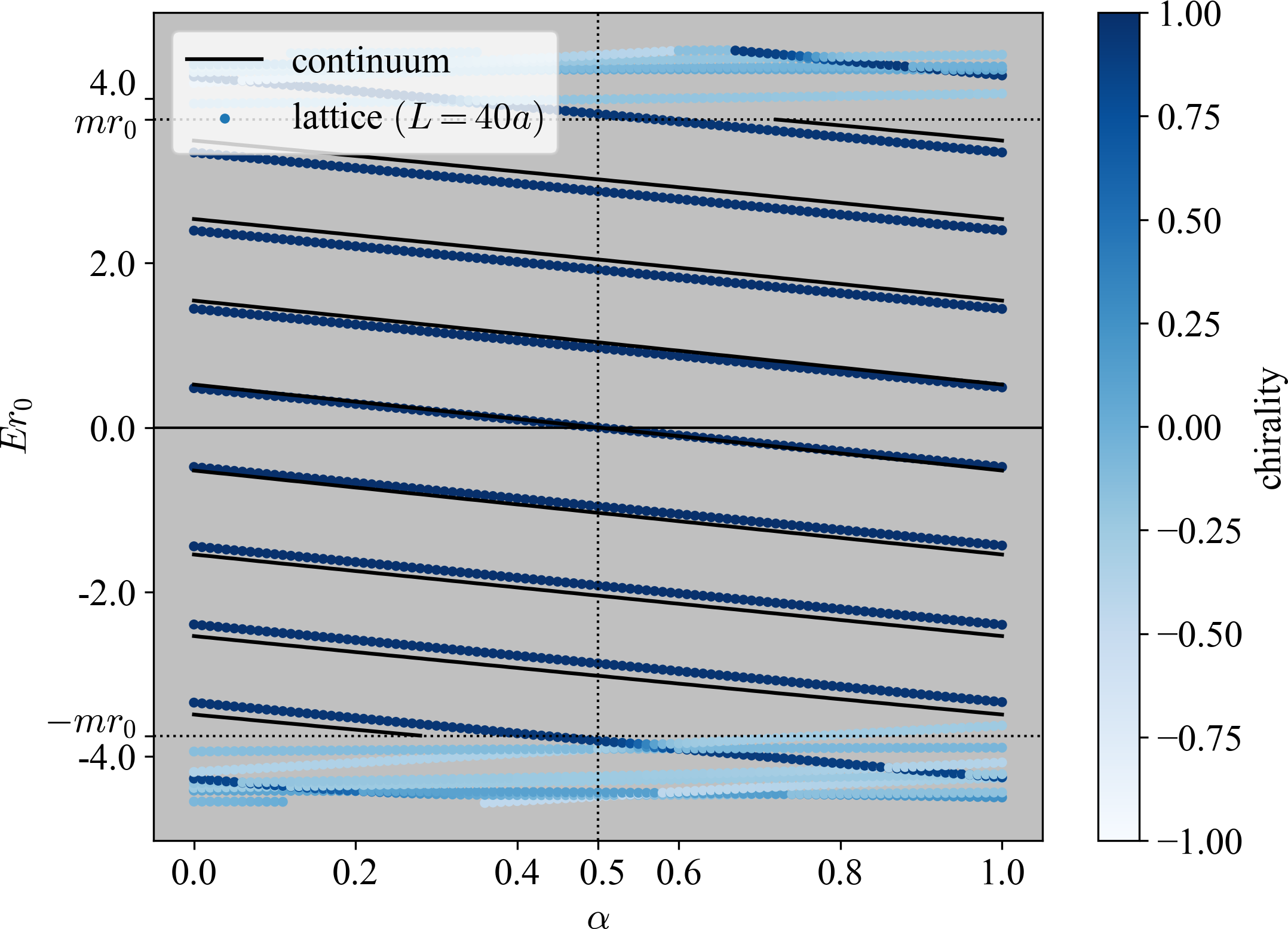} 
\caption{The eigenvalue spectrum of the Wilson-Dirac operator \eqref{eq:Hermitian Wilson Dirac op of S^1 in R^2 with flux} as a function of the total flux $\alpha$ when $L=40a$, $m=15/L$, $r_0=L/4=10a$ and $r_1=r_0/2=5a$. The color gradation represents their chirality, which is the expectation value of $\gamma_\text{normal}$ \eqref{eq:gamma normal S1}. The solid lines express the continuum prediction \eqref{eq:condition of E S1}.}
\label{fig:AnomalyInflow}
\end{figure}

We solve the eigenvalue problem of $H$ for each $\alpha$. We plot the eigenvalue spectrum as a function of $\alpha$ in Figure \ref{fig:AnomalyInflow}. Here, we fix $L=40a$, $m=15/L$, $r_0=L/4=10a$ and $r_1=r_0/2=5a$. The color gradation shows their chirality. The black lines denote the continuum results predicted by the equation \eqref{eq:condition of E S1}. 

We can see all low-energy modes ($\abs{E}<m$) have a positive chirality. Their eigenvalues decrease as a linear function of $\alpha$ due to the Aharanov-Bohm effect. In the large mass limit, the effective Dirac operator for these modes is given by $H^{S^1}$ \eqref{eq:effective Dirac operator S1} and we have the eigenvalue
\begin{align}
    E_{n+\frac{1}{2}}=\frac{1}{r_0} \qty(n+ \frac{ 1}{2}-\alpha )
\end{align}
for $n\in \mathbb{Z}$. The $Spin^c$ connection $1/2$ comes from the periodicity of the edge modes. When $\alpha=1/2$, it is canceled and there is a zero mode. When $\alpha=1$, the eigenvalue spectrum is the same as that of $\alpha=0$ but $E_{n+\frac{1}{2}}$ shifts to $E_{n-\frac{1}{2}}$. In Figure \ref{fig:AnomalyInflow}, these behaviors are captured even when $m$ is a finite value.


The asymmetry of the spectrum reflects the $T$ anomaly of the one-dimensional Dirac operator. It is characterized by the eta invariant \cite{Atiyah1975spectral}
\begin{align}
    \eta( H^{S^1})=&\lim_{\epsilon \to 0} \sum_{\lambda \neq
0} \frac{\lambda}{ \abs{\lambda}^{1+\epsilon}}+ \dim \text{Ker}(
H^{S^1}),
\end{align}
 where $\lambda $ denotes the eigenvalue of $H^{S^1}$. The eta invariant appears as the phase factor of the partition function on $S^1$ regularized by the Pauli-Villars field:
 \begin{align}
    Z^{S^1}_{\text{reg}}[A]= \lim_{M_{\text{PV}} \to + \infty }\det \frac{ H^{S^1}}{ H^{S^1} + iM_{\text{PV}}} \propto \exp[ -i \pi \frac{\eta(H^{S^1})}{2} ]. 
 \end{align}
 $M_{\text{PV}}$ is a mass parameter of the Pauli-Villars field. For the partition function, the time-reversal transformation corresponds to the complex conjugate $Z^{S^1}_{\text{reg}}[A] \to (Z^{S^1}_{\text{reg}}[A])^\ast$. If $Z^{S^1}_{\text{reg}}[A]$ is a real function, it respects the time-reversal symmetry. Namely, the edge modes break the time-reversal symmetry when $\frac{\eta(H^{S^1})}{2} $ is not an integer. By using the zeta function regularization, we can compute the eta invariant
\begin{align}
     \frac{1}{2} \eta(H^{S^1})=& \frac{1}{2}\qty(
\lim_{\epsilon\to 0} \sum_{n\in \mathbb{Z}}
\frac{n+\frac{1}{2}-\alpha}{\abs{n+\frac{1}{2}-\alpha}^{1+\epsilon} }
+\#\set{ \text{zero modes}} ) \nonumber\\
    =&  \alpha  -\lfloor \alpha +\frac{1}{2} \rfloor .
\end{align}
The bracket $\lfloor \ast \rfloor$ denotes the Gauss symbol, which takes the integer part of the given real number.

This anomaly is canceled by the contribution from a two-dimensional space whose boundary is the one-dimensional circle with the radius $r_0$. The partition function of $\mathbb{D}^2$ is given by
\begin{align}
    Z^{\mathbb{D}^2 }_{\text{reg}}[A] = \det \frac{i \Slash{D}^{\mathbb{D}^2}-im  }{ i \Slash{D}^{\mathbb{D}^2}+iM_{\text{PV}}} \propto \exp[ i\pi \frac{1}{2\pi} \int_{r<r_0} F ]= \exp[ i\pi \alpha ].
\end{align}
Then the partition function of the composite system of $S^1$ and $\mathbb{D}^2$ is defined as the product of these partition functions:
\begin{align}
    Z[A]&=Z^{S^1 }_{\text{reg}}[A] \times Z^{\mathbb{D}^2 }_{\text{reg}}[A] \\
    &\propto \exp[ i\pi \qty( \frac{1}{2\pi} \int_{r<r_0} F - \frac{1}{2} \eta(H^{S^1}) ) ]= \exp[ i\pi  \lfloor \alpha +\frac{1}{2} \rfloor ].
\end{align}
Since $  \lfloor \alpha +\frac{1}{2} \rfloor$ is an integer, the partition function function $Z[A]$ is a real function and has a time-reversal symmetry. The $T$ anomaly on the boundary is canceled by the topological term on the bulk system and there is no anomaly in the total system. This cancellation mechanism is called the anomaly inflow \cite{Witten:2015Fermion}. In general, this is justified by the Aityah-Patodi-Singer (APS) index theorem. Here, $  \lfloor \alpha +\frac{1}{2} \rfloor$ is the APS index on the two-dimensional disk.

In recent works \cite{Fukaya_2017Atiyah-Patodi-Singer,fukayaFuruta2020physicistfriendly}, a novel formulation of the APS index has been presented. They put a domain-wall on an even-dimensional closed manifold $X$ and considered the Hermitian Dirac operator with the domain-wall mass: 
\begin{align}
    H=\bar{ \gamma} ( i \Slash{D}+ m \epsilon),
\end{align}
where $\epsilon $ is a step function defining the domain-wall and $\bar{\gamma}$ is the chirality operator. They mathematically proved that $-\eta(H)/2$ is always an integer and equal to the APS index for the negative mass region. In \cite{FukayaKawai2020TheAPS}, they considered a flat domain-wall system on a square lattice space and showed that $-\eta(H)/2$ is consistent with the APS index. Here they assumed that the link variables are sufficiently weak.

The Dirac spectrum on our square lattice indicates that the anomaly inflow mechanism is well reproduced even with a curved domain-wall. From Fig.~\ref{fig:AnomalyInflow}, the eta invariant of $H$ \eqref{eq:Hermitian Wilson Dirac op of S^1 in R^2 with flux} is obtained by
\begin{align}
     -\frac{1}{2} \eta(H)= \left\{ \begin{array}{cc}
    0  & ( 0<\alpha < \frac{1}{2}) \\
    1 & (  \frac{1}{2} \leq \alpha <1)
    \end{array} \right. .
\end{align}
This result agrees with the continuum prediction $\lfloor \alpha +\frac{1}{2} \rfloor $.

\section{Domain-wall Creation}
\label{subsec:Domwain-wall_Creation}

In the previous section, the gauge field was taken perturbatively weak. In this section, we consider a strong gauge field on a square lattice. 
Let us take a small $r_1$, while the total flux $\alpha$ is kept unchanged. When $r_1 < a/2$, the flux fits into one plaquette including the origin and it represents a strong gauge field. It is a non-trivial problem whether or not the anomaly inflow based on the APS index theorem is still present with such a high-energy configuration. As will be shown below, a dynamical change occurs but the $T$ symmetry is protected. 

\begin{figure}
    \centering
    \subfigure{	
    \includegraphics[bb=0 0 517 373,height=0.4 \textheight]{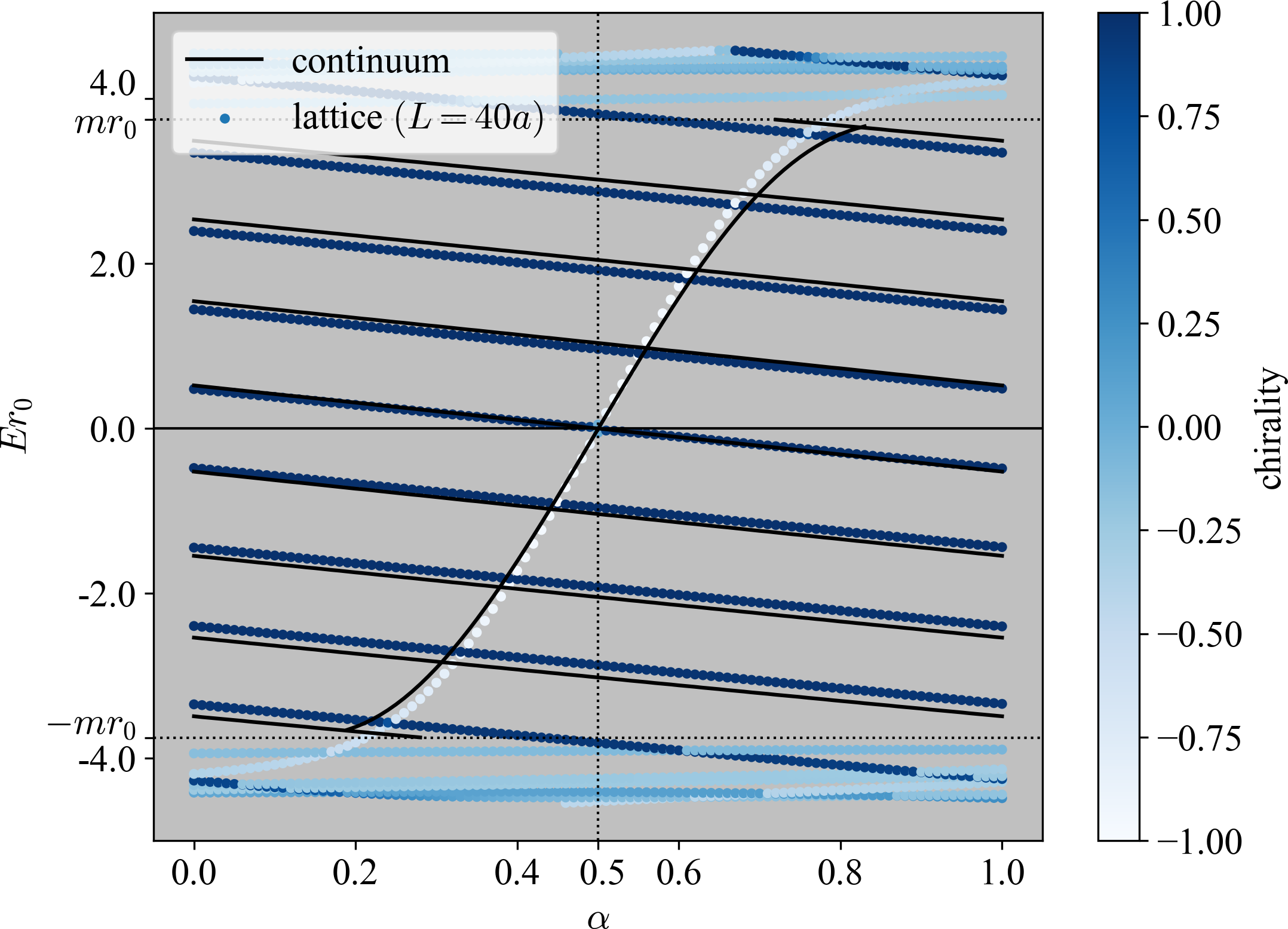}
    }\\ 
    \subfigure{	
    \includegraphics[bb=0 0 517 373,height=0.4 \textheight]{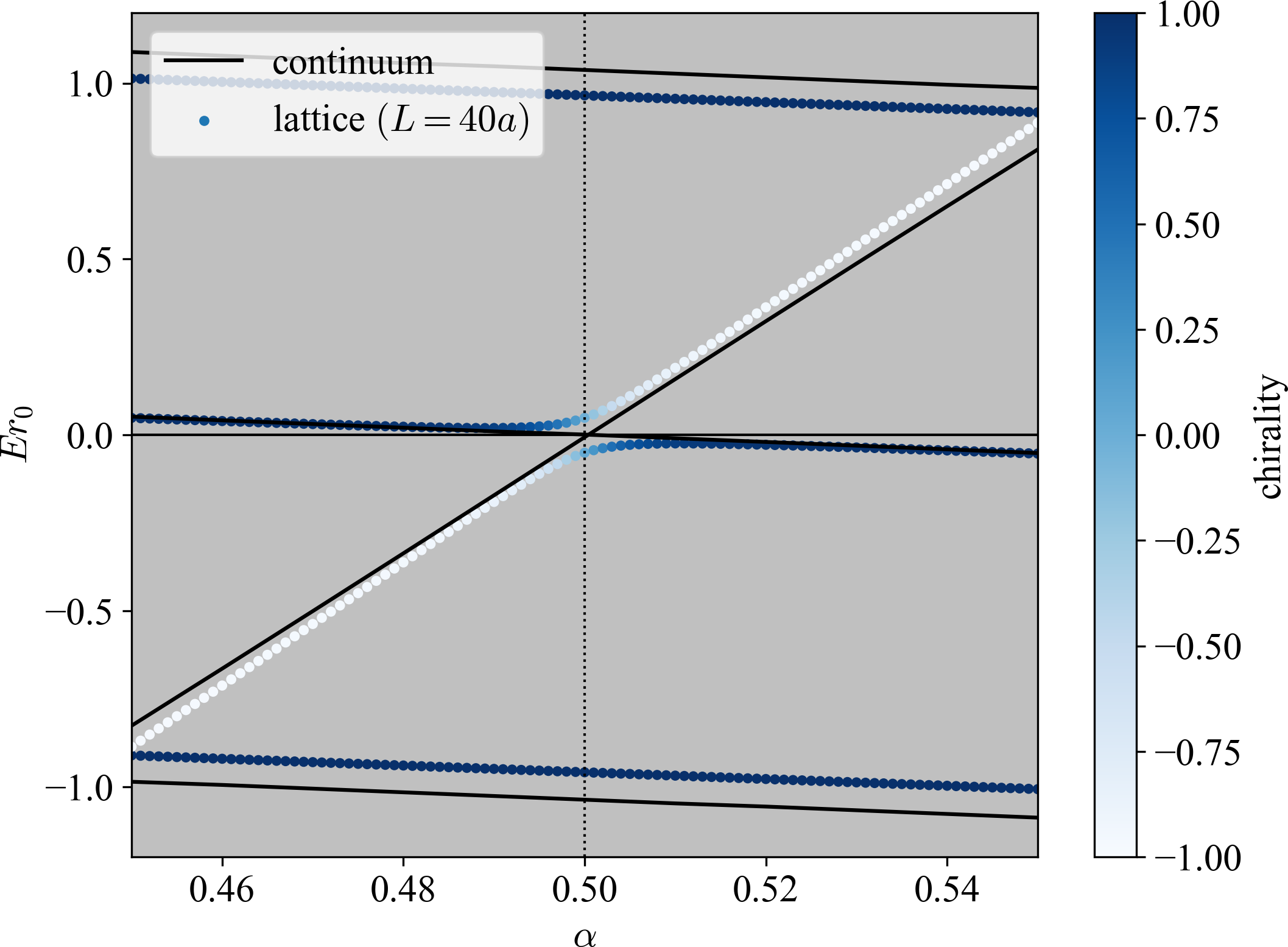}
    }
    \caption{Top panel: The eigenvalue spectrum of the Wilson-Dirac operator \eqref{eq:Hermitian Wilson Dirac op of S^1 in R^2 with flux} as a function of the total flux $\alpha$ when $L=40a$, $m=15/L$, $r_0=L/4=10a$ and $r_1$ is smaller than $1/2$. The color gradation represents their chirality, which is the expectation value of $\gamma_\text{normal}$ \eqref{eq:gamma normal S1}. The black lines denote the continuum prediction. The straight lines are computed by the equation \eqref{eq:condition of E S1} and the curved line is by the equation \eqref{eq:condition of E S1 center r1 is small}. Bottom: Enlarged view of the left panel near $E=0$.}
    \label{fig:AnomalyInflow_singular}
\end{figure}

We plot the eigenvalue spectrum as a function of the total flux $\alpha$ in Figure \ref{fig:AnomalyInflow_singular}. The linear change of the edge-localized modes is seen as in the previous section. However, we find another mode with the negative chirality, which appears when $\alpha=0.2$ at the energy level $\sim -m$ and disappears when $\alpha=0.8$ at the energy level $\sim +m$. When $\alpha=1/2$, the eigenvalue crosses $E=0$\footnote{To be precise, the two near zero modes around $E=0$ do not cross zero due to mixing via the tunneling effect and are split apart (see the bottom panel of Fig.~\ref{fig:AnomalyInflow_singular}).}. 

\begin{figure}[]
    \begin{center}
     \subfigure{	
     \includegraphics[bb=0 0 448 301, height=0.3 \textheight]{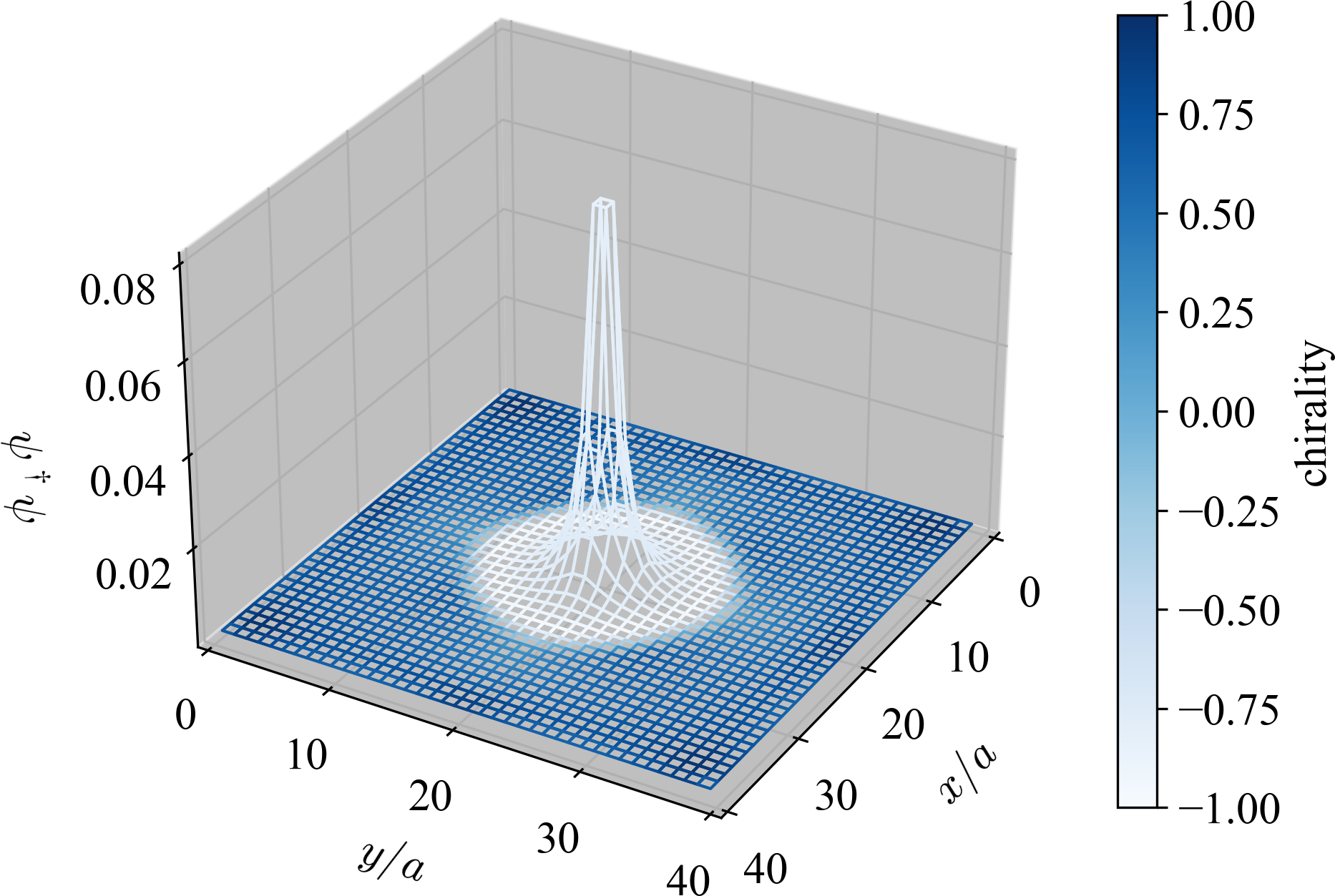}
     }\\ 
     \subfigure{
        \includegraphics[bb=0 0 448 301, height=0.3 \textheight]{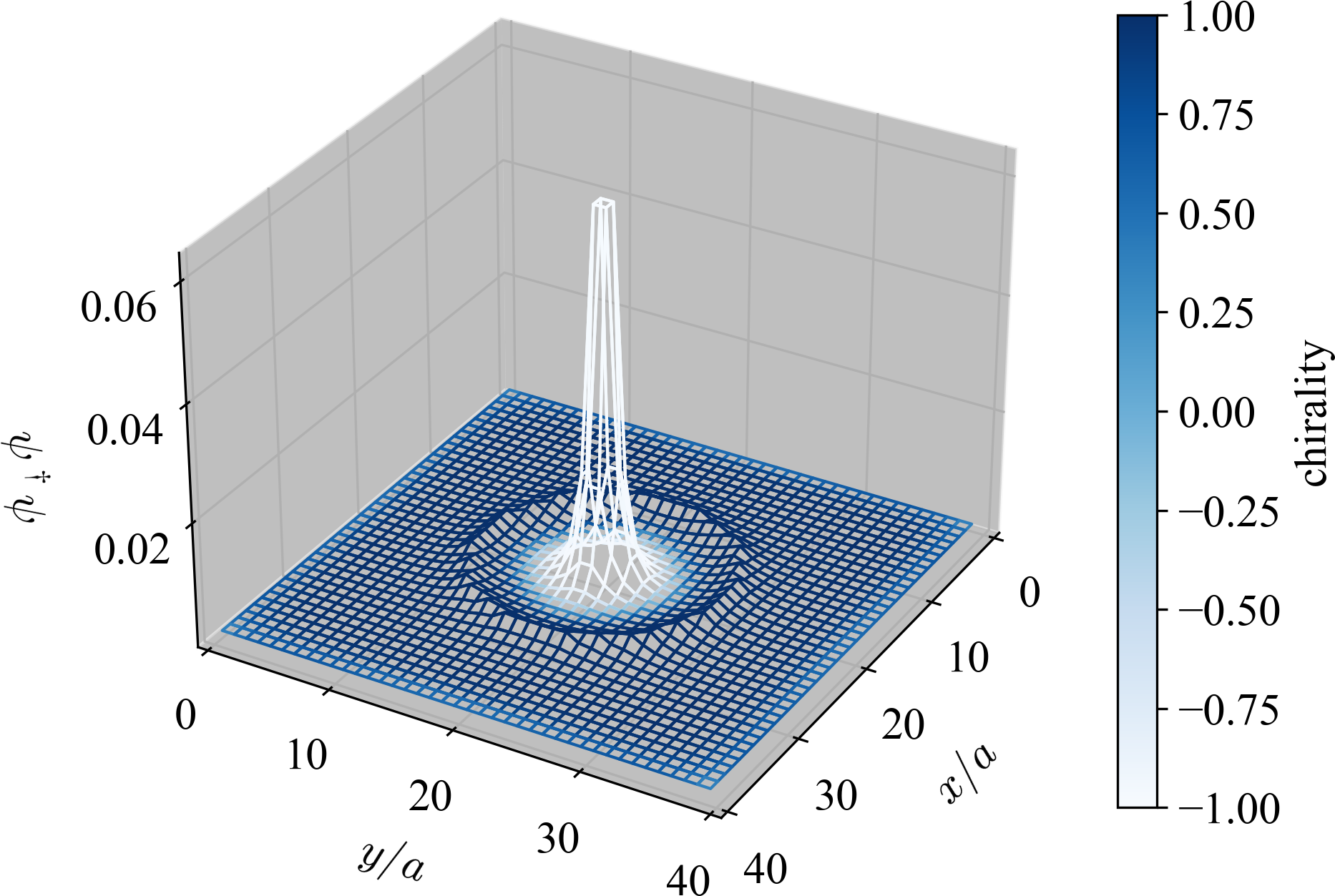}
     }\\
     \subfigure{
        \includegraphics[bb=0 0 448 301,height=0.3 \textheight]{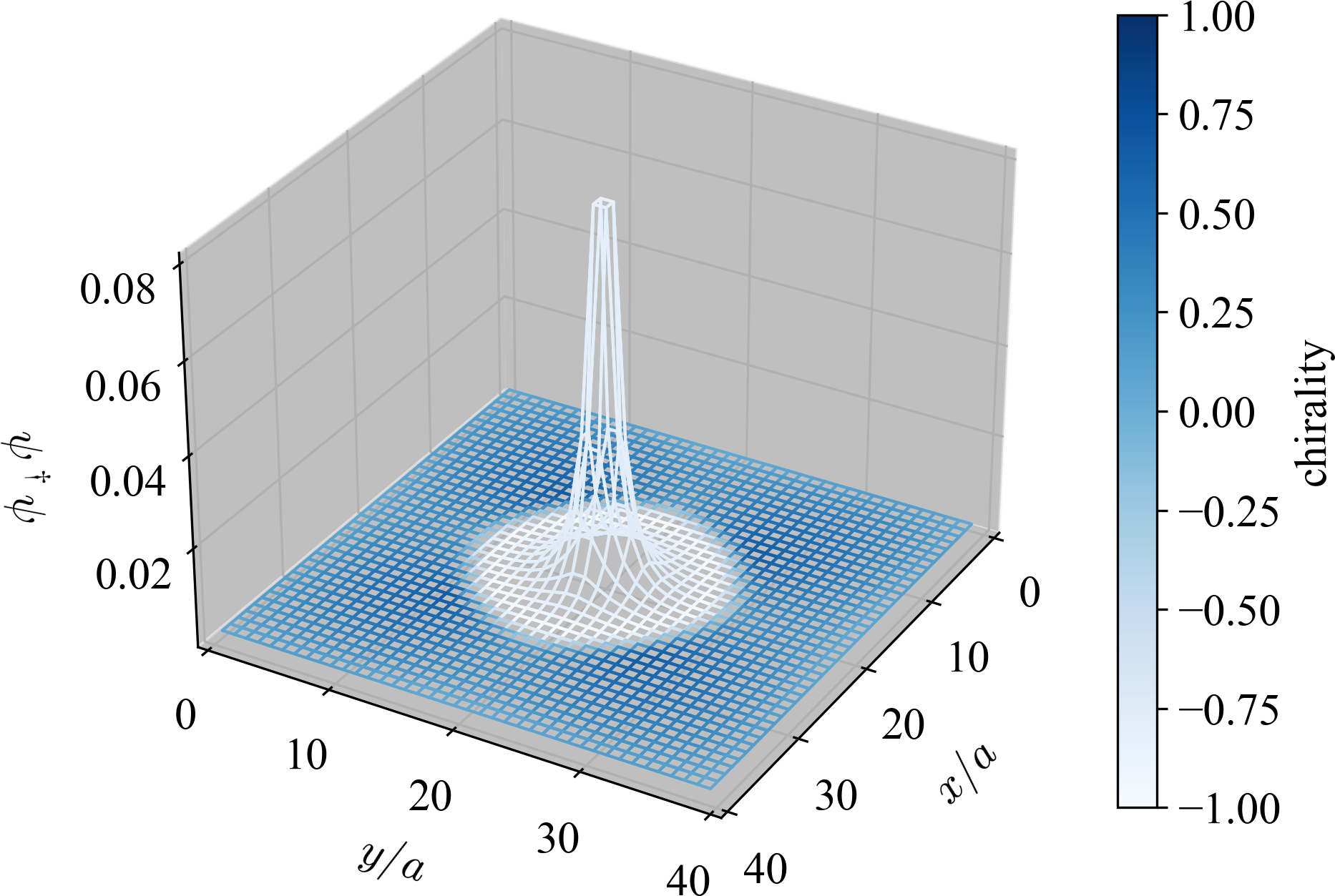}
     }
     \caption{The center-localized mode with $L=40a$, $m=15/L$, $r_0=L/4=10a$ and $r_1=0.001a$. The results for $\alpha=0.3$ (top panel), $\alpha=0.5$ (middle) and $\alpha=0.7$ (bottom) are presented. 
     The color gradation represents the chirality at each lattice point.} 
     \label{fig:S1_eigenstate_center}
    \end{center}
\end{figure}

The amplitude of this state is depicted in Figure \ref{fig:S1_eigenstate_center} when $\alpha=0.3,~0.5$ and $0.7$. The color gradation represents the chirality at each lattice point. We can see that a negative chiral mode appears at the center. Especially when $\alpha=1/2$, the tunneling effect between two near-zero modes at the center and the wall are mixed and the eigenvalues are slightly shifted from $E=0$ as shown in the bottom panel of Figure \ref{fig:AnomalyInflow_singular}.


\begin{figure}
\centering
\includegraphics[width=\textwidth,bb=0 0 361 311]{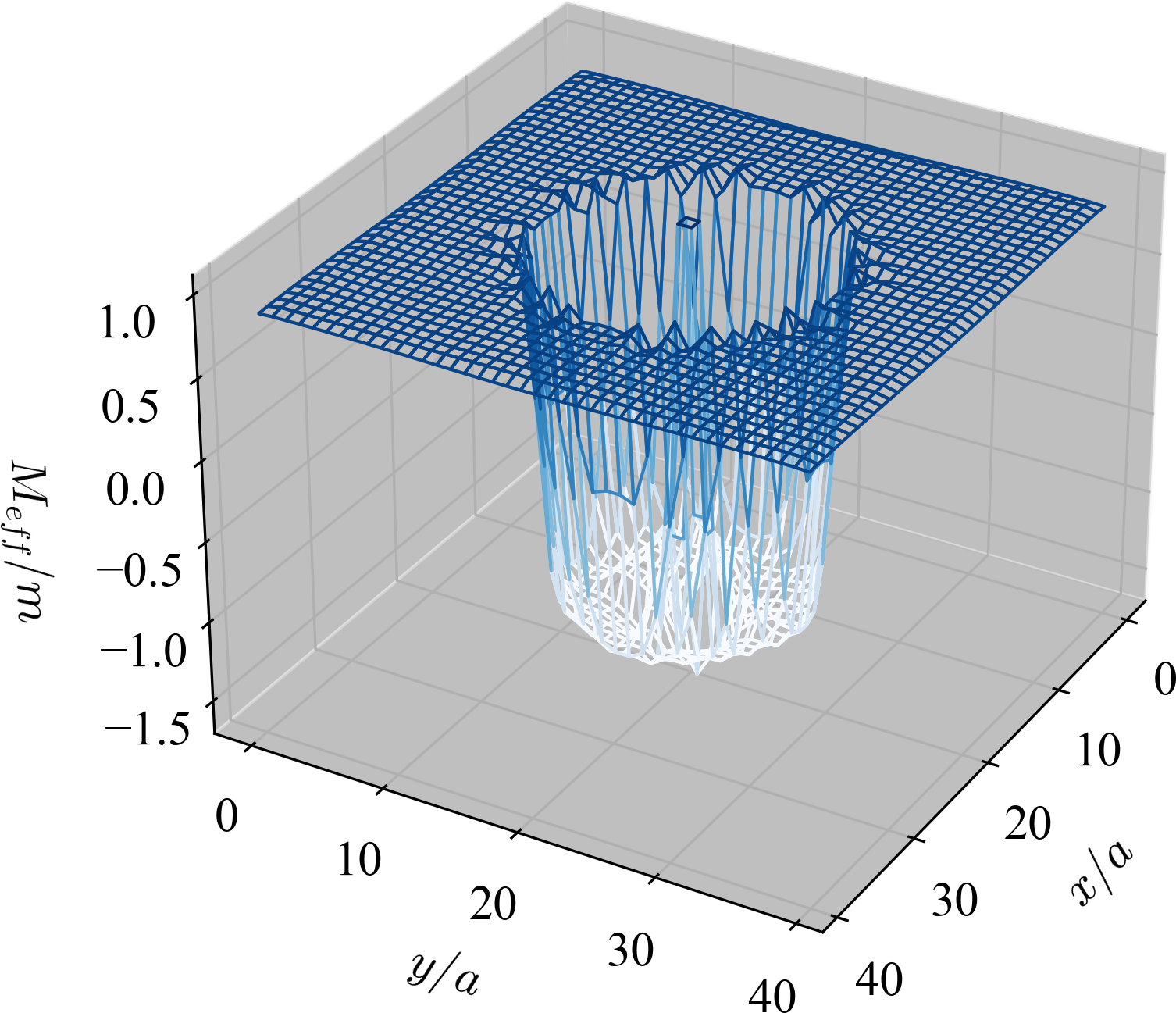}
\caption{The effective mass \eqref{eq:effectivemass} normalized by $m$ for one of the zero modes with $L=40a$, $r_0=10a$, $m=15/L$ and $\alpha=0.5$. The blue color represents the positive mass region and the white color denotes the negative mass region.}
\label{fig:S1effectivemass}
\end{figure}

The reason why the negative mode appears at the singular point of the gauge connection is that the Wilson term makes a positive mass region around the point. It is well-known that the Wilson term gives additive mass renormalization violating the chiral symmetry. 
We confirm the same phenomenon but it gives a positive mass only around the singular point. We define an ``effective mass''
\begin{align}\label{eq:effectivemass}
    M_{eff}= \frac{1}{ \psi(x)^\dagger \psi(x)}  \psi(x)^\dagger
 \qty(\epsilon m+\sum_{i=1,2}\frac{1}{2a}\nabla_i \nabla^\dagger_i  ) \psi(x),
 \end{align}
for a fermion field $\psi$. The second term is the Wilson term, which depends on the gauge field. We plot the effective mass for one of the zero modes in Figure \ref{fig:S1effectivemass}. The additive mass makes a positive mass region and a new domain-wall around the singular point. Since the orientation of the new domain-wall is opposite to the given one, the chirality of the center-localized mode is $-1$.

To find the center-localized mode in the continuum theory, we consider a modified Hermitian Dirac operator 
\begin{align} 
    H=\sigma_3 \qty(\sum_{j=1}^2 \sigma_j D_j -\frac{1}{M_{PV} } D^2 -m)    =\sigma_3 \mqty( -m-\frac{D^2}{M_{PV}} & e^{-i\theta} \qty( \pdv{}{r}-i \frac{1}{r} \pdv{}{\theta} -\frac{\alpha}{r} ) \\ e^{i\theta} \qty( \pdv{}{r}+i \frac{1}{r} \pdv{}{\theta} +\frac{\alpha}{r} ) & -m -\frac{D^2}{M_{PV}} )
\end{align}
Here we add a second derivative term in the Dirac operator as a Pauli-Villars regularization. $M_{PV}$ is the mass of the Pauli-Villars field. For simplicity, we ignore the positive mass region and focus on the area around the flux in the negative mass region. We find the eigenstate localized at the vortex with eigenvalue $E$,
\begin{align}
    \psi_j(r,\theta)=&A \mqty(\qty(-m -\frac{\kappa_{+}^2}{M_{PV}} +E) K_{j-\frac{1}{2}-\alpha}(\kappa_{+} r) e^{i(j-\frac{1}{2})\theta} \\ \kappa_{+} K_{j+\frac{1}{2}-\alpha}(\kappa_{+} r) e^{i(j+\frac{1}{2})\theta} ) \nonumber \\
    &+B\mqty(\qty(-m -\frac{\kappa_{-}^2}{M_{PV}} +E) K_{j-\frac{1}{2}-\alpha}(\kappa_{-} r) e^{i(j-\frac{1}{2})\theta} \\ \kappa_{-} K_{j+\frac{1}{2}-\alpha}(\kappa_{-} r) e^{i(j+\frac{1}{2})\theta} ),
\end{align}
where $j \in \mathbb{Z}+\frac{1}{2}$ is a total angular momentum, $A$ and $B$ are complex coefficients and $\kappa_{\pm}$ is defined as
\begin{align}
    \kappa_{\pm}= M_{PV} \sqrt{\frac{1}{2}\qty(1 \pm \sqrt{ 1-4 \frac{m}{M_{PV}} + 4 \frac{E^2}{M_{PV}^2}} )-\frac{m}{M_{PV}}  }. 
\end{align}
We assume $\psi(r) \to 0$ at $r=0$. In the large $M_{PV}$ limit, the solution $\psi$ converges to
\begin{align}
    \psi(r,\theta )\simeq & A  \mqty( -M_{PV} K_{j-\frac{1}{2}-\alpha}(M_{PV} r) e^{i(j-\frac{1}{2})\theta} \\ M_{PV} K_{j+\frac{1}{2}-\alpha}( M_{PV} r) e^{i(j+\frac{1}{2})\theta} ) \nonumber \\
    &+B \mqty( (-m+E) K_{j-\frac{1}{2}-\alpha}(\sqrt{m^2 -E^2} r) e^{i(j-\frac{1}{2})\theta} \\ \sqrt{m^2 -E^2} K_{j+\frac{1}{2}-\alpha}(\sqrt{m^2 -E^2} r) e^{i(j+\frac{1}{2})\theta} )
\end{align}
and we obtain the equation for $E$, 
\begin{align}
    \frac{-m+E}{\sqrt{m^2-E^2} } \qty(\frac{\sqrt{m^2-E^2}}{M_{PV}})^{ \abs{j+\frac{1}{2}-\alpha }-\abs{j-\frac{1}{2}-\alpha } } +1=0 .
\end{align}
Note that the chirality of this solution is almost $-1$. Only when $m>0$ and $\abs{j -\alpha}<1/2$, there exists the solution $E$ and it is expressed as the inverse function of
\begin{align}\label{eq:condition of E S1 center r1 is small}
    \alpha -[\alpha]= \frac{1}{2} \frac{ \log ( \frac{m-E}{2 M_{PV} })}{ \log ( \frac{ \sqrt{m^2 -E^2 }}{2 M_{PV} }) }.
\end{align}
It describes the same behavior of the white line in Figure \ref{fig:AnomalyInflow_singular}. Here, we put $M_{PV}=2/a$. See the appendix \ref{app:CLM} for details.

In the domain-wall fermion formulation, the eta invariant of $H$ describes the APS index on the negative mass region. Note that the topology of the negative mass region changes from $\mathbb{D}^2$ to $\mathbb{D}^2-\qty{0}$, which has the two $S^1$ boundaries at $r=r_0$ and $r=r_1$. Since there is no $U(1)$ gauge flux through $\mathbb{D}^2-\qty{0}$, the APS index turns into the difference of the eta invariant on each boundary
\begin{align}
    \text{ind}= \int_{r_1<r<r_0} \frac{F}{2\pi} -\frac{1}{2}  \eval{\eta}_{r=r_0} + \frac{1}{2}  \eval{\eta}_{r=r_1}=-\frac{1}{2}  \eval{\eta}_{r=r_0} + \frac{1}{2}  \eval{\eta}_{r=r_1}.
\end{align}
This expression provides a new description of the anomaly inflow that the $T$ anomaly on the new domain-wall at $r=r_1$ cancels that on the boundary at $r=r_0$.

In particular, Figure \ref{fig:AnomalyInflow_singular} indicates a mathematically interesting phenomenon at $\alpha=0.5$, where two near-zero modes appear. When $\alpha=0.5$, the Dirac operators on the two-dimensional disc and $S^1$ boundaries become real operators. The eta invariant of the Dirac operator on $S^1$ does not change as long as it retains the real structure. It means that the eta invariant is a spin bordism invariant of a one-dimensional manifold. The spin bordism group is equivalent to
\begin{align}
    \Omega_{1}^\text{spin}(\qty{pt}) \simeq \mathbb{Z}_2, 
\end{align}
where the identity element denotes an anti-periodic spinor and the generator expresses a periodic spinor on $S^1$. They appear at the wall when $\alpha=0$ and $\alpha=1/2$, respectively. The eta invariant corresponds to the number of the zero modes modulo two, which is called the mod two index \cite{AtiyahSinger1971TheIndex5} and only depends on the spin structure of $S^1$\footnote{The mod two index is related to the global anomaly \cite{Witten:1982fp}.}. In our system, there exists one zero mode at $r=r_0$ so we need another zero mode in the negative mass region. If one violates the real structure on the bulk like in the previous section, the another zero mode disappears. It reflects a mathematical equation
\begin{align}
    \Omega_{1}^{\text{spin}^c}(\qty{pt}) =0.
\end{align}


\chapter{Fractional Charge}
\label{sec:FractionalCharge}

In this section, we discuss an application of the curved domain-wall fermions to condensed matter physics. In the previous sections, we have considered Euclidean spacetimes, but here, we regard our lattice as the spacial part in a Minkowski space. Then the Hermitian Dirac operator we consider is a Hamiltonian of the Dirac fermions. In our domain-wall fermion formulation, the negative mass region represents a topological insulator and the positive mass region corresponds to the normal phase of matters. We introduce the Wilson term to eliminate the UV divergence around a magnetic vortex or monopole. The Wilson term also distinguishes the two phases as long as it is weakly perturbed. However, when a magnetic monopole or vortex is in the domain-wall, the Wilson term makes a positive mass shift due to a strong magnetic field near the monopole and generates a normal insulator region around such a defect. Then an electron zero mode is localized at its boundary and the defect acquires a fractional electric charge. In contrast to the standard argument \cite{Kazama:1976fm,Goldhaber:1977xw,Callias:1977cc,Yamagishi:1982wpTHE,Yamagishi1983Fermion-monopole, Grossman:1983yf,Zhao2012Amagneticmonopole,Tyner:2022qpr} where an artificial boundary condition is imposed around the defect, a chiral boundary condition arises as a result of the dynamical creation of the domain-wall. We solve the Hamiltonian in the presence of the vortex/monopole and show that our curved domain-wall fermion microscopically explains how the electric charge is captured by these solitons.
\

\section{Witten Effect}
\label{subsec:WittenEffect}

We first review the Witten effect in a four-dimensional flat continuum spacetime, which describes that a monopole becomes a dyon. We consider a Dirac fermion with a negative mass $-m$, then the regularized partition function is 
\begin{align}
    Z= \det \frac{\Slash{D}-m}{\Slash{D}+M_{PV}},
\end{align}
where $\Slash{D}= \gamma^\mu (\partial_\mu -i A_\mu)$ is the Dirac operator with a $U(1)$ gauge connection $A_\mu$ and $M_{PV}$ is the mass of the Pauli-Villars field. By an axial $U(1)$ transformation, the negative mass $-m$ turns into $+m$ and we obtain a topological term as a $U(1)$ phase from the chiral anomaly
\begin{align}
    Z= \det \frac{\Slash{D}+m}{\Slash{D}+M_{PV}} \exp(i \pi \frac{1}{32 \pi^2} \int d^4x \epsilon^{\mu\nu \rho \sigma} F_{\mu \nu } F_{\rho \sigma} ),
\end{align}
where $\epsilon^{\mu\nu \rho \sigma} $ is the anti-symmetric Levi-Civita tensor. This $U(1)$ phase is generated by the chiral anomaly and is well-known as the $\theta$ term at $\theta=\pi$.

In the presence of the $\theta$ term, the Maxwell equations change to
\begin{align}
    \partial_\mu F^{\mu \nu} = -\frac{\theta}{ 8\pi^2} \partial_\mu \epsilon^{\mu\nu \rho \sigma}  F_{\rho \sigma} 
\end{align}
for arbitrary $\theta$. The $\nu=0$ component gives the relation between the divergence of the electric field $\mathbf{E}$ and the magnetic field $\mathbf{B}$. When we put a magnetic monopole with a total flux $n$, then the electric charge $q_e$ is given by the Gauss law 
\begin{align}
    q_e= \int d^3 x \nabla \cdot \mathbf{E}=-\frac{\theta}{ 4\pi^2} \int d^3 x \nabla \cdot \mathbf{B} = -\frac{\theta}{ 2\pi}  n,
\end{align} 
where $n$ is an integer called topological charge. Since $\theta$ is equal to $\pi$ in our system, a fractional charge $q_e= - 1/2$ is obtained with a single magnetic monopole $n=1$. The partition function with the $\theta$ term has a $2\pi$ periodicity of $\theta$ so that $ \pmod{\mathbb{Z}}$ ambiguity is allowed to $q_e$.

A similar discussion applies to the $(2+1)$-dimensional system. The topological term is given by the Chern-Simons term with an integer level $k$ and it modifies the Maxwell equations to 
\begin{align}
    \partial_\mu F^{\mu \nu} = -\frac{k}{ 8\pi^2}  \epsilon^{\nu \rho \sigma}  F_{\rho \sigma} .
\end{align}
By the Gauss law, the magnetic flux obtains an electric charge
\begin{align}
    q_e= \int d^2 x \nabla \cdot \mathbf{E}=-\frac{k}{ 2\pi} \int d^2 x \nabla \cdot \mathbf{B} = -k  \alpha.
\end{align}
When $\alpha=0.5$ and $k=1$, the electric charge is fractional $q_e=-1/2$.

\section{$S^1$ Domain-wall System with Magnetic Flux $\alpha=\frac{1}{2}$}
\label{subsec:S1_vortex}

\begin{figure}
    \centering
    \includegraphics[bb=0 0 488 296,width=\textwidth]{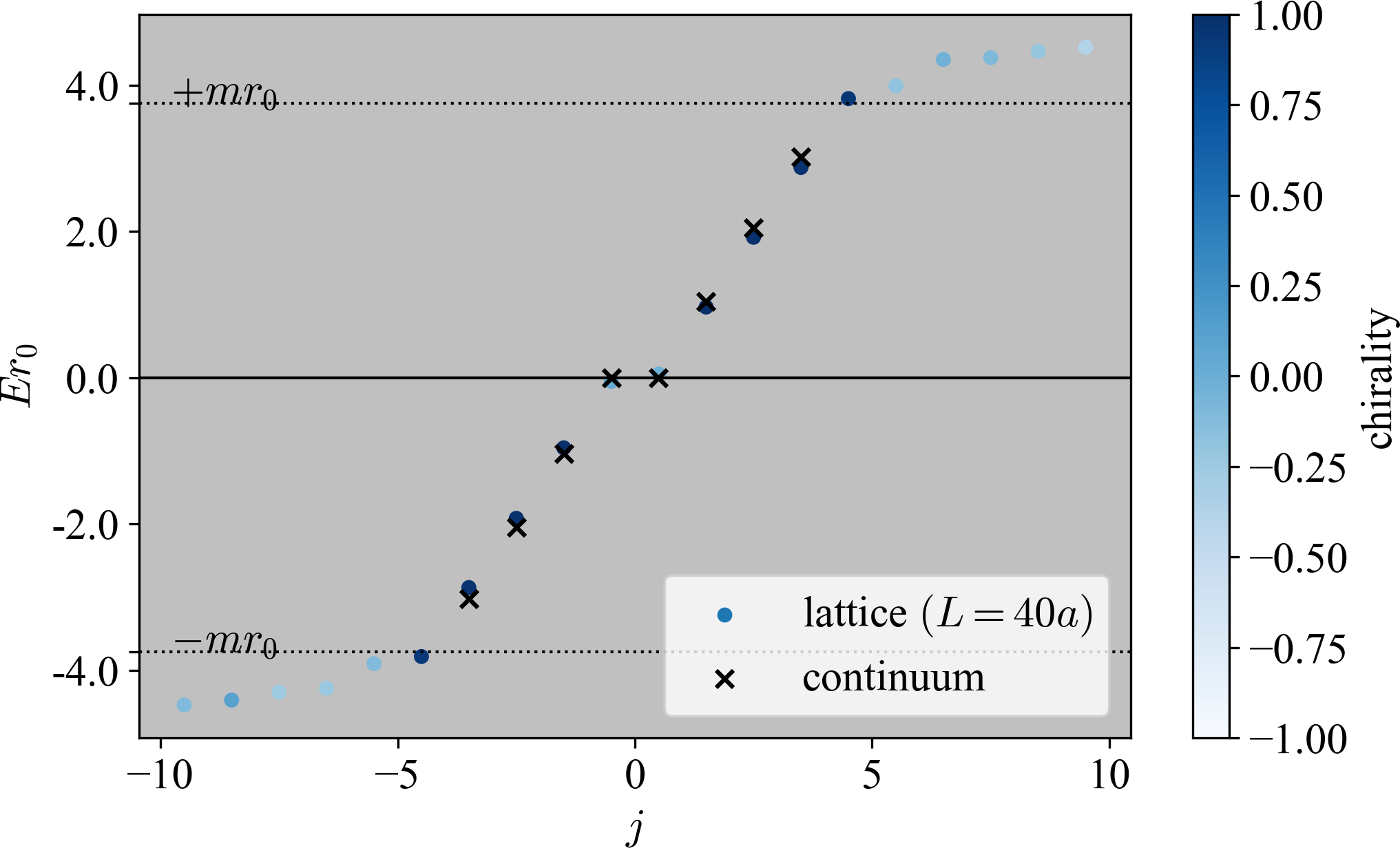} 
    \caption{The eigenvalue spectrum of the Wilson-Dirac operator \eqref{eq:Hermitian Wilson Dirac op of S^1 in R^2 with flux} when $L=40a$, $r_0=10$, $m=15/L$ and $A=0.5$. The color gradation represents the chirality, which is the expectation value of $\gamma_\text{normal}$. The crosses are their continuum predictions \eqref{eq:condition of E S1}.} 
    \label{fig:eigenvalue_A=0.5}
\end{figure}

Let us set the Fermi energy at $E=0$ so that all negative modes are occupied or half-filled. Figure \ref{fig:AnomalyInflow_singular} indicates that one valence band at the vortex and one conduction band on the wall meet at $E=0$ when $\alpha=0.5$. We plot the eigenvalue spectrum of the Wilson-Dirac operator \eqref{eq:Hermitian Wilson Dirac op of S^1 in R^2 with flux} in Figure \ref{fig:eigenvalue_A=0.5}. The two near-zero modes are mixed and the eigenvalues are slightly split from $E=0$. When $\alpha>0.5$, one valence band at the vortex becomes a conduction band and one conduction band at the wall turns into a valence band. Then one valence electron at the vortex is pumped to the wall.

\begin{figure}
    \begin{minipage}[b]{0.45\linewidth}
    \centering
    \includegraphics[bb=0 0 412 298,width=\textwidth]{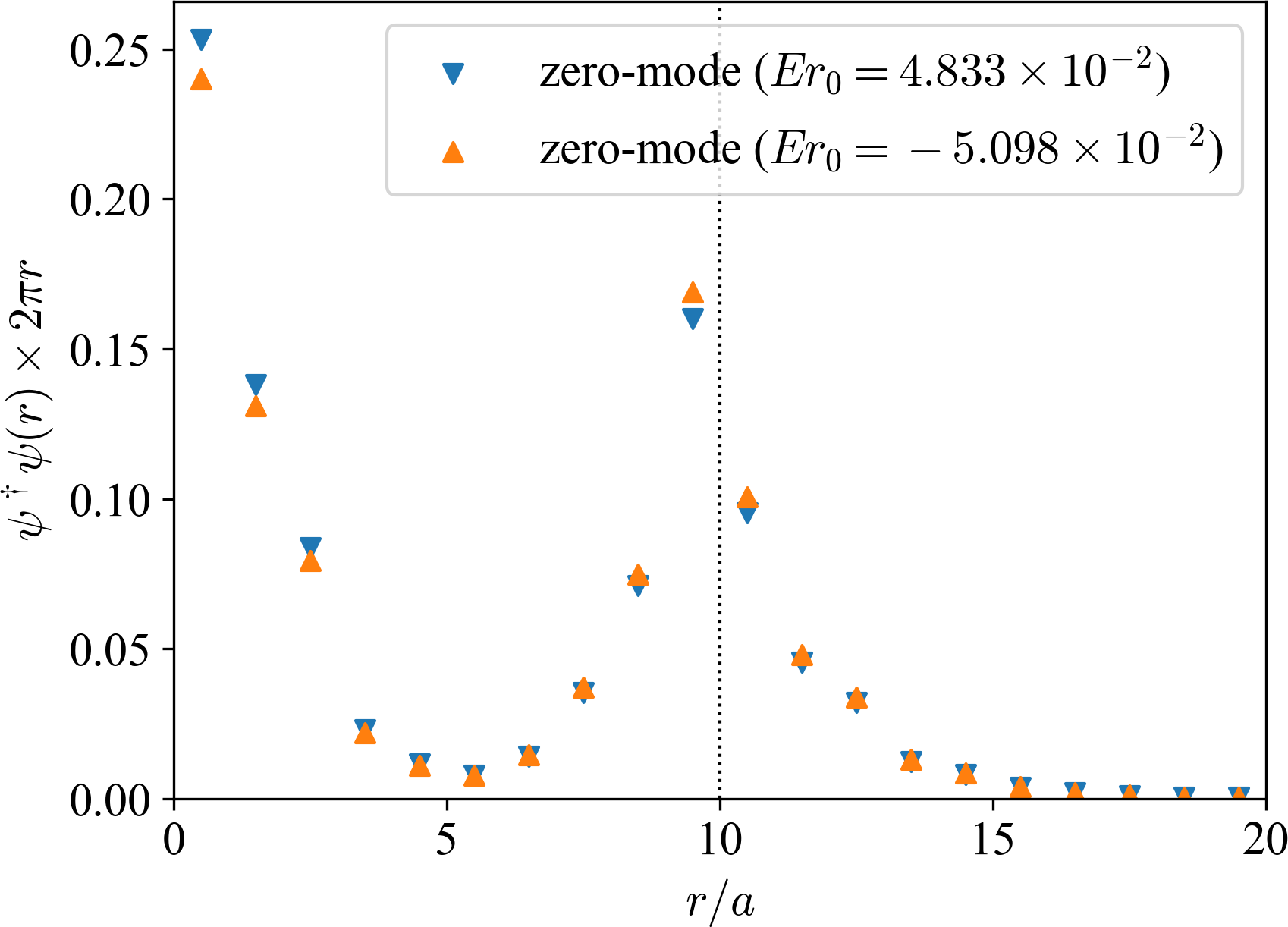}
    \end{minipage}
    \hfill
    \begin{minipage}[b]{0.45\linewidth}
        \centering
        \includegraphics[bb=0 0 412 298,width=\textwidth]{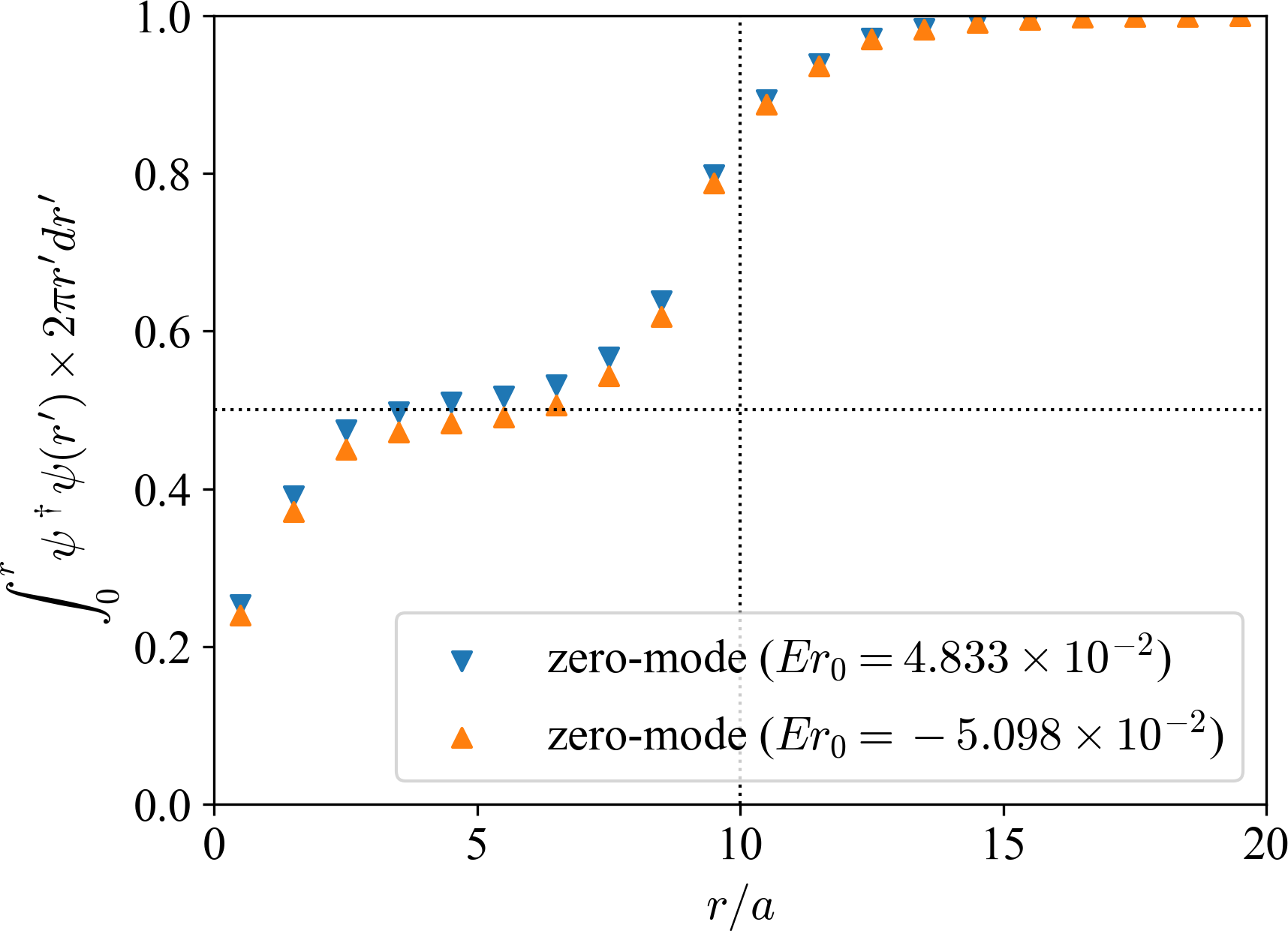}
        \end{minipage}
    \caption{Left panel: The amplitude of the two near-zero modes in the $r$ direction when $L=40a$, $r_0=10$, $m=15/L$ and $A=0.5$. Right: Cumulative distribution of them.}
    \label{fig:zeromodes_A=0.5}
\end{figure}

When $\alpha=0.5$, we plot the amplitude $ \psi^\dagger \psi(r ) \times 2 \pi r$ and the integral up to $r$ of two near-zero modes in Figure \ref{fig:zeromodes_A=0.5}. There are two near-zero modes localized at the vortex and the wall, and they are mixed. The $50 \%$ of the eigenstate is observed at the vortex and the remaining $50 \%$ stands at the wall.

This gives us a microscopic description of the vortex acquiring a fractional charge $1/2$. In our lattice setup, we have seen that the Wilson term in the presence of the intense gauge field makes a new domain-wall around the vortex and it confines an electron. When $\alpha=1/2$, the time-reversal symmetry is restored and it does not cost any energy to catch the electron by the cobordism argument. Then there exist two near-zero modes and they maximally mix due to the tunneling effect. This gives a microscopic description of how the vortex obtains the fractional charge $1/2$.





\section{$S^2$ Domain-wall System with Magnetic monopole}
\label{subsec:S2_monopole}

\begin{figure}
\centering
\includegraphics{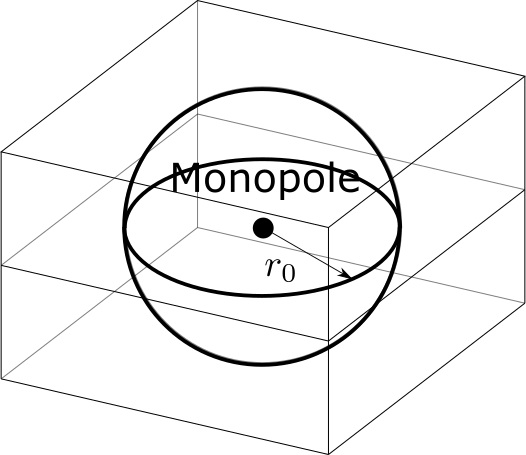}
\caption{Schematic illustration of our lattice set-up. We put a $S^2$ domain-wall with the radius $r_0$ and assign the monopole at the center.}
\label{fig:S2DW_Mono}
\end{figure}

We consider a $S^2$ domain-wall fermion in a three-dimensional system in the presence of a Dirac monopole. In the continuum space, we have calculated the Hamiltonian and obtained the eigenvalues of the edge modes in \ref{subsec:Curved_conti_S2}. In this section, we numerically solve the lattice version of the Dirac equation \eqref{eq:Hermitian Dirac operator for S^2 in R^3}. This system is illustrated in Figure \ref{fig:S2DW_Mono}.

The monopole at the center of the $S^2$ generates the chiral anomaly on the boundary system, which is described by Atiyah-Singer (AS) index of the Dirac operator on $S^2$. The AS index gives a relationship between the number of the zero modes and the topological charge of the gauge connection. However, a singularity at the monopole arises since the AS index is one of the cobordism invariants. Similar to the previous section, we expect a dynamical creation of the domain-wall around the monopole and it changes the topology of the negative mass region. We find extra zero modes around the monopole and they cancel the chiral anomaly on the wall.

On the square lattice, the $U(1)$ gauge connection generated by the monopole is assigned as link variables
\begin{align}\label{eq:LinkVariables_Mono}
    U_\mu(x)= \exp( i \int_{x+a \hat{\mu}}^x A )~(A= n\frac{1-\cos \theta}{2} d\phi).
\end{align}
Here we take a polar coordinate $(r,\theta,\phi)$ with $(x_0,y_0,z_0)=( a(N-1)/2, a(N-1)/2, a(N-1)/2)$ as a center. We also impose the periodic boundary condition for all directions. It gives non-trivial link variables on the links between $x=a(N-1)$ and $x=aN=L \sim 0$ for all $y $ and $z$, and those between $y=a(N-1)$ and $y=L \sim 0$ for all $x $ and $z$. Note that $A$ is not defined on the string between $(x_0,y_0,z_0)$ and $(x_0,y_0,0)$, which is the so-called Dirac string. On the $(x,y)$-plain, this string looks like a magnetic vortex with a total flux $n$. However, the Aharanov-Bohm effect disappears when $n$ is an integer.

Let us consider the Wilson-Dirac operator as a Hamiltonian in the presence of the Dirac monopole 
\begin{align}\label{eq:Hermitian Wilson Dirac op of S^2 in R^3 with Mono}
    H =\frac{1}{a}\bar{\gamma} \qty(\sum_{i=1}^3\qty[\gamma_i\frac{\nabla_i-\nabla^\dagger_i}{2} +\frac{1}{2}\nabla_i \nabla^\dagger_i ]+\epsilon am ), 
\end{align}
where $\nabla_i$ is the covariant forward difference operator in the $i$-direction given by the link variables \eqref{eq:LinkVariables_Mono}. The gamma matrices $\bar{\gamma}= \sigma_3 \otimes 1 ,~\gamma_i=\sigma_1 \otimes \sigma_i$ are the same as those in the continuum theory. $\bar{\gamma} $ is interpreted as a gamma matrix in the time direction. Since the Wilson-Dirac operator $H$ anti-commutes with $ \tau=\sigma_1 \otimes 1$, the spectrum is symmetric in the positive and negative directions. The gamma matrix in the normal direction to the $S^2$ domain-wall is given by \eqref{eq:S2 gamma matrix on lattice}.

\begin{figure}
    \subfigure{	
    \includegraphics[bb=0 0 501 309, height=0.4 \textheight]{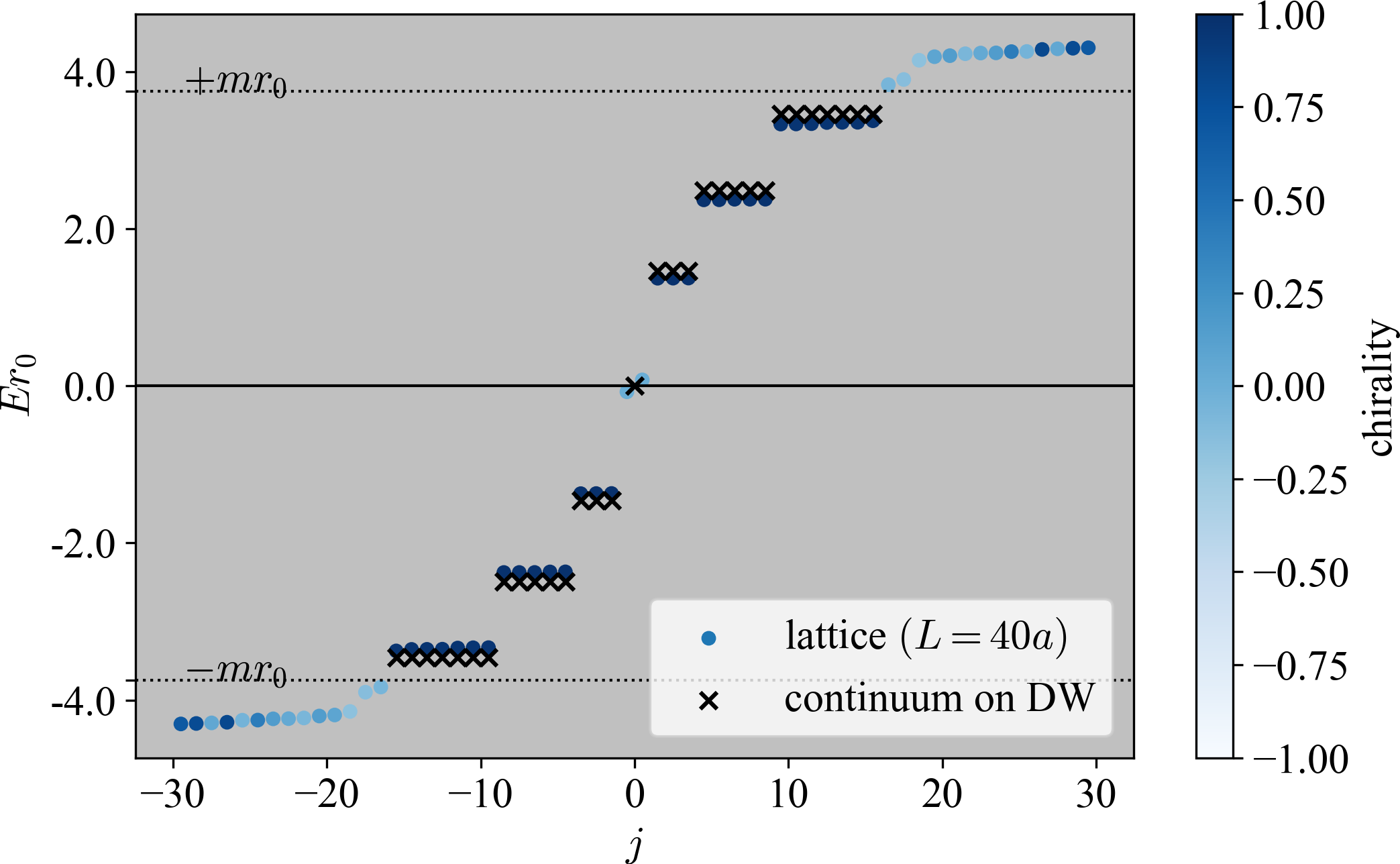}
    }\\
    \subfigure{	
    \includegraphics[bb=0 0 501 309, height=0.4 \textheight]{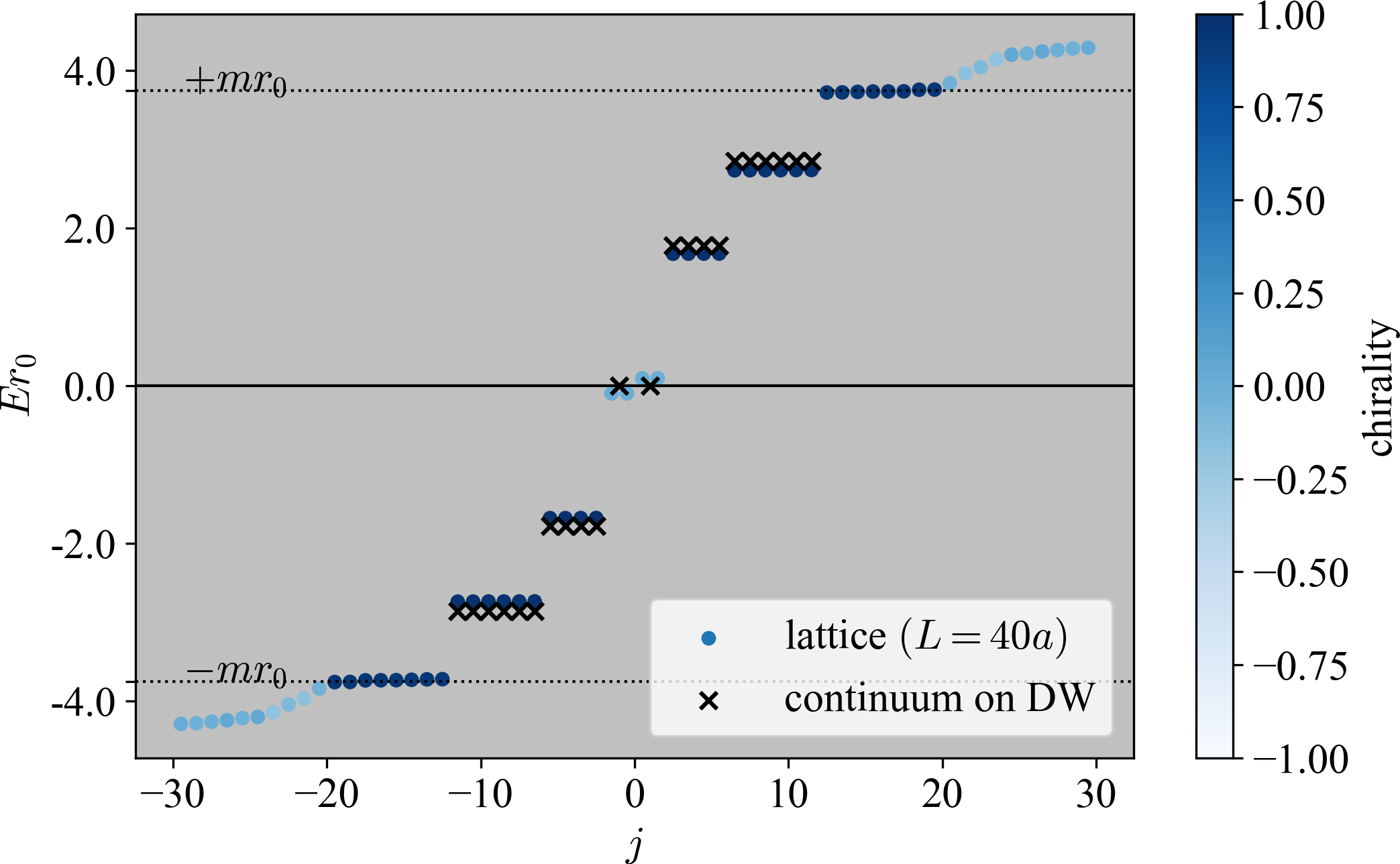}
    }
\caption{The eigenvalue spectrum of the Wilson-Dirac operator \eqref{eq:Hermitian Wilson Dirac op of S^2 in R^3 with Mono} when $L=40a$, $r_0=10$, $m=15/L$. The cases with $n=1$ and $n=-2$ are plotted in the top and bottom panels, respectively. The color gradation represents the chirality, which is the expectation value of $\gamma_\text{normal}$. The crosses are their continuum predictions \eqref{eq:condition of E S2}.}
\label{fig:S2_Monopole_eigenvalue_chi}
\end{figure}

We plot the energy spectrum of the Hamiltonian \eqref{eq:Hermitian Wilson Dirac op of S^2 in R^3 with Mono} with the topological charge $n=1$ and $n=-2$ in Figure \ref{fig:S2_Monopole_eigenvalue_chi}. Here we fix $L=40a$, $m=15/L$ and $r_0=L/4=10a$. The blue gradation shows their chirality and the crosses are the continuum predictions \eqref{eq:condition of E S2}. In the case with the $S^1$ domain-wall, the eigenstates with $\gamma_\text{normal}=+1$ appear between $E=m$ and $-m$ and are localized at the wall\footnote{Precisely, if we impose an appropriate boundary condition at the monopole by hand, we find the zero mode localized there \cite{Yamagishi:1982wpTHE, Yamagishi1983Fermion-monopole, Yamagishi1984Magnetic}. However, in the first-order differential system, how the boundary condition is chosen was not discussed. We need a second-order derivative to determine the boundary condition \cite{Zhao2012Amagneticmonopole, Aoki:2023lqp}.}. However, we can see an extra zero mode, which is absent from the spectrum in the continuum theory.


\begin{figure}
\centering
\subfigure{	
\includegraphics[bb=0 0 415 305, width=.45\columnwidth]{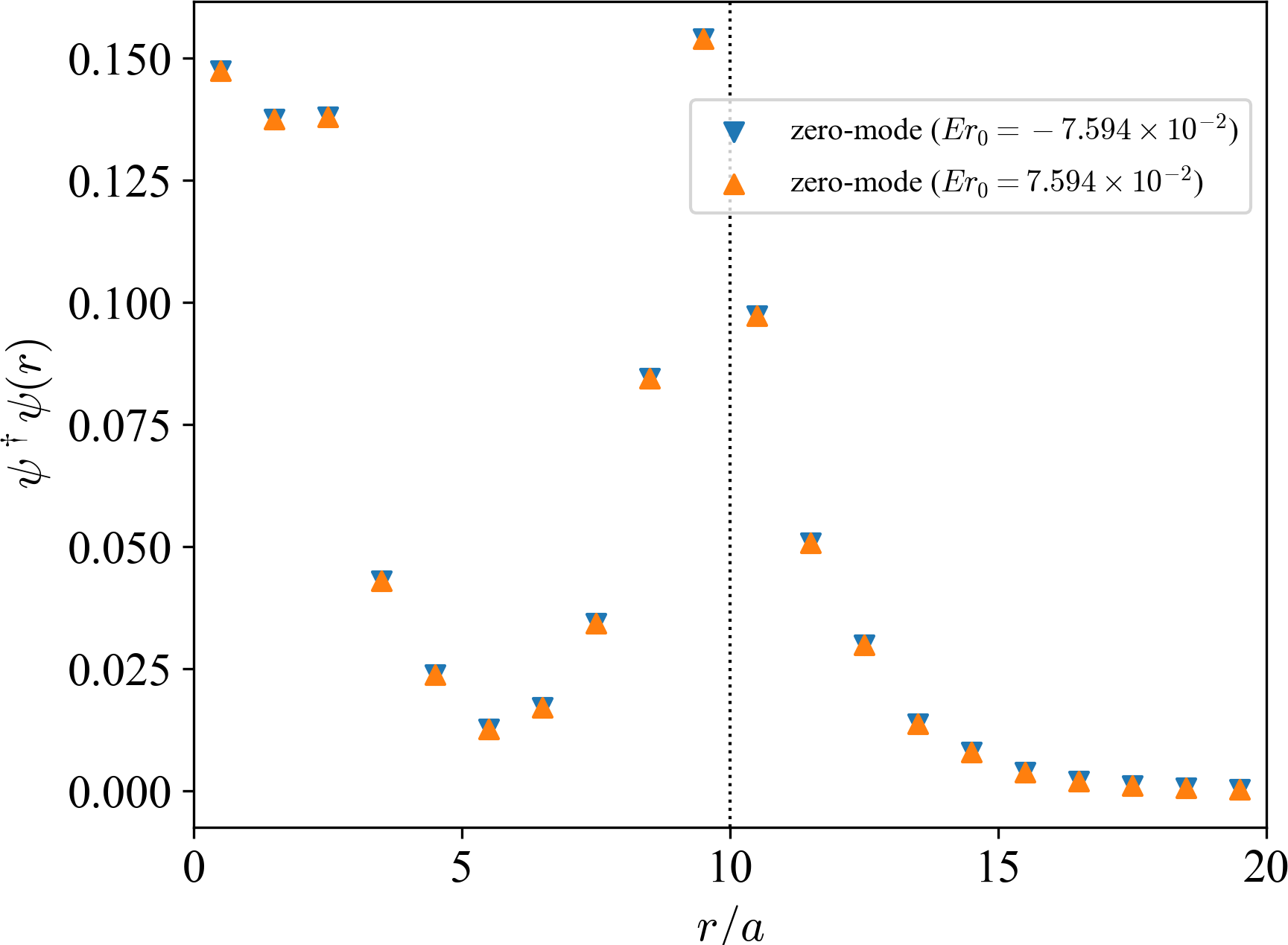} 
}
\subfigure{	
\includegraphics[bb=0 0 415 305, width=.45\columnwidth]{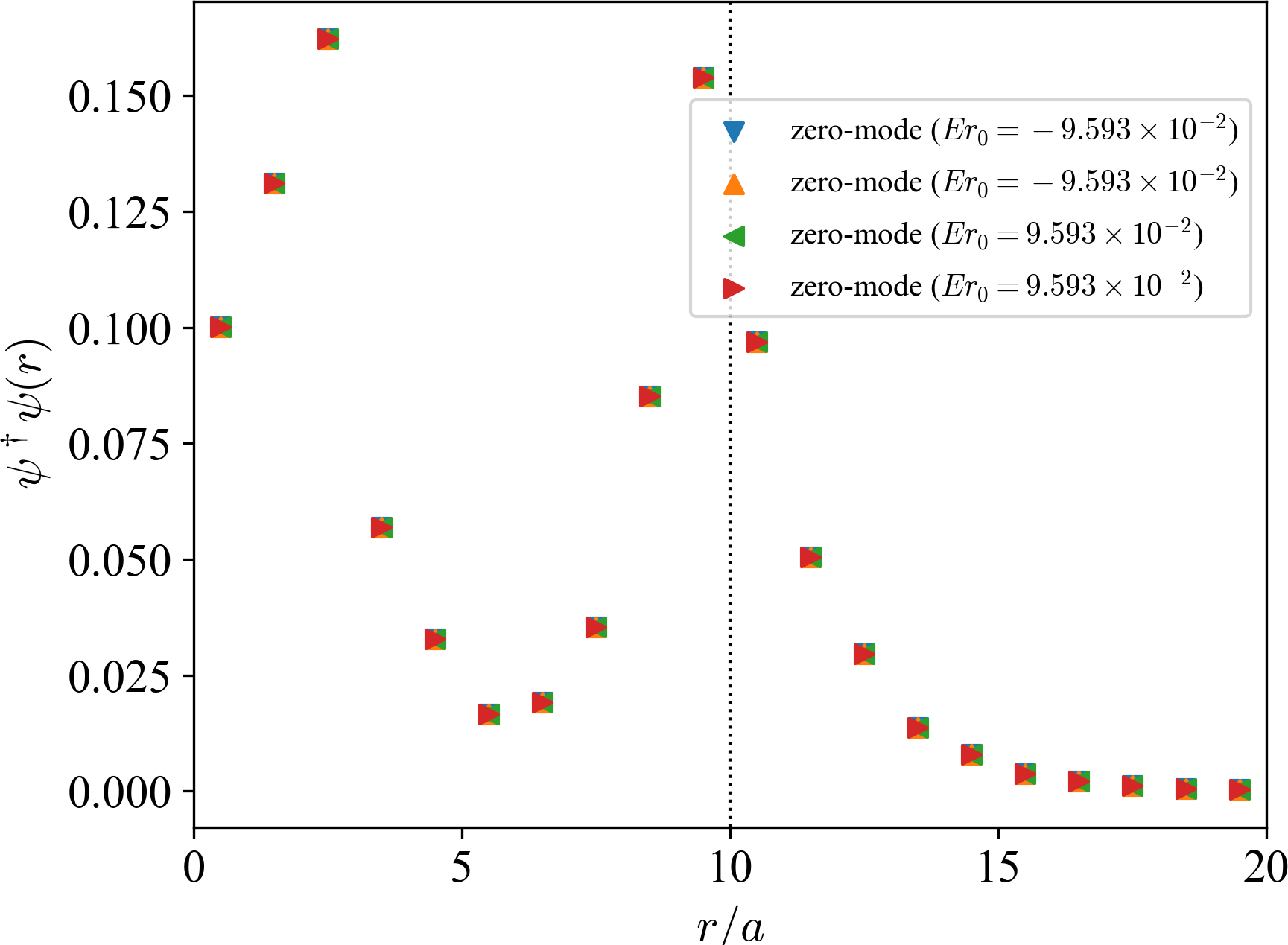}
}
\caption{The amplitude of the near-zero modes in the $r$ direction when $L=40a$, $r_0=10$, $m=15/L$. The left panel is the case with $n=1$, where two near-zero modes appear. The right panel is that with $n=-2$ and we can see four near-zero modes.}
\label{fig:S2_Monopole_eigenstate}
\end{figure}

\begin{figure}
    \centering
    \subfigure{	
    \includegraphics[bb=0 0 415 305, width=.45\columnwidth]{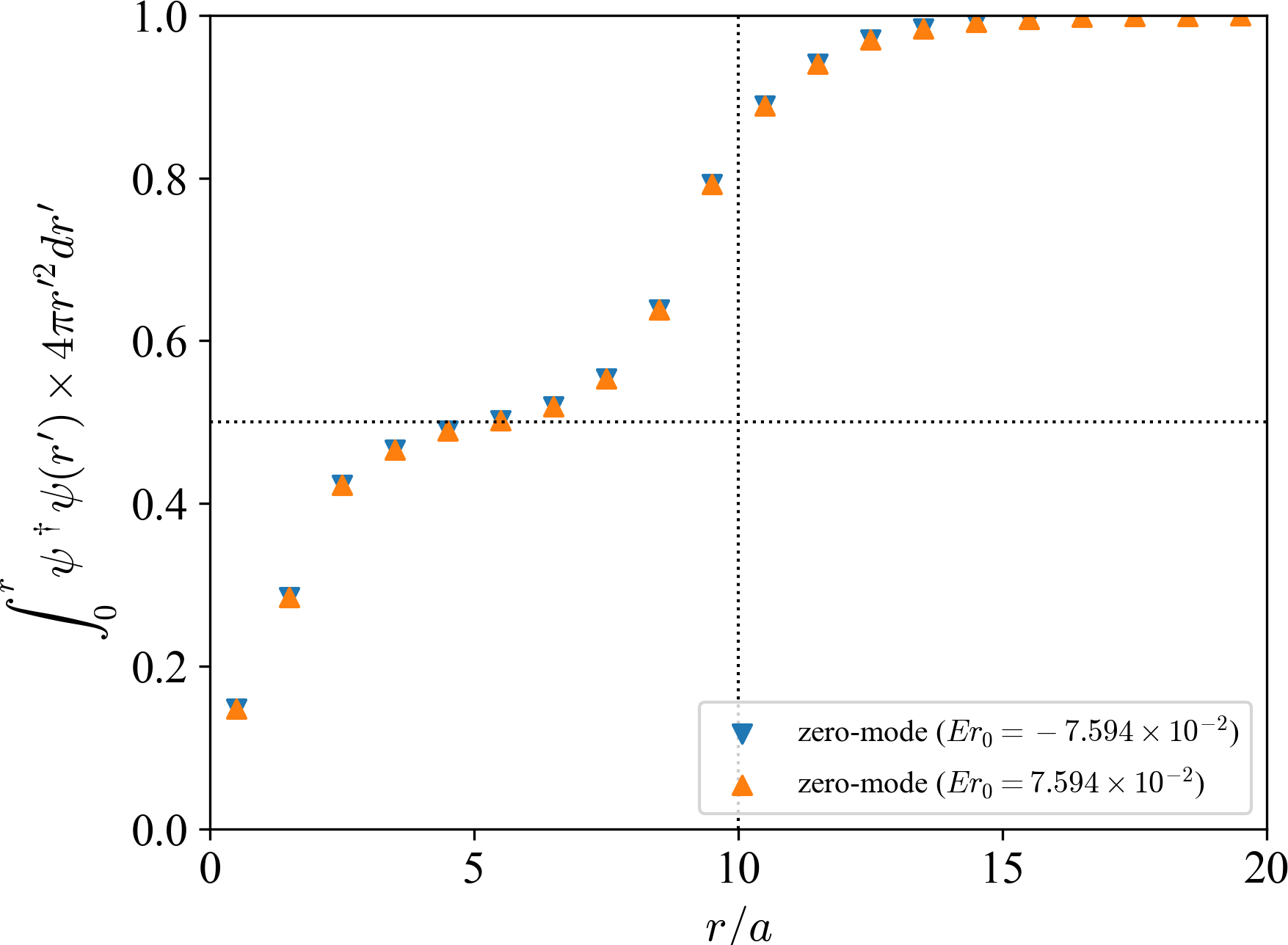}
    }
    \subfigure{	
    \includegraphics[bb=0 0 415 305, width=.45\columnwidth]{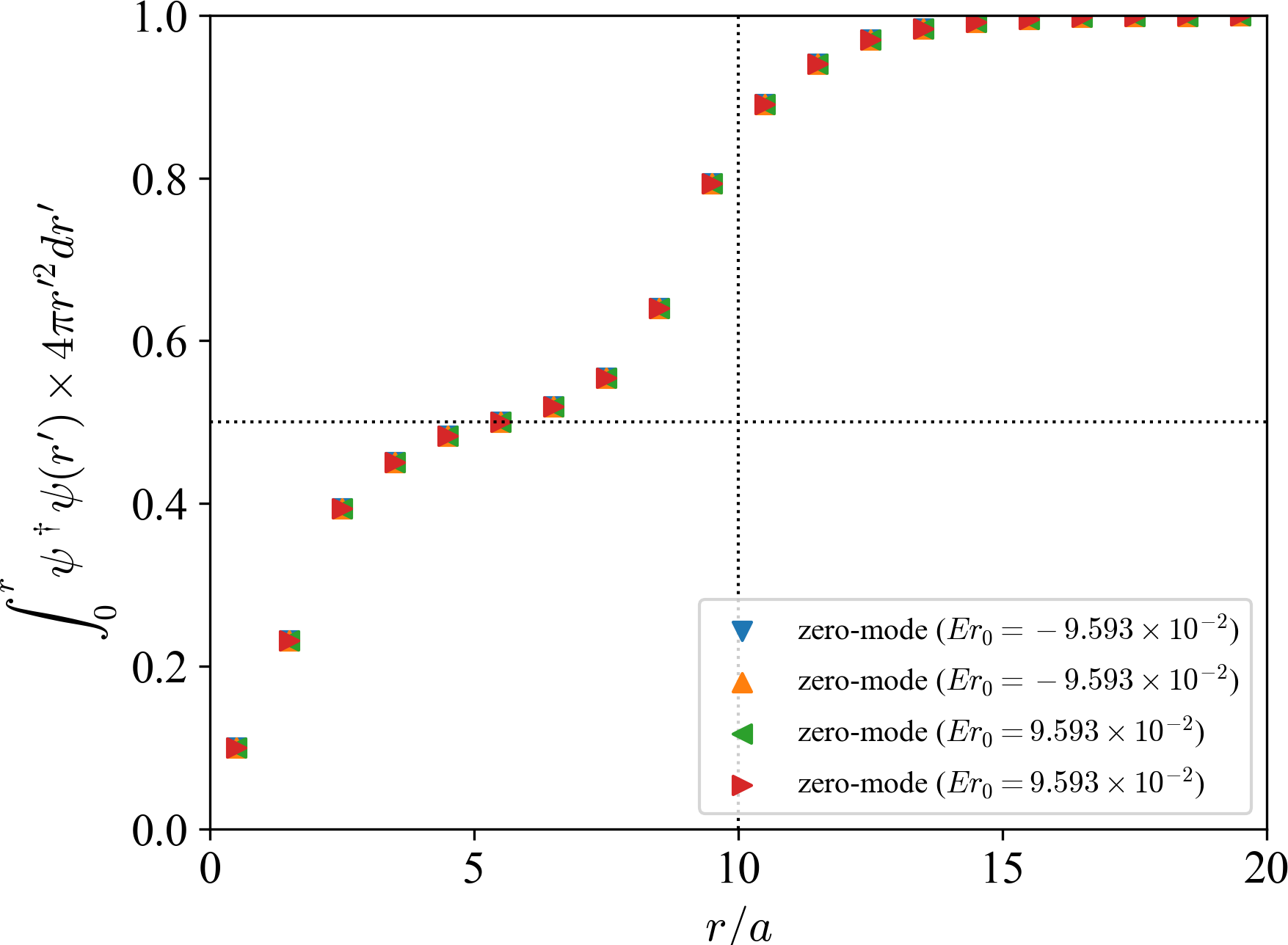}
    }
    \caption{The cumulative distribution of the near-zero modes in the $r$ direction when $L=40a$, $r_0=10$, $m=15/L$. The left panel is the case with $n=1$, where two near-zero modes appear. The right panel is that with $n=-2$ and we can see four near-zero modes.}
    \label{fig:S2_Monopole_eigenstate_cumsum}
\end{figure}

The amplitude of the zero modes in the $r$-direction is plotted in Figure \ref{fig:S2_Monopole_eigenstate}. We can see peaks at $r=0$ and $r=r_0$, which displays the tunneling effect of the zero modes between the monopole and the $S^2$ domain-wall. Figure \ref{fig:S2_Monopole_eigenstate_cumsum} indicates that the $50~\%$ of the zero mode amplitude is localized at the wall. 

\begin{figure}[]
    \begin{center}
     \subfigure{	
     \includegraphics[bb=0 0 353 298, width=.45\columnwidth]{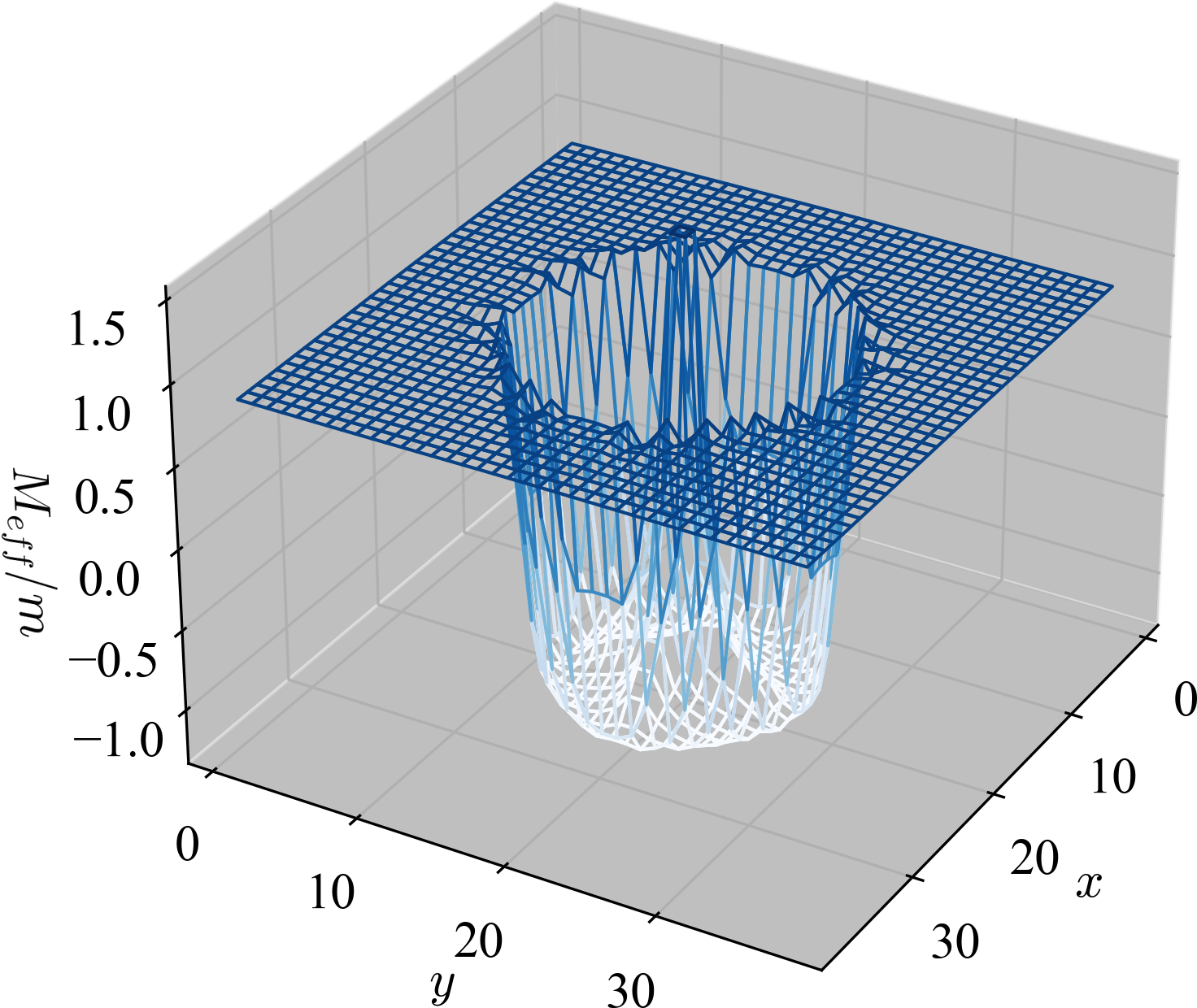}
     }
     \subfigure{	
     \includegraphics[bb=0 0 353 298, width=.45\columnwidth]{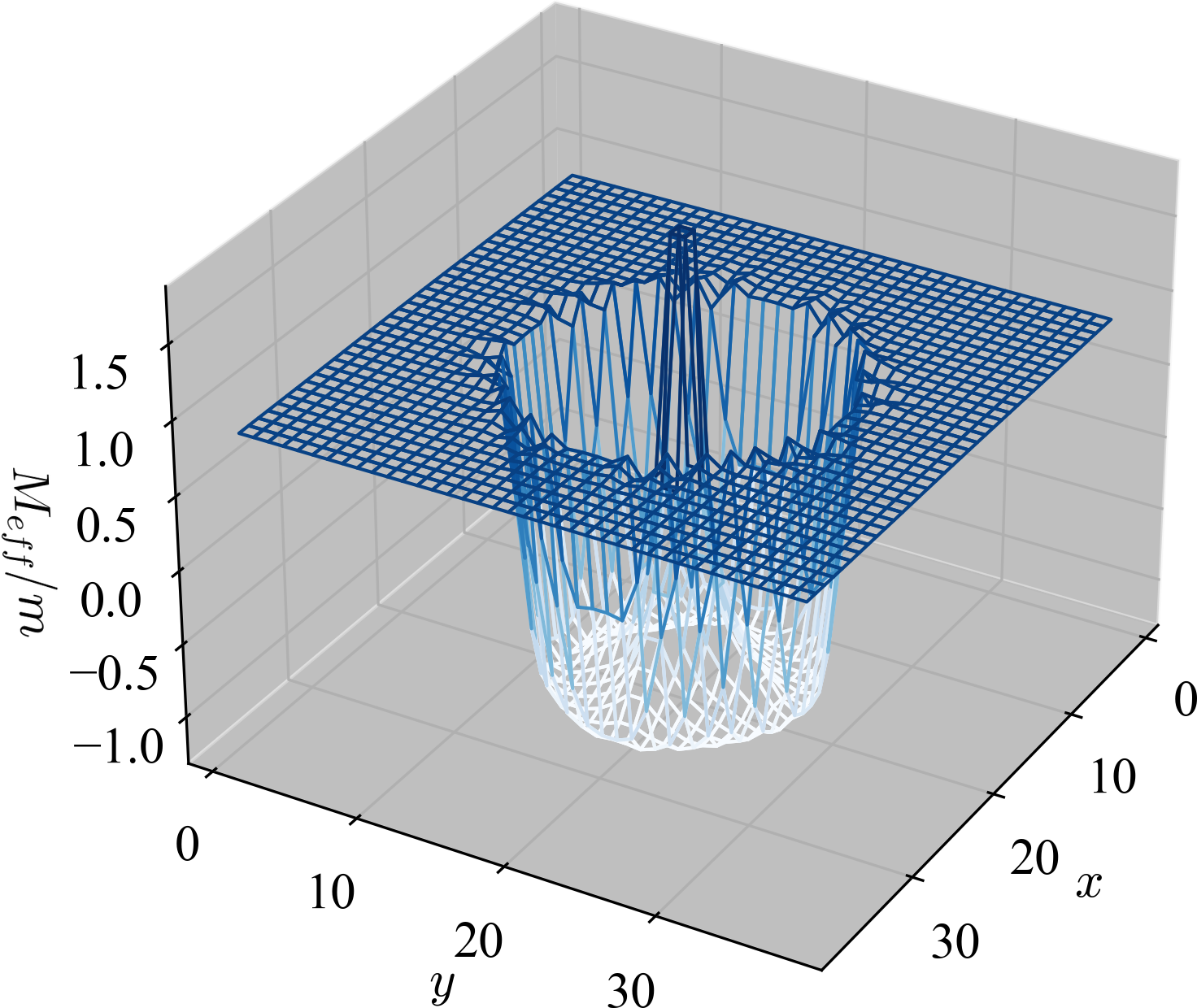}
     }
     \caption{The effective mass for the zero mode \eqref{eq:effective mass for three D} normalized by $m$ when $L=40a$, $r_0=10$ and $m=15/L$ on the $z=aN/2$ plane. The left panel is for the topological charge $n=+1$ and the right one is for $n=-2$. Inside the domain-wall $r<r_0$, there is a small region where $M_\text{eff}/m$ is positive.
     }
     \label{fig:S2_Monopole_EffectiveMass}
    \end{center}
\end{figure}

Let us evaluate the local effective mass around the monopole,
\begin{align}\label{eq:effective mass for three D}
    M_{\text{eff}}=\frac{1}{ \psi(x)^\dagger \psi(x)} \psi(x)^\dagger
 \qty(\epsilon m+\sum_{i=1}^3\frac{1}{2a}\nabla_i \nabla^\dagger_i  ) \psi(x).
\end{align}
The numerical result is plotted in Figure \ref{fig:S2_Monopole_EffectiveMass}. We can see another domain-wall at the origin. The Wilson term shifts the mass from negative to positive there. Then the zero modes with $\gamma_\text{normal}=-1$ appear since the orientation is opposite to the original $S^2$ domain-wall at $r=r_0$. This gives the microscopic description of the Witten effect and the monopole dresses the fractional charge.

The center-localized mode is protected by the nature of the cobordism invariance of the AS index. To count the index, we consider the product of $\gamma_\text{normal}$ and $\tau$, 
\begin{align}
\gamma_{\text{normal}}^{S^2}= \gamma_\text{normal} \tau=1 \otimes \frac{x^i \sigma^i}{r}.
\end{align}
On the domain-wall, the trace of $\gamma_{\text{normal}}^{S^2}$ is related to the AS index on $S^2$ \cite{Aoki:2023lqp}. We define the $S^2$-chirality as the expectation value of $\gamma_\text{normal}^{S^2}$.

\begin{figure}
    \subfigure{	
    \includegraphics[bb=0 0 501 309, height=0.4 \textheight]{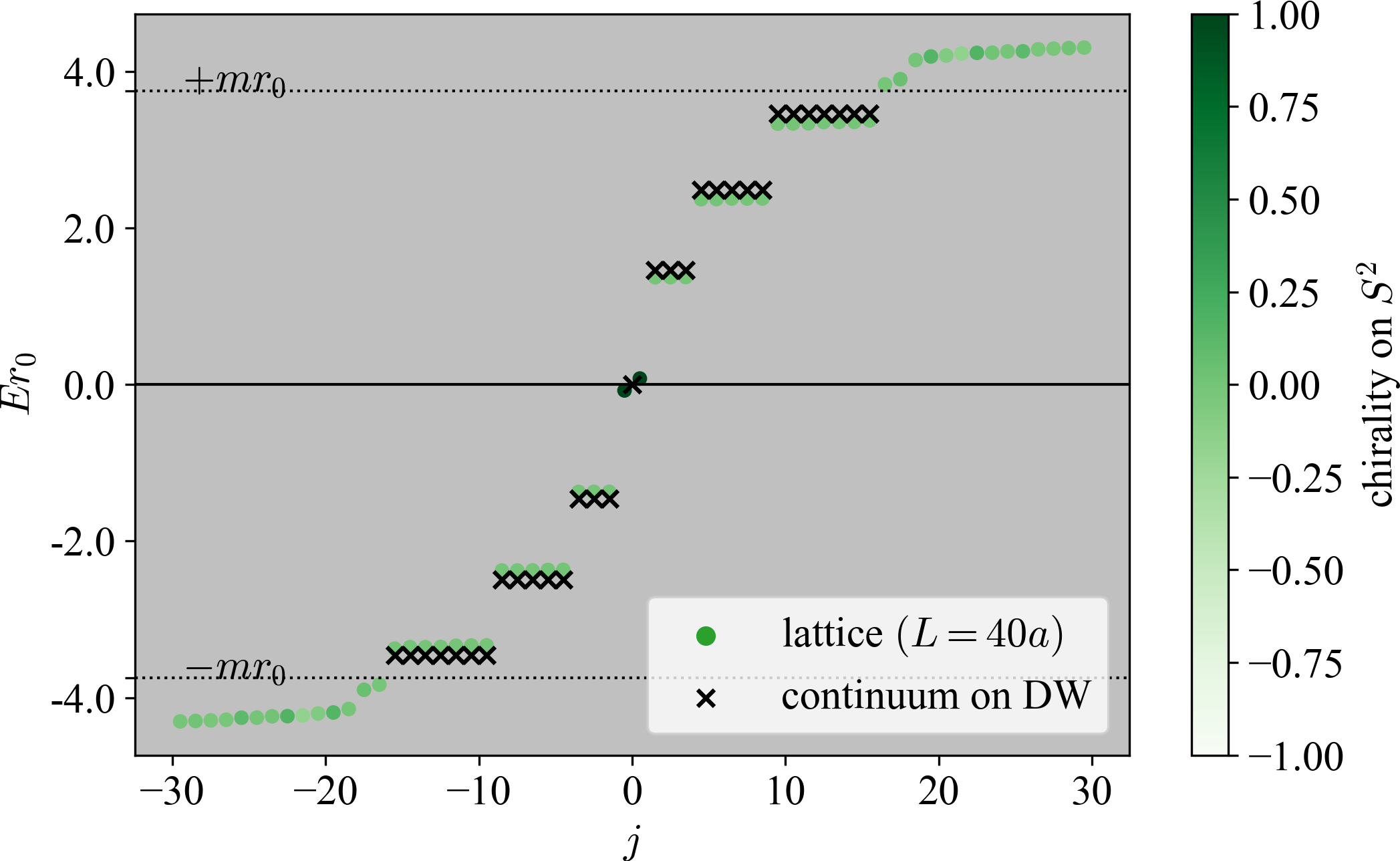}
    }\\
    \subfigure{	
    \includegraphics[bb=0 0 501 309, height=0.4 \textheight]{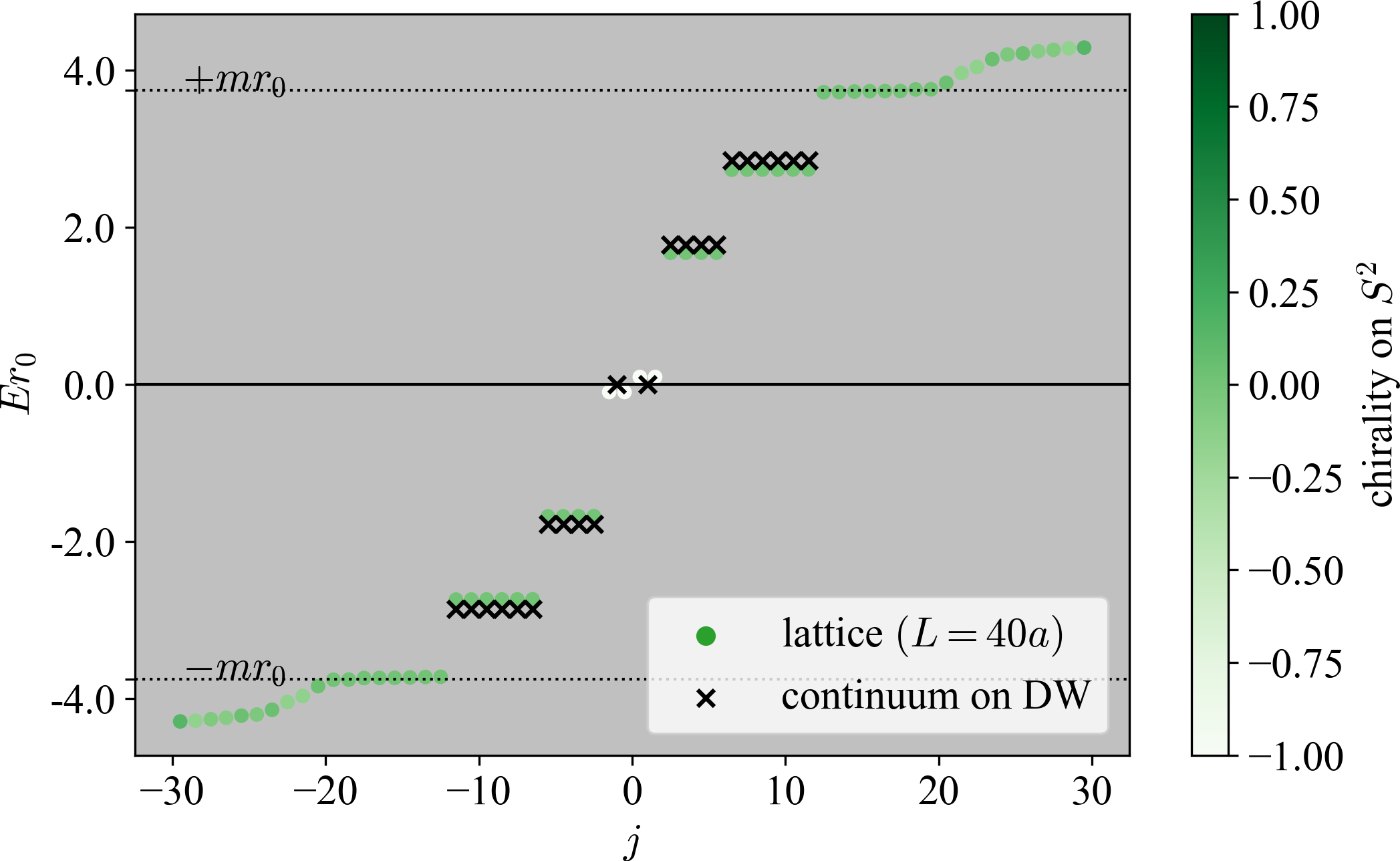}
    }
\caption{The same plots as Fig.~\ref{fig:S2_Monopole_eigenvalue_chi} but the green color gradation represents the expectation value of $\gamma^{S^2}_\text{normal}$. 
}
\label{fig:S2_Monopole_eigenvalue_chiS2}
\end{figure}

We depict the $S^2$-chirality of the eigenstate of $H$ as a green gradation in Fig.~\ref{fig:S2_Monopole_eigenvalue_chiS2}. We can see that all the zero modes are eigenstates of $\gamma_{\text{normal}}^{S^2}$.
Since half of the zero modes appear at the wall, the AS index on $S^2$ is equal to the topological index $n$ and the monopole gains the electric charge $\abs{n}/2$.   

Let us derive the monopole-localized mode in the continuum theory. We consider the regularized Dirac operator
\begin{align}
    H_\text{reg}=\bar{\gamma} \qty(\sum_{i=1}^3 \gamma^i\qty( \pdv{}{x^i}-iA_i )-\frac{1}{M_{PV}}D^2 -m ) = \mqty( -m-\frac{D^2}{M_{PV}} &  \sigma_i\qty( \pdv{}{x^i}-iA_i )\\  -\sigma_i\qty( \pdv{}{x^i}-iA_i ) & +m +\frac{D^2}{M_{PV}} )
\end{align}
in the negative mass region. As well as in the previous section, we add the second derivative term as the Wilson term and ignore the positive mass region. We find the zero mode localized at the monopole,
\begin{align}
    \psi_{j,j_3}(r,\theta,\phi)= \frac{e^{-M_{PV}r/2}}{\sqrt{r}}I_\nu (\kappa r) \mqty(1 \\ - \text{sign}(n))\otimes  \chi_{j,j_3,0} (\theta,\phi),
\end{align}
where $\nu=\frac{1}{2}\sqrt{2 \abs{n}+1},~\kappa= \frac{1}{2}M_{PV}  \sqrt{ 1- 4m/M_{PV}}$ and $I_\nu$ is the modified Bessel function of the first kind. $\chi_{j,j_3,0}$ is defined in the section \ref{subsec:Curved_conti_S2}. The chirality of this mode is exactly $-1$. Since $\nu$ is larger than $1/2$, $\psi_{j,j_3}(r,\theta,\phi)$ becomes zero at $r=0$. In the large $M_{PV}$ limit, the localized mode converges to
\begin{align}
    \psi_{j,j_3}(r,\theta,\phi)\sim \frac{e^{-mr}}{r\sqrt{\pi M_{PV}}} \mqty(1 \\ - \text{sign}(n))\otimes  \chi_{j,j_3,0} (\theta,\phi),
\end{align}
which coincides with the naive solution of the Dirac equation without the second derivative term. Furthermore, there is a peak at $r=\frac{\abs{n}}{2M_{PV}}$. It implies that the second derivative term makes a domain-wall around the monopole and improves the behavior of the localized mode. We plot the amplitude of the monopole-localized state with and without the Wilson term in Fig. \ref{fig:S2_Naive_vs_Wilson}. Here, we set $n=+1,~m=0.1$ and $M_{PV}=10$. See the appendix \ref{app:MLM} for details.

\begin{figure}
\centering
\includegraphics{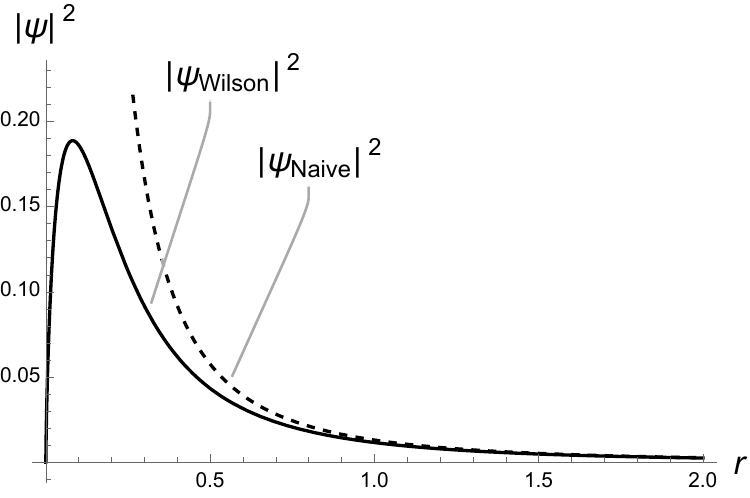}
\caption{The amplitude of the monopole-localized mode of $H_\text{reg}$ with and without the Wilson term. The parameters are set with $n=+1,~m=0.1$ and $M_{PV}=10$ \cite{Aoki:2023lqp}.}
\label{fig:S2_Naive_vs_Wilson}
\end{figure}

\chapter{Summary and Outlook}
\label{sec:Summary}

\section{Summary}




In this thesis, we have investigated curved domain-wall fermion systems on square lattices. On the $S^1$ domain-wall in two-dimensional square lattice space and the $S^2$ domain-wall in three-dimensional space, we have revealed that edge-localized modes having a positive chirality appear and feel gravity through the induced connection. We have also put $U(1)$ link variables as a $U(1)$ gauge field and analyzed the Aharanov-Bohm effect on the edge states.

In order to clarify the gravitational effect, we have first considered the domain-wall system on the lattice space without a $U(1)$ gauge field. The domain-wall on the square lattice has corners but the edge-localized states emerge as well as the continuum theory. The gravitational effect is observed in the spectrum of the Hermitian Dirac operator. There is a gap due to the positive scalar curvature of the spherical domain-wall. The spectrum agrees well with the continuum prediction. Our numerical data on the lattice converges to the continuum value as a linear function of the lattice spacing $a$. The finite volume effect is also saturated in the large volume limit. As expected, the rotational symmetry is automatically restored in the standard continuum limit.


We also add a weak and smooth $U(1)$ gauge field by link variables on the $S^1$ domain-wall system. Here, the flux is located at the center of the wall. The $U(1)$ gauge field influences the boundary system as the Aharanov-Bohm effect and causes the linear shift of the eigenvalue spectrum. In particular, when $\alpha=1/2$, the gravitational gap is eliminated and one eigenvalue crosses $E=0$. The asymmetry in the positive and negative eigenvalues spectra generates the $T$ anomaly on the boundary system. On the other hand, the chiral anomaly on the bulk caused by the violation of $T$ symmetry cancels the $T$ anomaly. On the domain-wall system, this cancellation mechanism is described by the APS index: the eta invariant of the Hermitian Dirac operator with the domain-wall mass term. We have confirmed that our lattice eta invariant coincides with the APS index on the continuum two-dimensional disk.


When we squeeze the $U(1)$ flux into one plaquette to make a vortex, the Aharanov-Bohm effect on the boundary system is the same but the singular point of the vortex gives rise to a drastic change of the bulk fermion structure. We have discovered a new localized mode at the vortex, whose chirality is opposite to the edge-localized modes on the $S^1$ domain-wall. Microscopically, the Wilson term, which regularizes the UV region on the lattice space, generates a new domain-wall around the vortex and it captures a fermion as the vortex-localized mode. Then, the negative mass region has two boundaries and the $T$ anomaly on the original wall is canceled by the vortex-localized mode rather than the topological term on the bulk.


We have applied our domain-wall method to condensed matter physics, where we have regarded the $d$-dimensional Hermitian Dirac operator as a Hamiltonian in $(d+1)$-dimensional spacetime. The negative mass region corresponds to a topological insulator phase and the positive mass region is a normal insulator phase. When we add a vortex or monopole as a defect, the Wilson term, which eliminates the UV divergence, causes the positive mass shift and locally makes a normal insulator phase around the defect. Then, by the tunneling effect, half of the amplitude of the electron is constrained there. This gives the microscopic description of how the defect obtains the fractional charge.


Mathematically, the domain-wall creation is explained by the cobordism invariant nature of the AS index or mod two index on the surface. Namely, if the disjoint union of two compact manifolds is realized as the boundary of a manifold of one higher dimension, the index of these two manifolds coincide. Now, zero modes are localized on the surface due to a defect and the index of the surface is non-zero, but a boundary must appear inside that cancels the index out. As a result, another boundary is created surrounding the defect, and the same numbers of zero modes as the outer index are localized.

\section{Outlook}

As we mentioned in sec. \ref{subsec:Higher_Codim}, the Schwarzschild spacetime can be embedded into some higher dimensional spacetimes \cite{Kasner1921}. Schwarzschild spacetime describes gravity around a massive object such as a neutron star or black hole. It is an interesting application of our domain-wall fermion formulation to a system with the Hawking radiation \cite{Hawking:1974rv}. 


According to the Nash embedding theorem, every Riemannian manifold can be isometrically embedded into some higher Euclidean space with a flat metric. However, its realization by the domain-wall is limited. 
For example, it is impossible to put an unorientable manifold\footnote{A global anomaly on unorientable manifolds has been reported in \cite{Witten:2015Fermion,Witten:2016cio}.} as a domain-wall since the orientation is manifest by the sign of the mass. 
One possibility of lattice formulation of an unorientable manifold is to construct an orbifold by identifying $x$ and $-x$ on a lattice space. This orbifold is expressed as $T^n/\mathbb{Z}_2$. Then, a spherical domain-wall on the square lattice would represent a real projective space on the orbifold lattice. When the dimension of the projective space is even, the space is unorientable. Furthermore, a projective hypersurface, which is defined as the zeros of a polynomial, could be realized as a domain-wall on the projective space.


Another interesting target is the $K3$ surface. By the Kummer construction \cite{MR0465761}, the $K3$ surface is given by blowing up $16$ fixed points to $S^2$ on $T^4/\mathbb{Z}_2$. A neighborhood of the fixed point looks like $T^\ast (\mathbb{C}P^1)$, which can be equipped with the Eguchi-Hanson (EH) metric \cite{Eguchi:1978gw}. Since the EH metric is asymptotically flat, we can construct a patchwork metric by gluing the EH metrics around the $16$ fixed points. This metric is called Page metric \cite{Page:1978plj} and the curvature is concentrated at the fixed points. Note that the method of gluing is not unique and there exists a Page metric that is closer to an Einstein metric. We will be able to naturally regularize $T^4 /\mathbb{Z}_2$ by a square lattice, but blowing up around the fixed points will be a nontrivial problem. 





\appendix

\chapter{Vortex-localized Modes in $\mathbb{R}^2$} 
\label{app:CLM}

In this appendix, we investigate the origin of the center-localized mode appearing near the vortex in a continuum analysis.

We assume that the domain-wall radius $r_0$ is infinity and put $r_1$ finite in which the $U(1)$ flux is focused. We note that we need a careful regularization of the short-range behavior of fermions near the vortex. In order to continue the dynamical creation of the domain-wall, the fermion partition function regularized by a Pauli-Villars field is given by 
\begin{align}
    Z_\text{reg}= &\lim_{M_{PV}\to \infty} \det \frac{i \sigma_j D_j -im}{ i \sigma_j D_j + iM_{PV}}
    \sim  \det \frac{1}{ M_{PV}}  ( \sigma_j D_j -m  - \frac{1}{M_{PV}} {D}^2 ),
\end{align}
where $D_j= \partial_j-iA_j$ is a covariant derivative in $j$-th direction. and we ignore $\order{1/M_{PV}^2, m/M_{PV}, F/M_{PV}} $ terms. Then, it is nontrivial to consider a modified Dirac operator 
\begin{align} \label{eq:approximation of H S1}
    H=\sigma_3 \qty( \sigma_j D_j -\frac{1}{M_{PV} } D^2 -m).
\end{align}
The second derivative term is interpreted as a contribution from the Pauli-Villars regulator. Compared to $M_{PV}$, the sign of fermion mass is well-defined \cite{Zhao2012Amagneticmonopole}. Since this operator also commutes with $J=-i\pdv{}{\theta}+ \frac{1}{2} \sigma_3$, the eigenfunction of $H$ can be written as $\psi= \mqty(f(r) e^{i(j-\frac{1}{2})\theta} \\ g(r) e^{i(j+\frac{1}{2})\theta})$.

First, we consider a solution of the eigenvalue problem in the region of $r>r_1$. Setting $A=\alpha d\theta$, we have
\begin{align}
    H=\sigma_3 \mqty( -m-\frac{D^2}{M_{PV}} & e^{-i\theta} \qty( \pdv{}{r}-i \frac{1}{r} \pdv{}{\theta} -\frac{\alpha}{r} ) \\ e^{i\theta} \qty( \pdv{}{r}+i \frac{1}{r} \pdv{}{\theta} +\frac{\alpha}{r} ) & -m -\frac{D^2}{M_{PV}} ),
\end{align}
where $D^2= \pdv[2]{}{r}+\frac{1}{r}\pdv{}{r}- \frac{1}{r^2} \qty( i\pdv{}{\theta} +\alpha)^2$. Then the equation for $f$ and $g$ is obtained 
\begin{align}
    E\mqty( f \\ g) = \mqty( -m + \frac{a_j^\dagger a_j}{M_{PV}} & a_j^\dagger \\ a_j & +m - \frac{a_j a_j^\dagger }{M_{PV}} )  \mqty( f \\ g),
\end{align}
where $E$ is the eigenvalue of $H$. $a_j$ and $a_j^\dagger$ are differential operators defined by
\begin{align}
    a_{j}=-\pdv{}{r} +\frac{j-\frac{1}{2} -\alpha}{r},~a_{j}^\dagger= \pdv{}{r} +\frac{j+\frac{1}{2} -\alpha}{r}.
\end{align}
We assume that $a_j f\propto g $ and $ a_j^\dagger g \propto f$, then the solutions are given by
\begin{align}
    f=a K_{j-\frac{1}{2}-\alpha}(\kappa r),~g=b K_{j+\frac{1}{2}-\alpha} (\kappa r),
\end{align}
which converges to zero at $r\to \infty$, exponentially. 

$a,~b$ and $\kappa $ are complex numbers and $\Re(\kappa)>0$. The coefficients $a$ and $b$ satisfy
\begin{align}
E\mqty(a\\ b)=  \mqty(  -m -\frac{\kappa^2}{M_{PV}} & -\kappa \\ +\kappa & m +\frac{\kappa^2}{M_{PV}}) \mqty(a\\ b).
\end{align}
and $\kappa$ satisfies 
\begin{align}
    0=E^2 -\qty(m+ \frac{\kappa^2}{M_{PV}})^2+ \kappa^2.
\end{align}
With its two solutions
\begin{align}
    \kappa_{\pm}= M_{PV} \sqrt{\frac{1}{2}\qty(1 \pm \sqrt{ 1-4 \frac{m}{M_{PV}} + 4 \frac{E^2}{M_{PV}^2}} )-\frac{m}{M_{PV}}  },
\end{align}
the two eigenstates are given by
\begin{align}
    \psi_{o\pm}=\mqty(\qty(-m -\frac{\kappa_{\pm}^2}{M_{PV}} +E) K_{j-\frac{1}{2}-\alpha}(\kappa_{\pm} r) e^{i(j-\frac{1}{2})\theta} \\ \kappa_{\pm} K_{j+\frac{1}{2}-\alpha}(\kappa_{\pm} r) e^{i(j+\frac{1}{2})\theta} )
\end{align}
for $r>r_1$.

Second, we solve the eigenvalue problem for $r<r_1$, where the $U(1)$ gauge flux is strong. In this region, the Dirac operator is given by
\begin{align}
    H=\sigma_3 \mqty( -m-\frac{D^2}{M_{PV}} & e^{-i\theta} \qty( \pdv{}{r}-i \frac{1}{r} \pdv{}{\theta} -\frac{\alpha r}{r_1^2} ) \\ e^{i\theta} \qty( \pdv{}{r}+i \frac{1}{r} \pdv{}{\theta} +\frac{\alpha r}{r_1^2} ) & -m -\frac{D^2}{M_{PV}} ),
\end{align}
where $D^2=\pdv[2]{}{r}+\frac{1}{r}\pdv{}{r} -\frac{1}{r^2} \qty(i\pdv{}{\theta}+ \alpha \frac{r^2}{r_1^2} )^2$. We put the same ansatz for $\psi$, then we get an equation for $f$ and $g$:
\begin{align}
    E\mqty( f \\ g) = \mqty( -m + \frac{1}{M_{PV}}\qty( a_j^\dagger a_j+ 2\frac{\alpha}{r_1^2}) & a_j^\dagger \\ a_j & m - \frac{1}{M_{PV}}\qty(a_j a_j^\dagger - 2\frac{\alpha}{r_1^2}) )  \mqty( f \\ g).
\end{align}
Here the derivative operators $a_j$ and $a_j^\dagger$ are defined by
\begin{align}
    a_{j}=-\pdv{}{r} +\frac{1 }{r} \qty(j-\frac{1}{2} - \alpha \frac{r^2}{r_1^2}) ,~a_{j}^\dagger= \pdv{}{r} + \frac{1 }{r}\qty(j+\frac{1}{2} - \alpha \frac{r^2}{r_1^2}).
\end{align}
Suppose that $f$ has no singularity and is written as $aF$, where $a$ is a complex constant value and $F$ is an eigenfunction of $a_j^\dagger a_j F =-LF$. Then $F$ is written as
\begin{align}
    F=\left\{ \begin{array}{ll}
        r^{j-\frac{1}{2}} e^{- \frac{ \alpha r^2}{ 2r_1^2}} {}_1 F_1(  \frac{r_1^2}{4 \alpha} L, j+ \frac{1}{2}; \alpha \frac{r^2}{r_1^2}) & (j=\frac{1}{2}, \frac{3}{2},\cdots) \\
       r^{-j+\frac{1}{2}} e^{- \frac{ \alpha r^2}{ 2r_1^2}} {}_1 F_1(-j+\frac{1}{2} + \frac{r_1^2}{4 \alpha} L, -j+ \frac{3}{2}; \alpha \frac{r^2}{r_1^2})  & (j=-\frac{1}{2}, -\frac{3}{2},\cdots)
    \end{array} \right. ,
\end{align}
where ${}_1 F_1(a,b;z)=\sum_{n=0}^\infty \frac{\Gamma( a+n)}{ \Gamma( a)} \frac{\Gamma( b)}{\Gamma( b+n)} \frac{z^n}{ n!}$ is a confluent hypergeometric function. It is a solution of a confluent hypergeometric function:
\begin{align}
    \qty(z \dv[2]{}{z} + (c_2-z) \dv{}{z} -c_1) {}_1 F_1(c_1,c_2;z) =0
\end{align}
for complex constants $c_1$ and $c_2$.

We also assume that $g=b a_j F$, then $a_j,~a_j^\dagger, a_j^\dagger a_j$ and $a_j a_j^\dagger$ turn out to be constants and we acquire an equation for $a$ and $b$:
\begin{align}
    E\mqty( a \\ b) = \mqty( -m + \frac{1}{M_{PV}}\qty( -L+ 2\frac{\alpha}{r_1^2}) & -L\\ 1 & +m + \frac{1}{M_{PV}}\qty(L + 2\frac{\alpha}{r_1^2}) )  \mqty( a \\ b).
\end{align}
$E$ is an eigenvalue of the matrix of the left-hand side so $E$ and $L$ must satisfy
\begin{align}
    0=\qty(E - \frac{2\alpha}{ M_{PV} r_1^2})^2 -\qty(m+ \frac{L}{M_{PV}})^2+ L.
\end{align}
We find two solutions for $L$:
\begin{align}
    L_{\pm}=M_{PV}^2 \qty( \frac{1}{2} \qty( 1 \pm  \sqrt{ 1-4\frac{m}{M_{PV}} + 4 \frac{ \qty(E- \frac{2\alpha}{M_{PV} r_1^2})^2}{M_{PV}^2} }) - \frac{m}{M_{PV}}).
\end{align}
and
\begin{align}
    \psi_{i\pm}=\mqty(\qty(-m -\frac{L_\pm}{M_{PV}} +E- \frac{2\alpha}{ M_{PV} r_1^2}) F(r) e^{i(j-\frac{1}{2})\theta} \\ a_j F(r) e^{i(j+\frac{1}{2})\theta} ).
\end{align}

Finally, we solve the continuity problem of the eigenstate $\psi $ and its derivative in the $r$-direction $\pdv{}{r} \psi$ at $r=r_1$. This determines the eigenvalue $E$ as well as the coefficients of
\begin{align}
    \psi= \left\{ \begin{array}{cc}
        c_1 \psi_{i+}+ c_2 \psi_{i-}  & (r<r_1) \\
        d_1 \psi_{o+}+ d_2 \psi_{o-} &  (r>r_1)
    \end{array} \right. ,
\end{align}
up to normalization.

From the assumption, $D^2\psi$ is also a continuous function at $r=r_0$ because $D^2 \psi$ is expressed by
\begin{align}
    D^2 \psi= M_{PV} \qty( \sigma_j D_j -m  -\sigma_3 H) \psi= M_{PV} \qty( \sigma_j D_j -m  -\sigma_3 E) \psi.
\end{align}
Therefore, we replace the continuation condition for $\partial_r\psi$ with that for $D^2 \psi$, with which the analysis is simpler. Since the $U(1)$ gauge field is singular, we assume a hierarchical limit $1/r_1 \gg M_{PV} \gg m$. For $j>0$, the functions $\psi_{i \pm },~\psi_{o \pm },~D^2\psi_{i \pm }$ and $D^2\psi_{o \pm }$ are approximated in this limit as





\begin{align}
    \psi_{i+}& \simeq \mqty( -\frac{4\alpha}{ M_{PV} r_1^2 }   {}_1 F_1(+\frac{1}{2}, j+ \frac{1}{2}; \alpha \frac{r^2}{r_1^2}) e^{i(j-\frac{1}{2})\theta} \\ -\frac{\alpha}{ r_1 (j+1/2) } {}_1 F_1(+\frac{3}{2}, j+ \frac{1}{2}; \alpha \frac{r^2}{r_1^2}) e^{i(j+\frac{1}{2})\theta} ) r^{ j -\frac{1}{2}}e^{- \frac{ \alpha r^2}{ 2r_1^2}}, \\
    D^2 \psi_{i+} & \simeq \mqty( \frac{\alpha}{ M_{PV} r_1^2 } M_{PV}^2  {}_1 F_1(+\frac{1}{2}, j+ \frac{1}{2}; \alpha \frac{r^2}{r_1^2}) e^{i(j-\frac{1}{2})\theta} \\ -\frac{\alpha}{ r_1 (j+1/2) } \frac{4\alpha}{r_1^2} {}_1 F_1(+\frac{3}{2}, j+ \frac{1}{2}; \alpha \frac{r^2}{r_1^2}) e^{i(j+\frac{1}{2})\theta} ) r^{ j -\frac{1}{2}} e^{- \frac{ \alpha r^2}{ 2r_1^2}},
\end{align}

\begin{align}
    \psi_{i-}& \simeq \mqty(-\frac{M_{PV}}{4}   {}_1 F_1(-\frac{1}{2}, j+ \frac{1}{2}; \alpha \frac{r^2}{r_1^2}) e^{i(j-\frac{1}{2})\theta} \\ \frac{\alpha}{ r_1 (j+1/2) } {}_1 F_1(+\frac{1}{2}, j+ \frac{1}{2}; \alpha \frac{r^2}{r_1^2}) e^{i(j+\frac{1}{2})\theta} )r^{ j -\frac{1}{2}} e^{- \frac{ \alpha r^2}{ 2r_1^2}}, \\
    D^2 \psi_{i-} & \simeq \mqty( -\frac{M_{PV}}{4} \frac{4\alpha}{r_1^2}  {}_1 F_1(-\frac{1}{2}, j+ \frac{1}{2}; \alpha \frac{r^2}{r_1^2}) e^{i(j-\frac{1}{2})\theta} \\ -\frac{\alpha}{ 4 r_1 (j+1/2) } M_{PV}^2 {}_1 F_1(+\frac{1}{2}, j+ \frac{1}{2}; \alpha \frac{r^2}{r_1^2}) e^{i(j+\frac{1}{2})\theta} ) r^{ j -\frac{1}{2}} e^{- \frac{ \alpha r^2}{ 2r_1^2}},
\end{align}

\begin{align}
\psi_{o+} &\simeq \mqty( -M_{PV} K_{j-\frac{1}{2}-\alpha}(M_{PV} r) e^{i(j-\frac{1}{2})\theta} \\ M_{PV} K_{j+\frac{1}{2}-\alpha}( M_{PV} r) e^{i(j+\frac{1}{2})\theta} ),\\
D^2\psi_{o+} &\simeq \mqty( (M_{PV})^3 K_{j-\frac{1}{2}-\alpha}(M_{PV} r) e^{i(j-\frac{1}{2})\theta} \\ -(M_{PV})^3 K_{j+\frac{1}{2}-\alpha}( M_{PV} r) e^{i(j+\frac{1}{2})\theta} ),
\end{align}

\begin{align}
\psi_{o-} &\simeq \mqty( (-m+E) K_{j-\frac{1}{2}-\alpha}(\sqrt{m^2 -E^2} r) e^{i(j-\frac{1}{2})\theta} \\ \sqrt{m^2 -E^2} K_{j+\frac{1}{2}-\alpha}(\sqrt{m^2 -E^2} r) e^{i(j+\frac{1}{2})\theta} ),\\
D^2\psi_{o-} &\simeq \mqty( -(-m+E) \sqrt{m^2 -E^2}^2 K_{j-\frac{1}{2}-\alpha}(\sqrt{m^2 -E^2} r) e^{i(j-\frac{1}{2})\theta} \\ -\sqrt{m^2 -E^2}^3 K_{j+\frac{1}{2}-\alpha}(\sqrt{m^2 -E^2} r) e^{i(j+\frac{1}{2})\theta} ).
\end{align}

The eigenvalue $E$ is determined by
\small
\begin{align}
    \det \mqty( -\frac{4\alpha}{ M_{PV} r_1^2 } u_{\frac{1}{2}}  & -\frac{M_{PV}}{4} v_{-\frac{1}{2}}   & -M_{PV} k_0( M_{PV} r_1)   & (-m+E) k_0( \sqrt{m^2 -E^2}  r_1) \\
-\frac{\alpha}{ r_1 (j+1/2) } u_{\frac{3}{2}}  & \frac{\alpha}{ r_1 (j+1/2) }  v_{\frac{1}{2}} & M_{PV}  k_1(M_{PV} r_1) & \sqrt{m^2 -E^2} k_1 (\sqrt{m^2 -E^2} r_1) \\
\frac{\alpha M_{PV}}{ r_1^2 }  u_\frac{1}{2} & -\frac{\alpha M_{PV}}{ r_1^2 } v_{-\frac{1}{2}} & (M_{PV})^3  k_0(M_{PV} r_1) & -(-m+E) \sqrt{m^2 -E^2}^2 k_0(\sqrt{m^2 -E^2} r_1) \\
-\frac{4\alpha^2}{ r_1^3 (j+1/2) } u_\frac{3}{2} & -\frac{\alpha M_{PV}^2}{ 4 r_1 (j+1/2) } v_\frac{1}{2} & -(M_{PV})^3 k_1(M_{PV} r_1)  &  -\sqrt{m^2 -E^2}^3 k_1(\sqrt{m^2 -E^2} r_1)  )=0,
\end{align}
\normalsize
where $u_{\frac{1}{2}+n}= {}_1 F_1(+\frac{1}{2}+n, j+ \frac{1}{2}+n; \alpha )e^{- \frac{ \alpha }{ 2}} r_1^{ j -\frac{1}{2}} \simeq r_1^{ j -\frac{1}{2}},~v_{-\frac{1}{2}+n} = {}_1 F_1(-\frac{1}{2}+n, j+ \frac{1}{2}+n; \alpha )e^{- \frac{ \alpha }{ 2}} r_1^{ j -\frac{1}{2}} \simeq r_1^{ j -\frac{1}{2}}$ and $k_n (z)= K_{j-\frac{1}{2}+n-\alpha}(z)\simeq \frac{ \Gamma( \abs{j-\frac{1}{2}+n-\alpha })}{2}\qty( \frac{2}{z })^{\abs{j-\frac{1}{2}+n-\alpha }}  ~(n=0,1) $. The determinant on the left-hand side is a polynomial of $1/r_1,~M_{PV}$ and $m$. Since $1/r_1$ is the largest parameter, we ignore all terms except for the most dominant term, which leads to a simpler equation
\begin{align}
    0= \frac{4\alpha^2}{ r_1^3 (j+1/2) } u_\frac{3}{2}  \frac{\alpha M_{PV}}{ r_1^2 } v_{-\frac{1}{2}} \det \mqty( -M_{PV} k_0( M_{PV} r_1)   & (-m+E) k_0( \sqrt{m^2 -E^2}  r_1) \\  M_{PV}  k_1(M_{PV} r_1) & \sqrt{m^2 -E^2} k_1 (\sqrt{m^2 -E^2} r_1) ).
\end{align}
This condition is equivalent to that $\psi_{o+}$ and $\psi_{o-}$ are parallel to each other. Since $M_{PV} r_1 $ and $ \sqrt{m^2 -E^2}r_1 $ are small enough, the eigenvalue $E$ is determined by
\begin{align}
    \frac{-m+E}{\sqrt{m^2-E^2} } \qty(\frac{\sqrt{m^2-E^2}}{M_{PV}})^{ \abs{j+\frac{1}{2}-\alpha }-\abs{j-\frac{1}{2}-\alpha } } +1=0 \label{eq:energy condition of localized mode}.
\end{align}
Only when $m>0$, there is a solution $E$. The same condition is obtained for $j<0$, too. Here, the $\psi_{i+}$ and $D^2\psi_{i+}$ are written as
\begin{align}
    \psi_{i+}& \simeq \mqty( -\frac{4\alpha}{ M_{PV} r_1^2 }   {}_1 F_1(-j+1, -j+ \frac{3}{2}; \alpha \frac{r^2}{r_1^2}) e^{i(j-\frac{1}{2})\theta} \\ -2\frac{-j+\frac{1}{2}}{r} {}_1 F_1(-j+1, -j+ \frac{1}{2}; \alpha \frac{r^2}{r_1^2}) e^{i(j+\frac{1}{2})\theta} ) r^{ -j +\frac{1}{2}}e^{- \frac{ \alpha r^2}{ 2r_1^2}} ,\\
    D^2 \psi_{i+} & \simeq \mqty( \frac{\alpha}{ M_{PV} r_1^2 } M_{PV}^2 {}_1 F_1(-j+1, -j+ \frac{3}{2}; \alpha \frac{r^2}{r_1^2}) e^{i(j-\frac{1}{2})\theta} \\ -2\frac{-j+\frac{1}{2}}{r} \frac{4\alpha}{r_1^2} {}_1 F_1(-j+1, -j+ \frac{1}{2}; \alpha \frac{r^2}{r_1^2}) e^{i(j+\frac{1}{2})\theta} ) r^{- j +\frac{1}{2}} e^{- \frac{ \alpha r^2}{ 2r_1^2}}
\end{align}
and the dependence of $r_1$ is the same as $ j>0$. 

Let us solve the equation \eqref{eq:energy condition of localized mode} and determine the eigenvalue $E$. $\abs{j+\frac{1}{2}-\alpha }-\abs{j-\frac{1}{2}-\alpha } $ takes three values
\begin{align}
    \abs{j+\frac{1}{2}-\alpha }-\abs{j-\frac{1}{2}-\alpha } =\left\{ \begin{array}{cc}
        +1 &  (j-\frac{1}{2}>\alpha) \\
        2(j-\alpha) & (j+\frac{1}{2} >\alpha> j-\frac{1}{2})  \\
        -1 & (j+\frac{1}{2}<\alpha)
    \end{array} \right. .
\end{align}
When $j -1/2> \alpha$ or $j+\frac{1}{2}<\alpha$, we get $E= M_{PV}+m$. However, it is contradict with $E^2<m^2$ so a  solution of \eqref{eq:energy condition of localized mode} exists only when $j+\frac{1}{2} >\alpha> j-\frac{1}{2}$, i.e. $j= [\alpha]+\frac{1}{2}$. $E$ is an odd function of $\alpha-[\alpha]-1/2$ and expressed as the inverse function of
\begin{align}\label{eq:condition of E S1 center}
    \alpha -[\alpha]= \frac{1}{2} \frac{ \log ( \frac{m-E}{2 M_{PV} })}{ \log ( \frac{ \sqrt{m^2 -E^2 }}{2 M_{PV} }) }.
\end{align}
Thus we obtain the eigenvalue $E$ as a function of $\alpha$:
\begin{align}
    E\simeq \left\{ \begin{array}{cc}
        -m & ( \alpha -[\alpha] \sim 0 ) \\
        \abs{ 2m \log( \frac{m}{ M_{PV}}) }\qty( \alpha-[\alpha] -\frac{1}{2}) & ( \alpha -[\alpha] \sim \frac{1}{2} )   \\
        m &  ( \alpha -[\alpha] \sim 1 )
    \end{array} \right. 
\end{align}
when $1/r_1 \gg M_{PV} \gg m$. Setting $M_{PV}=2/a$, this formula describes well the lattice result as shown in Fig.~\ref{fig:AnomalyInflow_singular}.

\chapter{Monopole-localized Modes in $\mathbb{R}^3$} 
\label{app:MLM}

Let us consider the $S^2$ domain-wall system in the presence of the monopole with $\abs{n} >0$ and find a monopole-localized mode in continuum theory. Here we use the same notation as in the section \ref{subsec:Curved_conti_S2}.

As well as the $S^1$ domain-wall case, we analyze a Hermitian Dirac operator regularized by a Pauli Villars field with the mass $M_{PV}$, 
\begin{align}
    H_\text{reg}=\bar{\gamma} \qty(\sum_{i=1}^3 \gamma^i\qty( \pdv{}{x^i}-iA_i )-\frac{1}{M_{PV}}D^2 -m ) = \mqty( -m-\frac{D^2}{M_{PV}} &  \sigma_i\qty( \pdv{}{x^i}-iA_i )\\  -\sigma_i\qty( \pdv{}{x^i}-iA_i ) & +m +\frac{D^2}{M_{PV}} ).
\end{align}
Here we ignore the positive mass region and add the second derivative term as the Wilson term. Using the equation \eqref{eq:sigma dot nabla S2}, we get 
\begin{align}
    D^2= \qty(\sigma_i\qty( \pdv{}{x^i}-iA_i ))^2 -\frac{n}{2r^2} \sigma_r =\qty( \pdv{}{r} +\frac{1}{r})^2 - \frac{D^{S^2}(D^{S^2}+1)}{r^2}-\frac{n}{2r^2} \sigma_r,
\end{align}
where $D^{S^2}$ is defined in the equation \eqref{eq:DS2}. $D^{2}$ and $H_\text{reg}$ commute with $J_i$ and $1 \otimes J_i$, respectively. However, $D^2$ does not commute with $D^{S^2}$:
\begin{align}
    [D^2, D^{S^2}] =- \frac{n}{r^2} \sigma_r D^{S^2} .
\end{align} 
For a zero mode of $\sigma_3 \otimes D^{S^2}$, it can be a eigenstate of $H_{\text{reg}}$.

The functional form of the zero mode of $\sigma_3 \otimes D^{S^2}=0$ is written by $\psi(r,\theta,\phi)= \mqty( f(r)\\g(r)) \otimes \chi_{j,j_3,0} (\theta,\phi)$, where $j= \frac{\abs{n}-1}{2}$ and $j_3= -j,\cdots ,j-1,j$. Then, the eigenvalue problem $H_{\text{reg}} \psi=E \psi$ turns into
\begin{align}
    E \mqty( f\\g) = \qty[-\sigma_3 \qty( m+ \frac{1}{M_{PV}}\qty( \qty(\pdv{}{r} +\frac{1}{r})^2 -\frac{\abs{n}}{2r^2} ) )+ i\sigma_2 \text{sign}(n) \qty( \pdv{}{r} + \frac{1}{r})]\mqty( f\\g). 
\end{align}
We focus on the zero mode localized at the monopole. The zero mode is also an eigenstate of $\sigma_1$, then we obtain 
\begin{align}
    \psi(r,\theta,\phi)=  \frac{e^{-M_{PV}r/2}}{\sqrt{r}}(A K_\nu (\kappa r)+B I_\nu (\kappa r) ) \mqty(1 \\ - \text{sign}(n))\otimes  \chi_{j,j_3,0} (\theta,\phi),
\end{align}
where $A$ and $B$ are complex coefficients, $\nu=\frac{1}{2}\sqrt{2 \abs{n}+1},~\kappa= \frac{1}{2}M_{PV}  \sqrt{ 1- 4m/M_{PV}}$, $I_\nu$ is the modified Bessel function of the first kind and $K_\nu$ is that of the second kind. In the limit of $M_{PV}\to \infty$, the asymptotic form of $\psi$ converges to 
\begin{align}
    \psi(r,\theta,\phi)\sim \frac{1}{\sqrt{  M_{PV} }r} \qty[ A \sqrt{\pi}e^{-M_{PV}r} + B \frac{1}{ \sqrt{\pi}} e^{-mr}] \mqty(1 \\ - \text{sign}(n))\otimes  \chi_{j,j_3,0} (\theta,\phi).
\end{align}
On the other hand, $\psi$ is divergent at $r=0$ as $r^{-1/2 -\nu}$ duo to the first term. Assuming that $\psi$ is a regular function around $r=0$, we acquire $A=0$ and
\begin{align}
    \psi(r,\theta,\phi)=&B \frac{e^{-M_{PV}r/2}}{\sqrt{r}}I_\nu (\kappa r) \mqty(1 \\ - \text{sign}(n))\otimes  \chi_{j,j_3,0} (\theta,\phi) \nonumber \\
    \simeq & B \frac{1}{\Gamma(\nu+1) \sqrt{r}} \qty(\frac{M_{PV}r}{4})^\nu  \mqty(1 \\ - \text{sign}(n))\otimes  \chi_{j,j_3,0} (\theta,\phi).
\end{align}
Here, $B$ is determined by the normalization. Since $\nu$ is larger than $1/2$, $\psi (r=0)$ is equal to zero. If $m<0$, this solution is not normalizable and there is no localized state at the monopole \cite{Zhao2012Amagneticmonopole}.

The effective mass for $\psi$ is given by
\begin{align}
    M_{\text{eff}}:= \psi^\dagger\qty(-m -\frac{D^2}{M_{PV}}) \psi/  \psi^\dagger \psi=  \frac{\frac{1}{2} +\nu}{r} -\frac{M_{PV}}{2}+\kappa \frac{I_{\nu+1} (\kappa r)}{ I_{\nu} (\kappa r)}.
\end{align}
Here, the last term $\frac{I_{\nu+1} (\kappa r)}{ I_{\nu} (\kappa r)}$ is a linear function around $r=0$. In the large $M_{PV}$ limit, $M_\text{eff}$ behaves as
\begin{align}
    M_{\text{eff}}\sim\left\{\begin{array}{cc}
        \frac{\frac{1}{2} +\nu}{r} & (r\sim 0)\\
        -m& (r\sim \infty)
    \end{array}\right. 
\end{align}
 and the new domain-wall appears near $r=\sqrt{ \frac{\abs{n}}{8 mM_{PV}} }$.

\bibliographystyle{JHEP}
\bibliography{ref}

\end{document}